\documentclass[lettersize,journal]{IEEEtran}
\usepackage{amsmath,amsfonts}
\usepackage{algorithmic}
\usepackage{algorithm}
\usepackage{array}
\usepackage[caption=false,font=normalsize,labelfont=sf,textfont=sf]{subfig}
\usepackage{textcomp}
\usepackage{stfloats}
\usepackage{url}
\usepackage{verbatim}
\usepackage{graphicx}
\usepackage{cite}
\usepackage{multirow}
\usepackage{booktabs}
\usepackage{xcolor}
\usepackage{enumitem}
\usepackage{comment}
\usepackage{float} 
\usepackage{array} 
\begin{document}

\title{Edge AI: A Taxonomy, Systematic Review and Future Directions}

\author{Sukhpal Singh Gill, Muhammed Golec, Jianmin Hu, Minxian Xu, Junhui Du, \\ Huaming Wu, Guneet Kaur Walia, Subramaniam Subramanian Murugesan, Babar Ali,  \\ Mohit Kumar, Kejiang Ye, Prabal Verma, Surendra Kumar, Felix Cuadrado, Steve Uhlig 

\thanks{{Sukhpal Singh Gill, Muhammed Golec, Subramaniam Subramanian Murugesan, Babar Ali and Steve Uhlig are with the School of Electronic Engineering and Computer Science, Queen Mary University of London, London, UK. Muhammed Golec is also with Abdullah Gul University, Kayseri, Turkey. Emails: \{s.s.gill, m.golec, s.subramanianmurugesan, b.ali, steve.uhlig\}@qmul.ac.uk}, {Jianmin Hu, Minxian Xu, Kejiang Ye are with Shenzhen Institute of Advanced Technology, Chinese Academy of Sciences, Shenzhen, China. Emails: \{jm.hu, mx.xu, kj.ye\}@siat.ac.cn}, {Junhui Du, Huaming Wu are with Center for Applied Mathematics, Tianjin University, Tianjin, China. Emails: \{dujunhui\_0325, whming\}@tju.edu.cn}, {Guneet Kaur Walia, Mohit Kumar are with Department of Information Technology, National Institute of Technology, Jalandhar, India. Emails: \{guneetkw.it.22, kumarmohit\}@nitj.ac.in}, {Prabal Verma is with Department of Information Technology, National Institute of Technology, Srinagar, India. Email: prabal.verma@nitsri.ac.in}, {Surendra Kumar is with the Department of Computer Engineering and Applications, GLA University, Mathura, India. Email: surendra.kumar@gla.ac.in}, {Felix Cuadrado is with the Technical University of Madrid (UPM), Spain. Email: felix.cuadrado@upm.es}. {All the authors contributed equally. Corresponding Author: Muhammed Golec.}}}

 \markboth{Accepted Preprint Version for Publication in Springer Cluster Computing, 2024}%
{Shell \MakeLowercase{\textit{\textit{et al.}}}: A Sample Article Using IEEEtran.cls for IEEE Journals}


\maketitle

\begin{abstract}
Edge Artificial Intelligence (AI) incorporates a network of interconnected systems and devices that receive, cache, process, and analyze data in close communication with the location where the data is captured with AI technology. Recent advancements in AI efficiency, the widespread use of Internet of Things (IoT) devices, and the emergence of edge computing have unlocked the enormous scope of Edge AI. Edge AI aims to optimize data processing efficiency and velocity while ensuring data confidentiality and integrity. Despite being a relatively new field of research from 2014 to the present, it has shown significant and rapid development over the last five years. This article presents a systematic literature review for Edge AI to discuss the existing research, recent advancements, and future research directions. We created a collaborative edge AI learning system for cloud and edge computing analysis, including an in-depth study of the architectures that facilitate this mechanism. The taxonomy for Edge AI facilitates the classification and configuration of Edge AI systems while examining its potential influence across many fields through compassing infrastructure, cloud computing, fog computing, services, use cases, ML and deep learning, and resource management. This study highlights the significance of Edge AI in processing real-time data at the edge of the network. Additionally, it emphasizes the research challenges encountered by Edge AI systems, including constraints on resources, vulnerabilities to security threats, and problems with scalability. Finally, this study highlights the potential future research directions that aim to address the current limitations of Edge AI by providing innovative solutions.
\end{abstract}

\begin{IEEEkeywords}
Edge Computing, Artificial Intelligence, Cloud Computing, Machine Learning, Edge AI 
\end{IEEEkeywords}

\section{Introduction}
 Recent advancements in Artificial Intelligence (AI), the growing adoption of Internet of Things (IoT) devices, and the rise of edge computing are converging to unleash the full potential of edge AI \cite{10335918}. Numerous analysts and businesses are conversing about and executing edge computing, which delineates its origins to the 1990s when edge servers positioned near customers were used to serve web and video content over content delivery networks \cite{ding2022roadmap}. Edge computing is a paradigm transformation in this edge AI that brings data storage and processing closer to the data source, improving response times and reducing bandwidth usage. Unlike traditional cloud computing, where centralized data centers process data, edge computing processes data at the network's edge \cite{golec2024computing}. This proximity reduces latency, enhances real-time data processing capabilities, and supports the expansion of IoT devices and services\cite{iftikhar2023ai}. The primary advantages of edge computing include improved agility of services, low latency, enhanced coherence, and the elimination of a single point of failure, making it highly relevant for applications in smart cities, self-sufficient vehicles, and industrial automation \cite{duan2023combining}. By distributing resources geographically, edge computing ensures that data processing occurs near the data source, satisfying the need for analytics and decision-making in real-time.

On the other hand, AI includes a wide array of technologies and methodologies that enable machines to carry out tasks that generally require human intelligence, such as learning, reasoning, and self-correction \cite{singh2023edge}. AI's applications span various domains, including healthcare, finance, transportation, etc, where it is used to analyze large datasets, automate tasks, and provide predictive insights \cite{shi2020communication}. Integrating AI into different sectors has revolutionized processes by enhancing efficiency, improving decision-making, and creating new opportunities for innovation. With betterment in Machine Learning (ML) or Deep Learning (DL), AI approaches have become increasingly competent in performing complex tasks that require human-like cognitive functions \cite{liu2022bringing}. AI algorithms, specifically those involving neural networks, have shown remarkable success in areas like image and speech recognition, autonomous driving, and predictive maintenance.

\subsection{Edge AI}
The fusion of edge computing and AI involves processing AI algorithms on users' devices, offering benefits like reduced latency, energy efficiency, and real-time applications. This integration allows for real-time data processing and decision-making at the source, significantly declining latency and bandwidth use \cite{rocha2024edge}. The combination of edge computing and AI enables the development of smarter and more responsive applications, such as autonomous vehicles, industrial IoT, smart home systems, etc. By leveraging edge AI, organizations can achieve greater efficiency, enhanced privacy, and faster insights, driving innovation across various sectors \cite{su2022ai}. Edge AI refers to integrating AI capabilities at the network edge, enabling distributed intelligence with edge devices. It intends to improve network connectivity, enable deployment of AI pipelines with defined quality targets, and allow adaption for data-driven applications.\cite{zhang2022edge}. Embedding AI functionalities at the edge addresses the limitations of cloud-based processing for IoT, such as privacy concerns and network connectivity issues. The deployment of AI at the edge enhances latency-sensitive tasks and reduces network congestion, improving efficiency and security in wireless networks.

Furthermore, AI-based technologies play a vital role in addressing Quality of Service (QoS)-aware scheduling and resource allocation challenges in edge environments, ensuring QoS and user experience. Edge AI enables the deployment of AI as a Service (AIaaS) with configurable model complexity and data quality, enhancing performance and reducing costs \cite{qureshi2023towards,golec2023qos}. This innovative approach supports smart security applications by leveraging AI capabilities at the edge and enhancing security measures for distributed systems. Edge intelligence, a promising technology, empowers real-time applications by moving computing from cloud servers to IoT edge devices, creating intelligent enterprises with vast possibilities \cite{shahriar2023survey}. The utilization of AI at the edge, instead of centralized locations, unlocks the potential of AI with IoT devices and edge computing, deploying AI algorithms on resource-constrained edge devices for various applications like autonomous vehicles, healthcare, and surveillance.

Edge AI's significance is underscored by its ability to provide immediate insights and actions without sending significant amounts of data to several centralized locations \cite{kumar2023ai}. This capability is particularly critical in scenarios where latency and bandwidth are significant constraints, such as in autonomous driving, where decisions must be made in real time, or in healthcare, where patient data must be processed quickly to provide timely interventions \cite{hoffpauir2023survey}. The rise of edge AI is also fueled by advancements in hardware, such as more powerful and energy-efficient processors, which make it feasible to run sophisticated AI models on devices like smartphones and IoT sensors \cite{ssgillquantum}.

\subsection{Need of Edge AI}

The motivation for integrating edge computing with AI is multifaceted, primarily driven by the imperative need for processing data in real time and navigating the inherent limitations of centralized cloud computing systems \cite{huang2023energy}. As we witness an exponential rise in the number of connected devices and a corresponding surge in data volume, traditional cloud-centric models increasingly grapple with issues such as latency, bandwidth constraints, and significant data privacy concerns. Edge AI emerges as a pivotal solution to these challenges, advocating for localized data processing \cite{verma2024data}. This shift not only diminishes the reliance on distant cloud infrastructures, thereby slashing latency, but also significantly bolsters the responsiveness of applications to real-time data inputs. This paradigm shift is particularly pivotal for fueling the development of next-gen technologies that necessitate instantaneous data analysis and decision-making, encompassing sectors like autonomous vehicles, smart city infrastructures, and cutting-edge healthcare systems.\\
Moreover, Edge AI empowers applications to operate remarkably efficiently, even in scenarios characterized by sparse connectivity, by facilitating data processing directly at the source. This capability is indispensable in remote or highly mobile environments where consistent and reliable internet access is only sometimes assured \cite{gill2024modern}. By processing data onsite, edge AI considerably amplifies data privacy and security measures, mitigating the need to transmit sensitive information over vast distances to central servers. This feature is exceptionally critical in domains such as healthcare and finance, where the confidentiality and integrity of data are of utmost concern.

Additionally, Edge AI champions bandwidth efficiency by mitigating the volume of data that needs to be transmitted over networks, making it an economical choice for data-intensive applications \cite{cloudAIBus2024}. This efficiency not only reduces operational costs but also relieves network congestion, facilitating smoother and more reliable data flows. Scalability is another significant advantage offered by edge AI \cite{singh2023edge}. As the network of devices expands, edge computing allows for seamless scalability without the bottleneck of centralized processing power, enabling businesses and technologies to grow without being hampered by infrastructure limitations.

Essentially, the combined use of edge computing and AI is not just a technical progression but also a tactical imperative to fulfill the dynamic requirements of contemporary applications. By championing lowered latency, enhanced privacy and security, bandwidth efficiency, and scalability, edge AI is set to revolutionize how data-driven decisions are made, ushering in a new era of intelligence that is both efficient and privacy-centric.

\subsection{Article Organization}
Section \ref{sec:background} offer present situation of Edge AI. Section \ref{sec:Methodology} details the methodology adopted for the review. Section \ref{sec:related} discusses a related surveys and studies focusing on different applications in terms of algorithms, optimization techniques, security, and privacy concerns integrated with Edge AI. Section \ref{sec:Taxonomy} outlines a taxonomy encompassing infrastructure, cloud computing, fog computing, services, use cases, machine learning and deep learning, and resource management. Section \ref{sec:Comparisons} compares existing Edge AI implementations based on taxonomy. Section \ref{sec:analysis} presents an analysis and the results obtained, and the future research directions are discussed in Section \ref{sec:future}. Finally, Section \ref{sec:summary} summarizes the survey. 

\section{Edge AI: Background and Current Status}\label{sec:background}

This section explains some concepts related to background and current status in Edge AI. Subsection \ref{subsec:emerge} explains edge computing and its historical emergence. Subsection \ref{subsec:integ} provides information on the integration of AI and edge technologies. This section is completed by explaining Edge AI applications and challenges in subsection \ref{subsec:apps} and subsection \ref{subsec:challenge}, respectively.

\subsection{Historical Emergence of Edge Computing} \label{subsec:emerge}

\begin{figure}[t]
	\centering
	\includegraphics[scale=0.27]{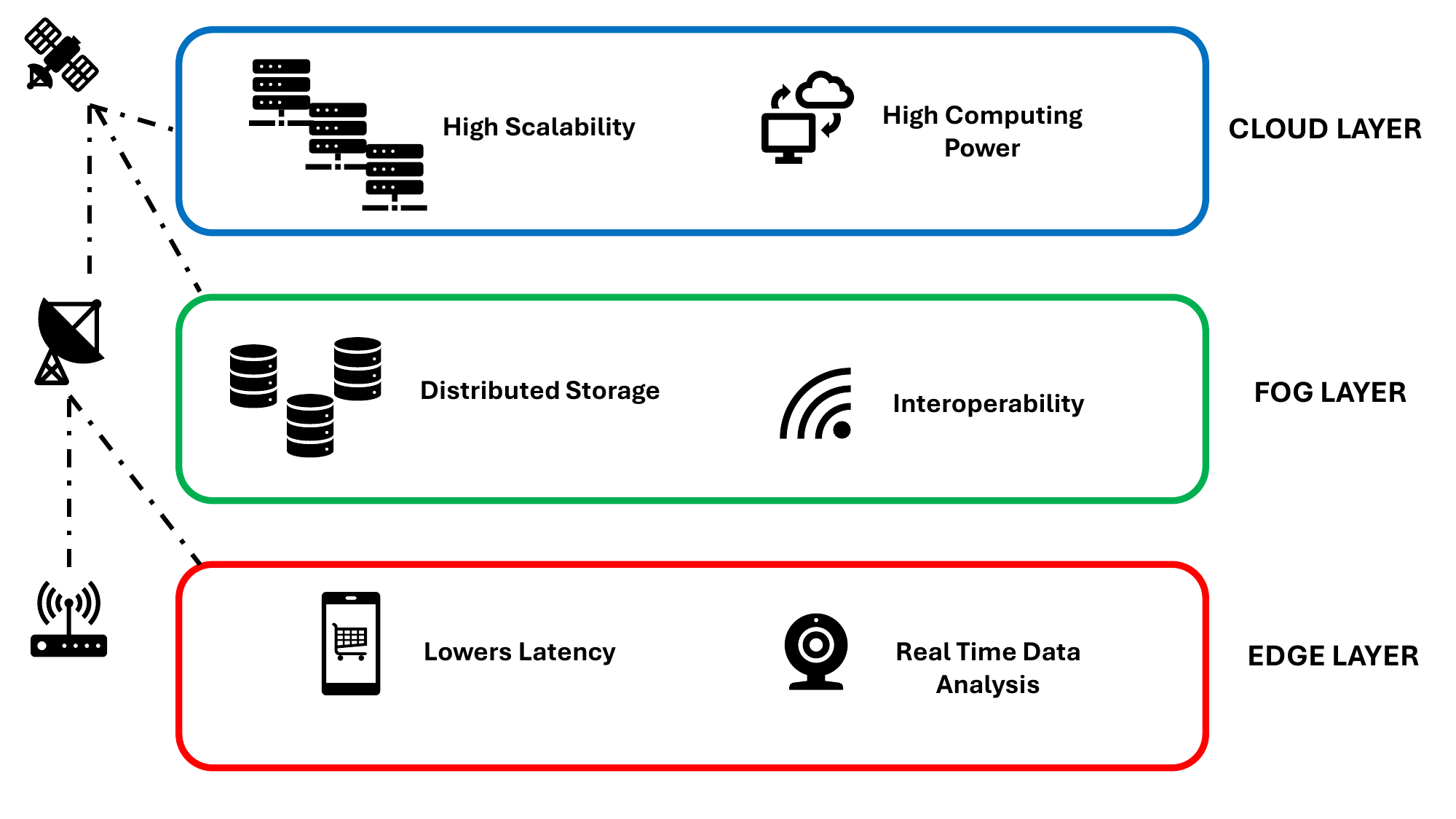}
	\caption{Computing Paradigms and their Objectives.}
	\label{fig:comparisonparadigm}
\end{figure}

\textcolor{black}{The concept of edge computing is a paradigm that brings computing resources closer to the data source, unlike the cloud, which provides services through a remote server \cite{golec1}. In this way, it is aimed to reduce problems such as unnecessary bandwidth occupation and latency in today's world where huge amounts of data that need to be processed are produced \cite{golec2}. To understand the emergence of edge computing, it will be more useful to examine previous paradigms such as cloud and fog computing. Fig.~\ref{fig:comparisonparadigm} shows the advantages of cloud, fog, and edge computing over each other and their layer arrangement. These concepts are discussed briefly below:}

\begin{itemize}

\item \textbf{Cloud Computing:} \textcolor{black}{It is a paradigm that dates back to the 1970s and refers to the use of common computing resources by users on a server via the Internet \cite{golec3}. Today, it is offered to users with various service models, especially by large companies such as Microsoft Azure, Google Cloud Platform and IBM Cloud. The advantages of cloud computing are as follows \cite{golec4}:}

\begin{itemize}
\item \textcolor{black}{High processing power and central storage, so users can easily access resources from anywhere there is the Internet. This reduces the user's risk of data loss and provides the user with the freedom to work from any location with Internet access.}

\item \textcolor{black}{Scalability, in case the need for computing resources increases (demand fluctuations), cloud computing provides services such as more processing power and storage by scaling the resources. In this way, performance measures such as service-level agreement (SLA) and QoS are ensured.}

\item \textcolor{black}{Pay as you go, with the serverless (Function as a Service (FaaS) + Backend as a service (BaaS)) service model provided by cloud computing, users are charged only for the amount they use their computing resources. In this way, an economical model is provided and appealed to more users.}
\end{itemize}

\item \textbf{Fog Computing:} \textcolor{black}{The concept of fog computing was introduced by Cisco in 2012 \cite{peter2015fog}. This paradigm recommends moving computing resources closer to the endpoints of the network (such as routers and gateways) to reduce the latency and bandwidth problems that occur in cloud computing. When Fig.~\ref{fig:comparisonparadigm} is examined, fog computing acts as a layer between the cloud and the edge.  The advantages of fog computing are as follows \cite{iftikhar2022fog}:}

\begin{itemize}

\item  \textcolor{black}{Fog computing has lower latency than cloud because it brings computing resources closer to the edge of the network.}

\item  \textcolor{black}{By acting as a layer between the cloud and end devices, it reduces unnecessary bandwidth usage by processing some of the huge amounts of data to be sent to the cloud.}

\end{itemize}

\item \textbf{Edge Computing} \textcolor{black}{The development of IoT and sensor technologies has increased the amount of data that needs to be processed to enormous levels. Processing all this data on cloud computing resources may cause unnecessary bandwidth occupation and latency problems. For this reason, the concept of edge computing has emerged as a paradigm that aims to optimize latency and bandwidth usage by processing data close to the data source \cite{golec2024priceless}. Additionally, edge computing is a good solution to address the complexity, security, and management challenges posed by fog computing, an extra layer \cite{golec2024master}. The advantages of edge computing are as follows \cite{gill2021manifesto}:}

\begin{itemize}
    
\item \textcolor{black}{Reduces latency and bandwidth usage by moving data processing to the edge of the network,}

\item \textcolor{black}{Compared to fog computing, it offers advantages such as less complexity and better security.}
 
\end{itemize}

\end{itemize}

\subsection{Integration of AI with Edge Technology} \label{subsec:integ}

\begin{figure*}[t]
	\centering
	\includegraphics[scale=0.47]{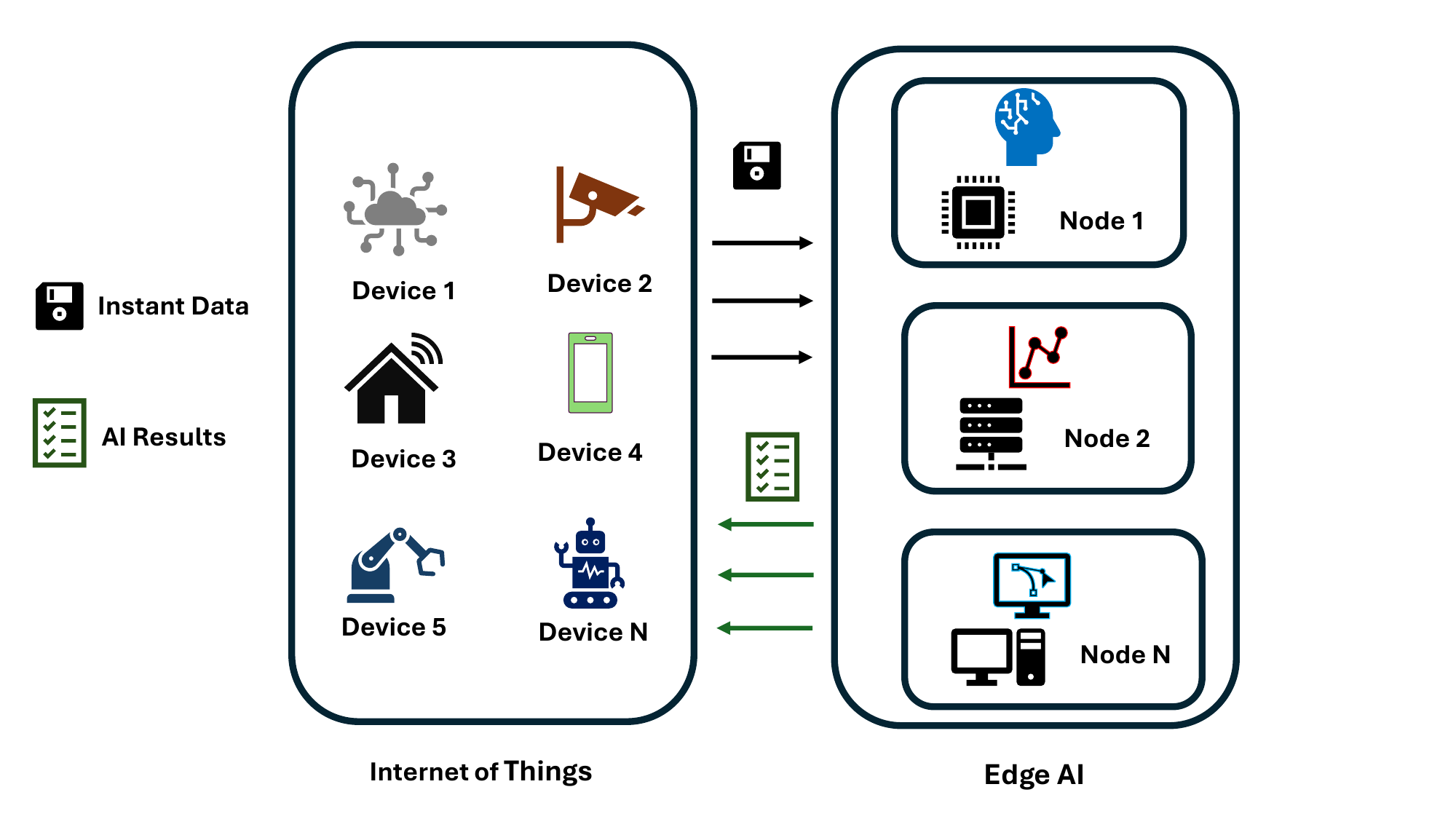}
	\caption{Architecture of Edge AI.}
	\label{fig:edgeai}
\end{figure*}

\textcolor{black}{Developing AI applications have begun to show themselves in many areas. One of these areas is EdgeAI, which is the combination of AI and Edge concepts \cite{golec2024master}. EdgeAI is based on the principle of processing data on edge nodes such as mobile devices and IoT instead of processing it on cloud servers \cite{nandhakumar2024edgeaisim}. This is achieved by distributing AI algorithms to edge nodes close to the data source, as shown in Figure~\ref{fig:edgeai}, which shows how data is processed in the EdgeAI concept and how it performs fast and efficient computation. The advantages offered by these two technologies can be listed as follows \cite{singh2023edge}}:

\begin{itemize}

\item \textcolor{black}{Low Latency: In delay-sensitive scenarios such as e-health and autonomous vehicle applications where patients are monitored instantly, millisecond delays are critical \cite{golec2022aiblock}. In traditional cloud-based systems, data must be sent between the user and the cloud to be processed in the AI model deployed in the cloud. This process will cause serious delay and unnecessary bandwidth usage \cite{golec2}. With Edge and AI integration, this problem can be overcome by processing data in real time. Because the data will be processed on the edge node closest to the source where it was produced, it will respond much faster than cloud-based systems.}

\item \textcolor{black}{Increased Security and Privacy: In cloud systems, data is sent from the source where it is produced to central servers. This expands the attack surface for hackers in the communication channels and storage areas of sensitive data such as biometric and health data \cite{golec2022aiblock}. In EdgeAI systems, since the data is processed and stored locally compared to cloud systems, it can be said that the overall security of the system is higher. Similarly, privacy issues that may occur in the event of theft of sensitive data such as biometric data are reduced \cite{golec1}.}

\item \textcolor{black}{Resource Optimization and Scalability: EdgeAI systems consist of heterogeneous devices such as laptops, network routers, and mobile devices with different processing power and storage capabilities. This means that EdgeAI can share the resources of devices in the edge network if external processing power and resources are needed. In addition, balanced load distribution can be achieved by using advanced resource allocation algorithms to optimize resources.}
 
\end{itemize}

\textcolor{black}{Future Directions and Limitations: Despite the above-mentioned advantages and high potential, EdgeAI also brings with it challenges such as (i) limited processing power of devices in the edge network, (ii) management difficulties due to the heterogeneous structure of the edge network, and (iii) energy constraints due to resource limitations. If future researchers solve these challenges, it is expected that the areas of use of EdgeAI will expand.}

\subsection{Edge AI Applications} \label{subsec:apps}

\begin{figure*}[hbt!]
	\centering
	\includegraphics[scale=0.47]{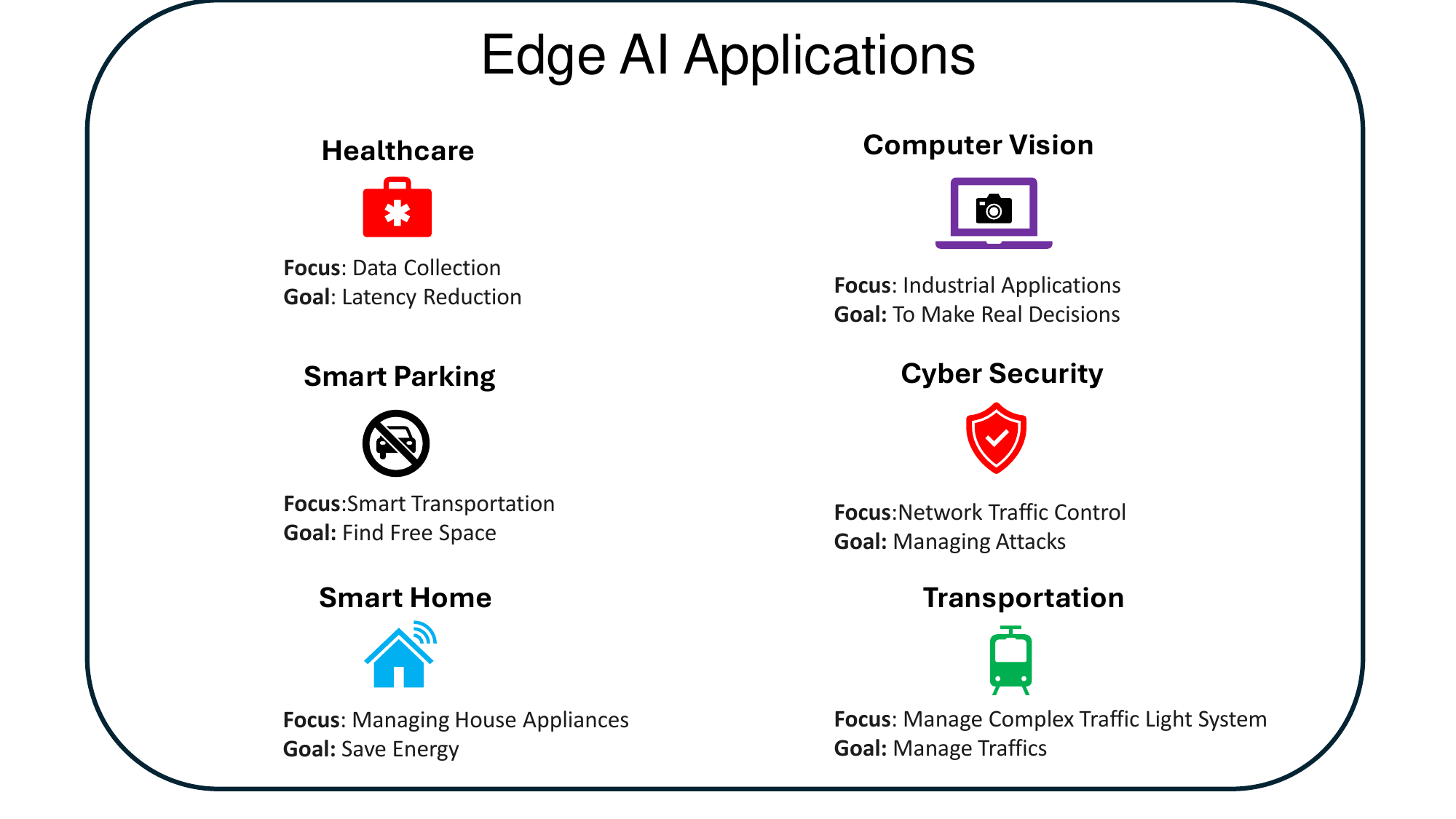}
	\caption{The Edge AI Applications.}
	\label{fig:edgeaiapp}
\end{figure*}

\textcolor{black}{Edge AI applications, created by combining the concepts of Edge and AI, provide lower latency and higher security than Cloud-based AI applications. Fig.~\ref{fig:edgeaiapp} shows popular applications of Edge AI, which are discussed briefly below:}

\begin{itemize}

\item \textcolor{black}{Healthcare: Edge AI applications are based on the processing of data collected from wearable devices in distributed AI models at the edge of the network. Additionally, early diagnosis studies using portable medical imaging techniques can be given as examples \cite{golec4}.}

\item \textcolor{black}{Smart Parking: With the increase in means of transportation, parking has become a big problem, especially in big cities. Edge AI-based solutions with the help of sensors and IoT devices can be used to solve these problems \cite{lee2022edge}.}

\item \textcolor{black}{Smart Home: Solutions used in modern homes such as home lighting systems and smart refrigerators can be given as examples of these applications. In this way, energy consumption can be optimized by preventing excess electricity consumption in cities \cite{iftikhar2022fog}.}

\item \textcolor{black}{Computer Vision: Edge AI can identify people using methods such as biometric authentication \cite{golec1}. Additionally, Edge AI provides great advantages in Industry applications that require real-time decisions \cite{golec2024master}.}

\item \textcolor{black}{Cyber Security: Unauthorized access, suspicious objects, and armed individuals can be detected with Edge AI-based security applications. Additionally, anomaly detection can be made by detecting suspicious traffic on a network to prevent cyber attacks \cite{patrikar2022anomaly}.}

\item \textcolor{black}{Transportation: Edge AI-based solutions can be used for today's complex traffic light operations \cite{wu2021edge}.}

\end{itemize}

\begin{figure}[hbt!]
	\centering
	\includegraphics[scale=0.45]{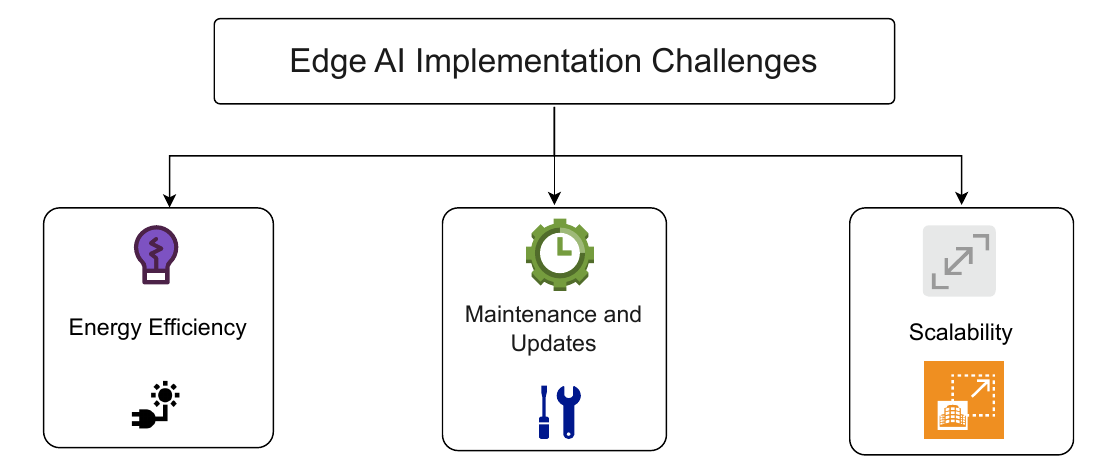}
	\caption{Edge AI Implementation Challenges.}
	\label{fig:challenges}
\end{figure}

\subsection{Edge AI Implementation Challenges} \label{subsec:challenge}

\textcolor{black}{EdgeAI, which emerges by combining Edge and AI, brings with it the advantages it offers, but also challenges that are still waiting to be solved. These challenges are shown in Fig.~\ref{fig:challenges}. These challenges are discussed briefly below:}

\begin{itemize}
   
\item \textcolor{black}{ Energy Efficiency: Edge devices generally consist of homogeneous and heterogeneous devices with low processing and storage capacity. Applications that require Natural Language Processing (NLP) and intensive image processing will cause excessive resource consumption on edge devices \cite{liu2022open}. For this reason, new solutions such as special AI chips or task engineering are needed. Therefore, new solutions have emerged, such as dedicated AI chips or task engineering. Examples include Google’s TPU and NVIDIA’s JetsonAI chips, which use low-power algorithms to achieve energy efficiency \cite{barekar2024object}. Another energy-efficient method is quantization and pruning techniques. These techniques reduce energy consumption by reducing the size of models in neural networks.}

\item \textcolor{black}{ Maintenance and Updates: Since edge devices consist of devices distributed in different locations, this means more attack targets for hackers \cite{golec2023blockfaas}. In addition, not all devices in the edge nodes have a homogeneous structure, which means separate system maintenance and updates for each node \cite{cao2020overview}. Measures such as automatic updating can be taken to solve these problems. In automatic update solutions, system incompatibilities may occur due to the heterogeneous structure of the devices. For this reason, containerization and orchestration solutions come to the fore in terms of service isolation and ease of management \cite{dolati2022layer}.}

\item \textcolor{black}{ Scalability: Since edge devices generally consist of heterogeneous devices, the distribution of a single application to different devices is still a challenge (task scheduling, etc.) \cite{satyanarayanan2017emergence}. Additionally, it is difficult to synchronize data across all devices. Effective microservice architectures and load-balancing algorithms that prevent a node from being overloaded can be used to solve this problem. Examples include Kubernetes and Istio tools \cite{larsson2019hands}. These tools can easily overcome load balancing and automating service discovery issues, while performance optimization should be considered for scenarios that require low latency.}
 
\end{itemize}

\section{Related Studies and Surveys}\label{sec:related}
In this section, we discuss related studies and surveys, as well as our main contributions.
\subsection{Related Studies}
Here, we discuss various studies, which are about the different applications consisting of smart cities, smart manufacturing, autonomous vehicles, the Internet of Vehicles (IoV), industrial automation, and healthcare monitoring systems. These are highlighted when edge computing meets AI and AI for edge computing. There are also considerations of traditional ML, computation offloading optimization, and concerns related to privacy and security, reflecting a comprehensive analysis of the challenges and strategies integrating AI and edge computing.

\subsubsection{Smart Cities}
In the case of innovative city applications, the intersection of AI and edge computing in smart cities emphasizes the importance of optimizing computation offloading and fostering a Federation between edge, cloud, and fog computing for efficient operations, which have been discussed in the following articles. In 2020, authors \cite{xu2020intelligent} presented an intelligent offloading method (IOM) that preserves privacy, boosts edge utility, and improves offloading efficiency for smart cities. The mechanism of information entropy is utilized in conjunction with edge computing to achieve an equilibrium between the maintenance of privacy and the facilitation of collaborative services. Further, authors \cite{huang2020collaborative} employed a cooperative compute offloading method to obtain the aforementioned trade-off in the cooperation of three ends: IoT device, cloudlet, and cloud. Offloading, on the other hand, can significantly reduce the processing strain on IoT devices; yet, it may incur high transmission costs and cloudlet resource use. Furthermore, authors \cite{li2020intelligent} outlined the introduction of a cyclic branch network, a DL-based intelligent offloading scheme that makes full use of network edge computing power and data traffic to reduce overall energy consumption in dual connectivity and nonorthogonal multiple access computation offloading systems. Moreover, authors \cite{zhang2022novel} suggested Robust Neural Networks From Coded Classification (CoDNN), a unique compute offloading method for multi-device collaborative pipelining processing of deep neural network (DNN) tasks. In \cite{fresa2022offloading}, authors focused on an approximation technique called Accuracy Maximization using LP-Relaxation and Rounding (AMR2), which is suggested and shown to produce a makespan of no more than 2T and a total accuracy that is less than the optimal total accuracy by a small constant. Another work \cite{khan2023deep} presents a revolutionary deep neural network-based energy-efficient offloading strategy that trains a smart decision-making model to select a reliable pool of application components. Finally, researchers \cite{du2020algorithmics} suggested method for the heterogeneous scenario is consistently effective in identifying a superior offloading scheme than the chosen existing algorithms, according to empirical findings. On the other hand, for the homogeneous scenario, the suggested solution can effectively accomplish the ideal approach.

\subsubsection{Smart Manufacturing}
In intelligent manufacturing, combining AI with Edge Computing significantly enhances efficiency, decision-making, and security, effectively addresses challenges, and improves data utilization. This integration streamlines operations, enables predictive maintenance, and supports autonomous decisions. The following papers are discussed to describe these advancements and subsequent challenges: In 2022, authors \cite{choudhury2022implementation} explored the scope of AI and its application in India's intelligent manufacturing industry, concentrating on the technology's current state, constraints, and recommendations for resolving issues. In \cite{plathottam2023review}, authors have discussed how ML and AI may boost productivity, sustainability, and manufacturing efficiency. However, there are several difficulties with implementing AI in manufacturing, including problems with infrastructure and human resources, security threats, trust, and data management and acquisition. Further, the authors \cite{yang2020big} suggested a new mode called ``AI-Mfg-Ops'' (AI-enabled Manufacturing Operations) with a supporting software-defined framework proposed as part of an open evolutionary architecture of the intelligent cloud manufacturing system. This mode can facilitate quick operation and upgrades of cloud manufacturing systems with intelligent assessment, analysis, planning, and execution in a closed loop. In 2021, authors \cite{moon2021smart} addressed the job shop scheduling challenge in the intelligent factory process while using a Deep Q-network (DQN). The suggested framework is contrasted and examined with other frameworks from the standpoint of offering an intelligent factory service.
Furthermore, the authors \cite{mishra2022smart} look at how it tends to integrate several productivity factors, such as big data analytics, Automation, and Operations Information, which connect machines via open platforms, resulting in real-time reactions to boost efficiency across the supply chain. Moreover, the authors\cite{zhang2022efficient} developed a service-oriented information model to standardized describe the functional characteristics and related operational data of heterogeneous manufacturing resources; additionally, a message middleware-based real-time transmission and integration method for high-volume operational field and sensor data is suggested to achieve the efficient distribution of related data and remote monitoring of distributed manufacturing resources. Finally, the authors \cite{mohanram2023architecture} discuss the possible advantages and difficulties of a federated learning architecture based on data gathered from 5G MSPs to enable predictive maintenance (PM) in industrial settings.
\subsubsection{Autonomous Vehicles and the Internet of Vehicles (IoV)}
In the context of autonomous and IoT-enabled vehicles, advancements in control and task optimization are propelled by AI integration and Edge Computing (EC). This synergy supports real-time decision-making, highlighting the importance of AI and EC in addressing challenges like real-time processing and security and optimizing communication and privacy within the IoV. The application of DL and Reinforcement Learning with Multiple Agents (MARL) Underscores the need for efficient solutions in autonomous driving technologies. In these scenarios, the following papers are described: In 2023, authors \cite{rizk2023model} offer a thorough technical overview of the most recent studies conducted in the areas of lateral, longitudinal, and integrated control strategies for self-driving cars. They also examine a variety of strategies and tactics used to attain accurate steering control while taking longitudinal factors into account. \cite{ning2023mobile} discusses key technologies, applications, solutions, and problems related to integrating Mobile Edge Computing (MEC) and ML in the Internet of UAVs and are covered in-depth by the author's thorough review. Further, authors \cite{ahmed2023vehicular} examined the most recent research on vehicular data offloading from the standpoint of communication, focusing on vehicle-to-vehicle (V2V), vehicle-to-roadside infrastructure (V2I), and vehicle-to-everything (V2X). The study also identified unresolved research issues in this area and forecasted future directions in the field. Furthermore, the authors \cite{xue2023cross} suggested a multi-access edge computing (MEC) framework to facilitate the cooperation of digital twins (DTs) into wireless networks and connected cars (CVs) to reduce the unreliability of long-distance communication between edge servers and CVs. Moreover, researchers \cite{ming2023exploration} outline a DL and edge computing-based vehicle intelligent control system that encourages the broad advancement of automation and intelligent technology. The findings show that the distance between the target and experimental vehicles is extremely close to the anticipated safe space. In \cite{firdaus2022joint}, authors suggested a safe edge intelligence that combined the advantages of blockchain, local differential privacy (LDP), and federated learning (FL) for automotive networks. The authors in \cite{atan2022ai} focus on a fast task execution technique in heterogeneous IoT applications that are powered by AI. This technique reduces decision latency by considering various system parameters, including the task's execution deadline, the device's battery level, the channel conditions between mobile devices and edge servers, and the capacity of the edge servers. Finally, the authors \cite{anees2023integration} combined edge computing and the Web of Things, compared their functions, and showed how edge computing improves the efficiency of real-time IoT applications by focusing on transmission, storage, and computation elements.

\subsubsection{Industrial Automation}
In industrial automation, several papers discuss revolutionary approaches to enhancing productivity by integrating AI, edge computing, robotics, and data analytics. The relevant papers look over the utilization of the Industrial Cyber Intelligent Control Operating System (ICICICOS), a cloud-edge computing-based system, for AI and industrial automation. It focuses on proposing AI with industrial processes at the edge. It emphasizes strategies for optimizing ML methods, deploying AI models on resource-constrained devices, and addressing security concerns through secure AI microservices at the edge. The relevant papers are described as follows: In 2022, the authors \cite{zhang2022edge} presented a flexible working mechanism by permitting the combined design of data quality ratios (DQRs) and model complexity ratios (MCRs) for the AI tasks and suggested a configurable model deployment architecture for edge AIaaS. Furthermore, the authors \cite{qureshi2023towards} provide a systematic overview of methods for addressing the paucity of training data for different kinds of data, and a methodology for addressing data scarcity in cellular networks is suggested. In \cite{shahriar2023survey}, the authors explore the privacy-enhancing solutions that are now in position, including the technologies, specifications, and process solutions to mitigate these risks. It also looks at privacy threats at various stages of the AI life cycle. Further, the authors \cite{hoffpauir2023survey} offer a thorough overview of edge intelligence and lightweight ML support for upcoming services and applications. The researchers have supplied a thorough analysis of cutting-edge intelligence applications, lightweight ML techniques, and their support for upcoming services and applications. In \cite{ajibuwa2023survey}, authors provide a thorough review of AI/ML-based IDS/MDSs and set baseline measurements pertinent to networked autonomous systems, emphasizing the gaps and assessment metrics in the existing research. In \cite{singh2023edge}, the authors wrapped up a thorough analysis of edge computing, covering both the shift to edge AI and related paradigms. Additionally, the history of every alternative put out for edge computing implementation, as well as the Edge AI strategy for putting AI models and algorithms on edge devices, were investigated.
\subsubsection{Smart Healthcare}
Intelligent healthcare systems focus on integrating AI and edge computing and the challenges related to privacy and security, decision-making, and optimization. These systems make use of technologies, including genetic-based encryption for data security, federated learning in the Internet of Medical Things, and nanosensor-equipped systems to improve efficiency and security. Mobile computing has played a vital role in healthcare, particularly during the COVID-19 pandemic, enabling telemedicine and contact tracing, emphasizing the significance of technology in tackling healthcare challenges. The following research papers delve deeper into the intersection of technology, healthcare, and privacy. In 2018, the authors
\cite{verma2018fog} proposed an intelligent home monitoring system based on edge-fog computing with AI capabilities. Latency issues and reliability are the main concerns for the authors in developing smart home real-time applications. Further, the authors \cite{shaikh2023machine} provide a comprehensive overview of the key elements of the MCPS from multiple perspectives, covering design, methodology, and significant supporting technologies such as cloud computing, edge computing, IoT, sensor networks, and systems with multiple agents. In \cite{ahmed2023edge}, authors provided a distinctive and specialized route resource recommendation (R3) protocol to handle resource management and connection problems in autonomous, connected ambulances (ACA) for route optimization. Furthermore, the authors \cite{misra2023kedge} provide a condition-aware analytical framework that may be used to recommend health conditions in IoT-based mobile healthcare systems. This framework corresponds to IoT devices that have limited resources, such as those with a memory utilization rate of 6.6\%. Moreover, the authors \cite{zhang2022edge} Present a flexible working mechanism that allows the combined configuration of data quality ratios (DQRs) and model complexity ratios (MCRs) for AI tasks and addresses a flexible model deployment architecture for edge AIaaS. In \cite{chakraborty2023intelligent}, researchers suggest using an edge-of-things (EoT) framework to implement centralized and federated transfer learning (CMTL) for cyberattack detection systems in the healthcare industry. Finally, the authors \cite{dvijotham2023enhancing} presents a new approach called CoDoC, which stands for complementary-driven deferral-to-clinical workflow. Its purpose is to decide when to rely on a diagnostic AI model and when to hand it off to a clinician.

\begin{table*}[]
\caption{Comparison of Our Survey with Related Surveys}
\resizebox{\textwidth}{!}{%
\begin{tabular}{llllllllllllllllllll}
\hline
 Work & Research Domain& \cite{ding2022roadmap} & \cite{iftikhar2023ai} & \cite{singh2023edge} &  \cite{shi2020communication} &  \cite{liu2022bringing} &  \cite{rocha2024edge} &  \cite{su2022ai} &  \cite{zhang2022edge} &  \cite{qureshi2023towards} &  \cite{shahriar2023survey} &  \cite{hoffpauir2023survey} & Our Survey (this paper) 
 \\ \hline
Year &   & 2022  & 2023& 2023 &2020  & 2022  &2024 & 2020 & 2023 &2024 & 2023 &2023 &2024 \\ \hline
\textbf{Edge AI }&  & \checkmark(*) &  & \checkmark(*)  & \checkmark(*) &  &  & & \checkmark(*) &   & &  &\checkmark \\ \hline
 
Taxonomy &  & \checkmark(*)  &  &  &  & \checkmark(*) & &  & &&&&\checkmark\\ \hline
\multirow{3}{*}{Infrastructure} & Cloud  &  &  &  & \checkmark &   & \checkmark  & \checkmark &  &  &\checkmark & &\checkmark \\ 

 & Fog  & & \checkmark &  &  &  &  &  &  &  & &  &\checkmark   \\
  & Edge &  \checkmark(*)  & \checkmark & \checkmark & \checkmark & \checkmark &\checkmark  &\checkmark & \checkmark & \checkmark & & \checkmark &\checkmark  \\
\hline
Application  & Monolithic  & \checkmark(*) & \checkmark &  &  &  &  &  & & &  &  &\checkmark   \\ 
Architecture & Microservice & & \checkmark(*) &  &  &  &  &  & & &  &  &\checkmark  \\
\hline

IoT  & Static  & \checkmark(*) & \checkmark &  & \checkmark  & \checkmark & \checkmark &  & & & &  &\checkmark \\ 
Use Cases & Mobile & & \checkmark & \checkmark & \checkmark & \checkmark & \checkmark & & & & & \checkmark(*) &\checkmark  \\
\hline

\multirow{5}{*}{Methods} & Heuristic  &  &  & \checkmark &  &  &  &  &  &  & \checkmark &  &\checkmark \\ 
& Meta-Heuristic  &  & \checkmark &  &  &  &  &  &  &  &  &  &\checkmark \\  
 & Machine & \multirow{2}{*}{\checkmark} & \multirow{2}{*}{\checkmark} & \checkmark  & \checkmark  &\checkmark   & \checkmark  & \checkmark  &\checkmark(*) & &  & \checkmark(*) &\checkmark   \\
 & Learning  &  &  &  &  &  &  &  &  \\
 & Deep Reinforcement  & \multirow{2}{*}{\checkmark} & \multirow{2}{*}{} &  &  &  &  & \checkmark  & \checkmark(*) & \checkmark  &   &  &\checkmark  \\
  & Learning  &  &  &  &  &  &  &  &  \\
\hline

 & Provisioning  &  &  &   &  & \checkmark  &  & \checkmark  & \checkmark(*)  & \checkmark  &  &   &\checkmark \\ 
Resource  & Resource Allocation  &  &  &\checkmark   &  & \checkmark(*)   &  & &  &  &   &   &\checkmark \\
Management &  Application Placement & \checkmark(*) &  &&  &   &  & &  &  & \checkmark(*)  & \checkmark  &\checkmark \\
 &  Workload Distribution & \multirow{2}{*}{} & \multirow{2}{*}{} &\checkmark & &  & &\checkmark &   &  &  &\checkmark &\checkmark  \\
  &  and Prediction &  & &  &  &  &  &  \\
\hline

\multirow{2}{*}{ML Model Sizing} & Reduced  & &  & \checkmark &  &  &  &  & & \checkmark & & \checkmark   &\checkmark \\ 
 & Full  &\checkmark  &  & & \checkmark &  &  & \checkmark & & & & &\checkmark  \\ 
\hline

\multirow{3}{*}{Heterogeneity} & Computational  & \checkmark(*) & \checkmark(*) & \checkmark(*) &  &  &  &  & & & &\checkmark &\checkmark\\ 
 & Hardware  &  &  &  &  &  &  &  & &&&&\checkmark\\ 
 & Platform  &  &  &  &  &  &  &  &  &&&&\checkmark\\
\hline

\multirow{3}{*}{Security} & Platform  &  &  &   &  & &  &  &  &&&&\checkmark\\ 
 & Host  &  &  &  & &  &  &  & &&&&\checkmark\\  
 & Network  &  &  & \checkmark & &  &  &  &   &&\checkmark(*)&&\checkmark\\
\hline

\multirow{4}{*}{Scheduling} & Container  &  & \checkmark(*) &  &  &  &  & \checkmark(*) &&&&&\checkmark  \\ 
 & Task  &  &  &  &  &  &  & &&&&&\checkmark  \\
  & Pod & &  &  &  &  &  & &&&&&\checkmark \\
  & Service & \checkmark &  &  &  &  &  & &&&&&\checkmark \\
\hline

\multirow{4}{*}{ Container Migration} & Stateful vs Stateless containers &  & \checkmark &  \ &  &  &  &  & &&& &\checkmark \\ 
 & Inter versus Intra cluster migrations  &  &  & & &   &  & &&&&&\checkmark  \\
  & Migrations at cloud/Edge/fog & \checkmark(*) &  & \checkmark(*) & \checkmark(*) &  &  & &&&&&\checkmark \\
  & Simulations versus real-world testbed migrations &  &  & \checkmark(*) &  &  & &&&&&&\checkmark \\
\hline

\multirow{2}{*}{Container Scaling} & Proactive versus reactive scaling decisions  &  & \checkmark &  &  &  &  &  & && & &\checkmark\\ 
 & Horizontal, Vertical and Hybrid scaling  &  &  &  &  &  &  & \checkmark(*) &&&&&\checkmark   \\

\hline

\end{tabular}%
}
Abbreviations: \checkmark:= method supports the property. *:= just an Overview/Visionary
\label{related}
\end{table*}

\subsection{\textcolor{black}{Related Surveys and Our Contributions}}

\textcolor{black}{Table \ref{related} shows the comparison of our systematic review with related surveys by focusing on the advancement of AI at the edge, optimizing algorithms in constrained environments, solving training data scarcity with AI techniques, and using AI/ML for resource management in fog and edge computing settings. There, the different columns are set on the basis of various applications, their methods of optimization, and fruitful utilization. This survey covers several edge computing concepts and emphasises the application of AI on edge devices with constrained resources. The survey examines the optimisation of ML algorithms for such restricted environments and covers current IoT applications across several sectors, including industrial automation, smart homes, and autonomous vehicles. It also highlights the difficulties and possible paths for edge computing and edge AI research, offering a strong basis for further investigation in the field. 
\subsubsection{Related Surveys}
Considering the dynamic, diversified, and resource-constrained character of fog and edge computing settings  \cite{ding2022roadmap}, this research emphasises the potential of AI/ML, particularly reinforcement learning techniques \cite{iftikhar2023ai}, in addressing resource management challenges in these systems  \cite{singh2023edge}. A thorough analysis of the literature was done to look at how AI/ML applications may be used to effectively manage resources in these kinds of situations \cite{zhang2022edge}. A taxonomy was established to categorise and contrast different approaches \cite{shi2020communication}. Enhancing explainability, reducing variance, and boosting online training of AI/ML algorithms are highlighted as critical future research directions to adapt to the constantly changing fog/edge computing landscapes \cite{hoffpauir2023survey}. The study emphasises the significance and changing challenges of resource management \cite{liu2022bringing}, whereas a presentation framework is described that addresses the issue of sparse training data in emerging radio access networks by utilizing a range of methods, such as interpolation, domain-knowledge techniques, generative adversarial networks (GANs), transfer learning, autoencoders, few-shot learning, simulators, and testbeds \cite{qureshi2023towards}. The challenges are highlighted and presented by insufficient training data, and the crucial role that Automation powered by AI plays in the operation, optimisation, and troubleshooting of cellular networks is described \cite{shahriar2023survey}. The technique suggests an integrated strategy to improve data availability in cellular networks and includes a survey and taxonomy of current approaches to lessen this scarcity. In addition, the study emphasises the necessity for scalable, reliable solutions that take conditional contexts into account for generating high-dimensional data in radio access network applications also \cite{rocha2024edge} \cite{su2022ai}.}
\textcolor{black}{Table \ref{related} reveals a dearth of reviews on Edge AI, with most existing surveys and review papers    \cite{ding2022roadmap}  \cite{iftikhar2023ai}  \cite{singh2023edge}   \cite{shi2020communication}   \cite{liu2022bringing}   \cite{rocha2024edge}   \cite{su2022ai}   \cite{zhang2022edge}   \cite{qureshi2023towards}   \cite{shahriar2023survey}   \cite{hoffpauir2023survey} presenting an overview or vision of the technology rather than a comprehensive survey, systematic review, and detailed taxonomy. To the best of the author's knowledge, this is the first survey paper on edge AI that provides a thorough taxonomy and a systematic review and highlights future research topics.}

\subsubsection{Our Contributions} 
The \textbf{\textit{main contributions}} of this paper are:

\begin{itemize}
    
\item We offer a thorough introduction to Edge AI, covering its history, challenges, and prospects.
\item We conduct a systematic review that provided a thorough examination of edge AI research based on many applications, highlighting current trends and possible directions for the future.
\item We propose a taxonomy for edge AI, which aids in the classification and arrangement of edge AI systems, and explore its potential impact across disciplines through various applications.
\item We emphasize how important edge AI is for processing data in real time at the network's edge. It also highlights the challenges faced by edge AI systems, such as resource limitations, security risks, and scaling issues.
\item We propose promising future directions that aim to address the current shortcomings of Edge AI by providing innovative solutions and opportunities for future research.
\end{itemize}

\section{Review Methodology} \label{sec:Methodology} 

\begin{figure*}[hbt!]
    \includegraphics[width=6.5in]{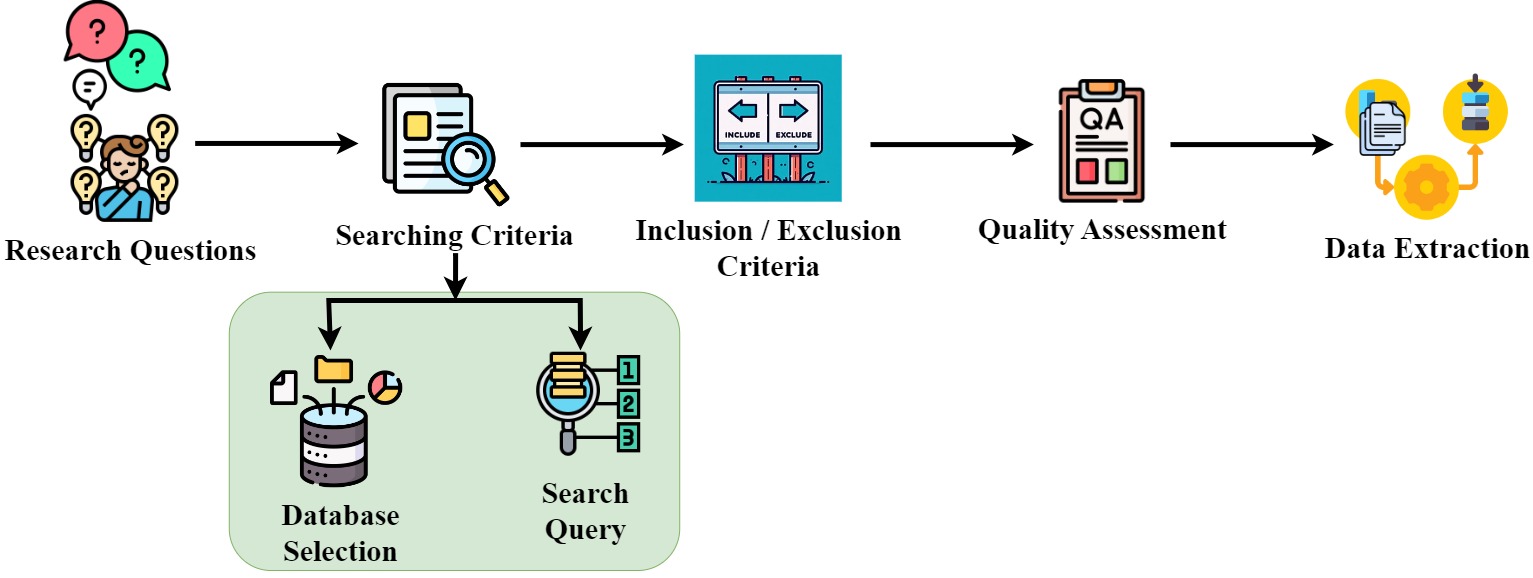}
    \caption{Road-map for Research Methodology}
    \label{fig:Road-map}
\end{figure*}

This article does a systematic review to categorize studies that are pertinent to this study domain or discuss specific research questions on "Edge AI". In this article, we used the guidelines established by Kitchenham \textit{et al.} \cite{keele2007guidelines, kitchenham2012systematic, brereton2007lessons} to provide a comprehensive review of Edge AI. The review is the optimal and reliable approach to documenting and analyse current research works. The systematic approach enables researchers to carefully analyse the positive and negative aspects of recent studies, conduct a thorough examination to identify potential gaps in research and future trends and difficulties, and provide a solid foundation and context for establishing a new study field. Furthermore, the complete research approach is presented in Fig.~\ref{fig:Road-map}, which represents the structure of the process that was used in the systematic review research.

\subsection{Design and Plan of Review}
The review procedure illustrates the methodologies applied to conduct a systematic review to minimize the potential for biased research. Therefore, possessing a pre-established process is crucial. In the absence of a systematic methodology, researchers' predispositions can influence the process of selecting and analyzing studies. This may result in the omission of crucial inquiries essential for a thorough analysis and comprehension of the subject matter. The review process encompasses the research inquiries, exploration approach, criteria for selecting studies, procedures for assessing quality, and techniques for extracting and synthesizing data \cite{tawfik2019step}.

\subsection{Research Questions}
Determining the research queries is crucial in the method of planning to develop a strong systematic review. The design of the research problem requires a thorough examination of existing literature studies. The primary aim of the present systematic review is to thoroughly examine and evaluate the various methods and strategies being employed for edge intelligence or edge AI. Furthermore, in order to emphasise the research findings and effectively showcase the useful consequences, the research questions that follow have been defined in Table \ref{review}.

 \begin{table*}[!t]
\caption {\textcolor{black}{Research Questions, Motivation, Category, and Mapping}}
\label{tab:research_questions}
\centering
\resizebox{15cm}{!}{
\begin{tabular}{p{1cm}p{5cm}p{5cm}p{3cm}p{2cm}}
\toprule
\textbf{Sr. No.} & \textbf{Research Question} & \textbf{Motivation} & \textbf{Category} & \textbf{Mapping Section} \\
\midrule
RQ1 & \textcolor{black}{What are the fundamental techniques and strategies in Edge AI, and how do they impact data processing efficiency for different applications?} & \textcolor{black}{This research question aims to explain the effect of various approaches on the efficacy of Edge AI in real-world applications.} & Methodologies and Strategies & \textcolor{black}{Sections 5.1, 5.2, and 5.4} \\
RQ2 & \textcolor{black}{What strategies can be used to optimize heuristic and meta-heuristic algorithms for enhanced performance and resource management in Edge AI applications?} & \textcolor{black}{The question evaluates the efficacy of heuristic and meta-heuristic approaches in improving performance and resource management inside Edge AI systems.} & Heuristic and Meta-Heuristic Methods & \textcolor{black}{Sections 5.4.1 and 5.4.2} \\
RQ3 & \textcolor{black}{What are the challenges and solutions for using machine learning methodologies in Edge AI for real-time data processing?} & \textcolor{black}{The research question emphasizes addressing the challenges encountered in the integration of machine learning techniques into Edge AI and possible solutions.} & Machine Learning Techniques & \textcolor{black}{Section 5.4.3} \\
RQ4 & \textcolor{black}{How can edge AI applications, especially those that use dynamic environments, implement deep reinforcement learning to improve decision-making?} & \textcolor{black}{This question examines the function of deep reinforcement learning in enhancing real-time decision-making inside Edge AI environments.} & Deep Reinforcement Learning & \textcolor{black}{Section 5.4.4} \\
RQ5 & \textcolor{black}{How can Edge AI increase scalability and resource efficiency with less computational power and storage size?} & \textcolor{black}{This question addresses the need for effective techniques for managing scalability and resources in Edge AI environments.} & Resource Management and Heterogeneity & \textcolor{black}{Sections 5.7 and 5.11.2} \\
RQ6 &\textcolor{black}{ What are the principal issues in resource allocation and task distribution for Edge AI applications, and how can they be mitigated?} & \textcolor{black}{This question tries to determine challenges in the deployment of Edge AI and provide methods for efficient resource management.} & Resource Provisioning and Workload Distribution & \textcolor{black}{Sections 5.5.1 and 5.5.4} \\
RQ7 & \textcolor{black}{What effects do architectural choices have on the scalability and performance of AI applications running on the edge?} & \textcolor{black}{This question evaluates the efficacy of various architectural methodologies in Edge AI systems.} & Architectural Comparisons & \textcolor{black}{Section 6.2} \\
RQ8 & \textcolor{black}{How does Edge AI resource management, including application placement and workload prediction, depend on important factors?} & \textcolor{black}{This question explores essential components that enhance efficient resource management techniques in Edge AI.} & Resource Management & \textcolor{black}{Section 5.5 and 6.5} \\
RQ9 & \textcolor{black}{What are the main challenges to federated learning and how might it enhance data sharing and communication among distributed Edge AI devices?} & \textcolor{black}{Federated learning is potential for distributed Edge AI applications, but data synchronization is addressed in this question.} & Federated Learning, Data sharing & \textcolor{black}{Section 8.2} \\
RQ10 & \textcolor{black}{What are the most important factors to consider when deciding between cloud, fog, and edge AI infrastructure, and how do these factors impact application performance?} & \textcolor{black}{This question examines factors impacting infrastructure decisions in Edge AI and its effects on performance.} & Infrastructure Selection & \textcolor{black}{Section 5.1 and 6.1} \\
RQ11 & \textcolor{black}{When it comes to real-time applications, how can edge AI provide data privacy and security, especially in vulnerable domains ?} & \textcolor{black}{The significance and possible solutions for data privacy and security in Edge AI applications are the primary aims of this question.} & Security and Privacy & \textcolor{black}{Sections 5.4 and 5.5} \\
\bottomrule
\end{tabular}}
\label{review}
\end{table*}

\subsection{Search Strategy}
\subsubsection{Database Selection}
The database selection includes conducting searches on various digital databases, such as IEEE Xplore, ACM Digital Library, Wiley, Taylor and Francis, Springer Link, Google Scholar, and Science Direct. These databases contain a wide range of impact factor journals, magazines, and significant conference proceedings, making them suitable for this systematic review. 
\subsubsection{Search Query}
A comprehensive search was conducted utilising Logical OR/AND operators to connect the keywords, concepts, synonyms, and abbreviations. The initial phase entails conducting an automated search using predetermined keywords that align with the study topics of this systematic literature review (SLR). The keywords used are [(((``Edge AI'' OR ``Edge Intelligence'' OR ``Edge Computing'') AND (``Machine Learning'' OR ``Deep Learning'' OR ``Reinforcement Learning'')) AND (``Resource Management'' OR ``Resource Allocation'')) OR (``Cloud Computing'' OR ``Fog Computing'') OR (``Application Placement'' OR ``Security'' OR ``Scheduling'' OR ``Simulation'')]. The search terms are obtained from the specified research topics and the framework of this systematic review in order to encompass the most significant and interrelated publications.

\begin{table*}[!htbp]
\caption{Overview of the criteria for determining inclusion and exclusion.}
    \label{tab:my_label}
    \centering
    \begin{tabular}{p{.9cm}p{6.3cm}p{6.3cm}}
  \hline
 \textbf{S. No.} & \textbf{Inclusion } & \textbf{Exclusion} \\
  \hline
  1 & English articles issued at conferences, journals, and book chapters. & Non-English articles. \\
  2 & Articles that are included in a database source and are available in their entirety. & Articles that are not available in their whole \\
  3 & Articles that specifically examine the process of choosing Edge AI infrastructure, such as Cloud, Fog, and Edge computing, and their effects on application efficiency and consumed resources. & Articles that explore diverse domains such as federated learning, IoT-based approaches, and other classic methods. \\
  4 & Articles published till 2024. & Articles that were not published during the designated search timeframe. \\
  5 &Relevant articles pertaining to the investigation queries. & Articles that fail to meet the research requirements or receive a score of 3.5 or lower in the quality assessment standards.\\
 6 & Systematic reviews often prioritise publications containing experimental or empirical research. & Articles that do not contain such research.\\
    \hline
   
\end{tabular}
\label{criteria}
\end{table*}

\subsection{Inclusion and Exclusion Criterion}
The systematic review needs to establish clear guidelines for inclusion and exclusion to guarantee that the chosen papers are relevant to the research topic and address the specific research objectives.
The primary purpose of establishing the criteria is to guarantee that the studies featured are appropriate and connected to AI-based methodologies in Edge computing. Hence, the chosen research must satisfy all the predetermined criteria. Table \ref{criteria} presents the specific phrases used to determine which criteria were included and excluded in this systematic review.

Furthermore, a method of screening is carried out to identify the appropriate research studies that are relevant to the context of this study shown in Fig.~\ref{RM-design}. The screening process consists of three distinct stages:
\begin{enumerate}[label=\alph*.]
\item \textbf{Title and abstract Phase:} During this stage, papers that were deemed irrelevant were excluded based on their title and abstract. Subsequently, the studies that satisfy at least some of the criteria listed in Table \ref{criteria} are chosen and advanced to the subsequent step for additional analysis.
\item \textbf{Full-text screening Phase:} During this step, studies were excluded while they failed to fulfill the criteria specified in Table \ref{criteria}, based on a thorough reading of the full-text or partial reading.
\item \textbf{Final selection Phase:} The next phase utilises the criteria terms outlined in Fig.~\ref{RM-design} to make the final selection and eliminates studies that do not meet any of the specified criteria.
\begin{figure*}[!ht]
\centering
\includegraphics[width=6.2in]{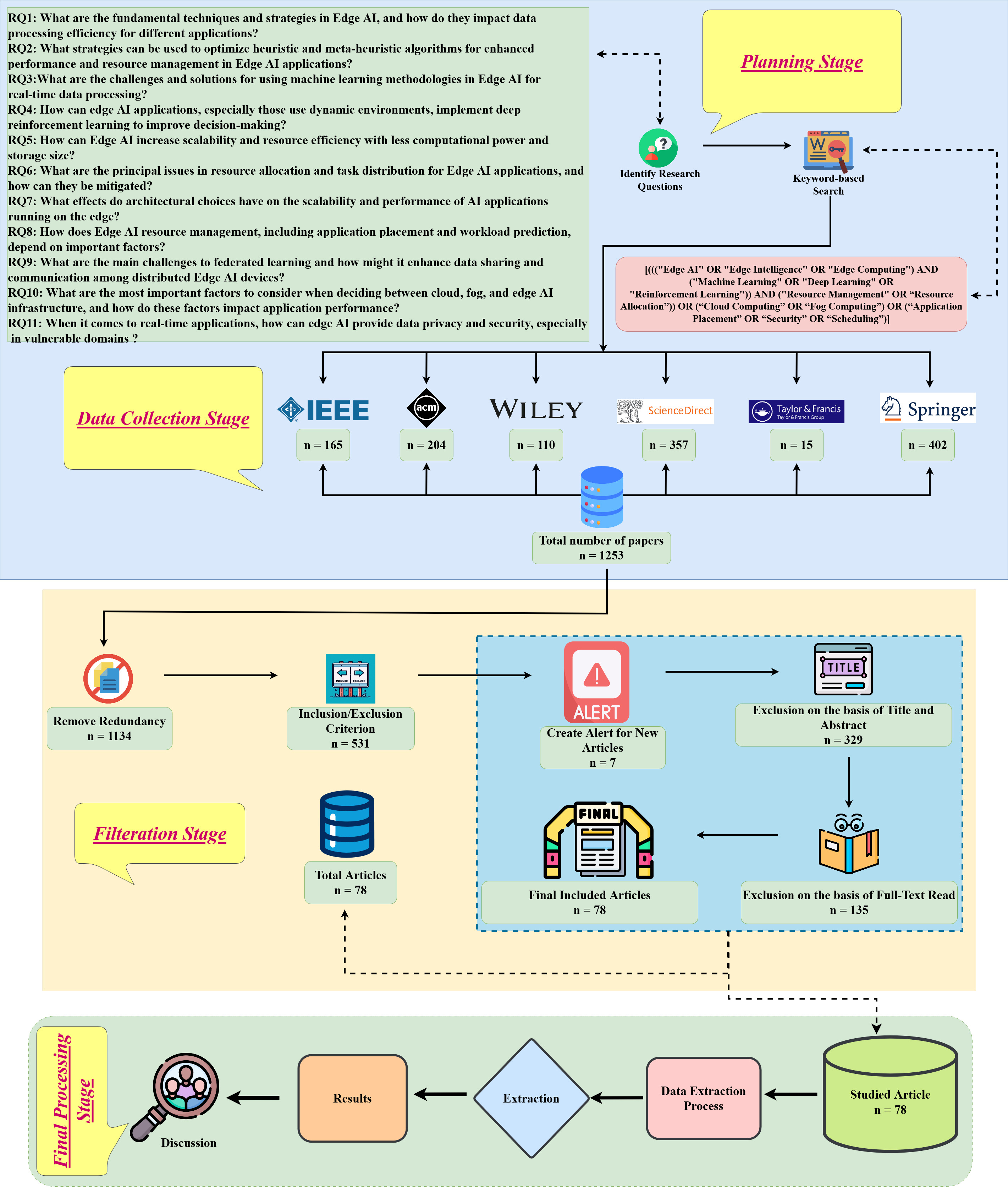}
\caption{\textcolor{black}{Research Methodology Protocol}}
\label{RM-design}
\end{figure*}

\begin{enumerate}[label=\roman*.]
\item The topic of the study must be pertinent and directly connected to the research topics.
\item The user did not provide any text. The research study examines the comprehensive solution for research advancements in edge intelligence and identifies four key components: monolithic and microservices architectures differ in terms of flexibility, performance, and resource utilisation in Edge AI for addressing practical challenges, finding solutions, and achieving optimisation goals.

\item The research paper presents essential factors to consider for managing resources in Edge AI, such as resource provisioning, allocation, deployment, and scheduling of workloads.
\item The user did not provide any text. The research study examines the factors that influence the choice of Edge AI infrastructure, such as Cloud, Fog, and Edge computing, and investigates their effects on application reliability and resource utilisation.
\item The user did not provide any text. How do Edge AI systems maintain reliable and intelligent tasks in dynamic and ambiguous instances?
\end{enumerate}
\end{enumerate}

\subsection{Quality Assessment}
In order to gather the most comprehensive and reliable information on this subject, we employed a systematic review methodology, following the standards outlined \cite{singh2020design}. Furthermore, a plethora of research papers and conference papers exist on the topic of AI applied to edge computing. Once we applied the criteria for inclusion/exclusion, researchers conducted a thorough evaluation of the articles that met the standards to identify the ones that were most worthy of further examination. Employed the standards established by the methodology to evaluate the overall quality of the research, taking into account its impartiality, internal consistency, and objectivity.

\subsection{Extraction and Synthesis}
This phase emphasises the process of extracting and combining data by thoroughly examining all 78 chosen studies and summarising and storing the relevant information. This stage involves the creation of a mechanism for extracting data items and compiling comprehensive reports that include all the information gathered from primary research studies \cite{tawfik2019step}. 

Furthermore, this study specifically chose items that relate to the research objectives as well as research questions. The data was taken from primary studies and meticulously recorded to determine the ultimate findings of the systematic review. The analysis step involved employing descriptive synthesis. The subsequent section discusses the results obtained from the synthesis.

\section{A Taxonomy of Edge AI} \label{sec:Taxonomy} 

\begin{figure*}[ht]
	\centering
	\includegraphics[scale=0.46]{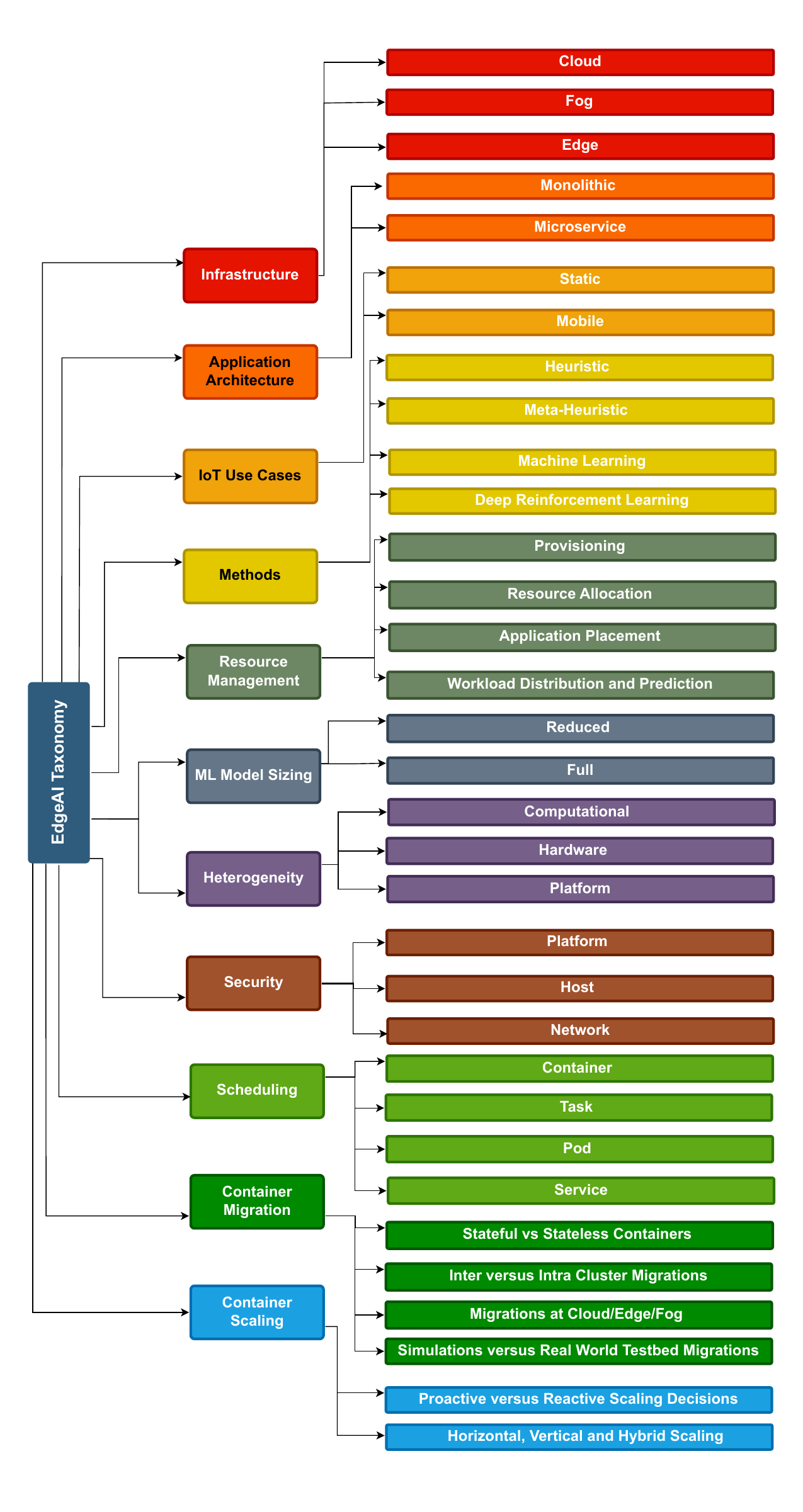}
	\caption{\textcolor{black}{The Taxonomy of EdgeAI.}}
	\label{fig:tax}
\end{figure*}

\textcolor{black}{In this taxonomy, we identify and categorize the current studies for Edge AI. To create a new taxonomy, the first three authors carefully examine the contents of the 60 papers available and then obtain the research summary as explained in Section \ref{sec:Methodology}. We classify the obtained solutions under the necessary headings. Based on the current advancements and existing literature on edge AI, we have proposed a taxonomy, as shown in Fig. \ref{fig:tax}. Leveraging edge AI-driven edge computing systems can be optimized. The best example of this is the EdgeAI-based latency-reducing studies found in the literature. We generally examine literature studies under 11 subheadings: infrastructure, application architecture, IoT use cases, methods, resource management, ML model sizing, heterogeneity, security, scheduling, container migration, and container scaling.}

\subsection{Infrastructure}
{AI models and applications can be deployed in different infrastructures based on their target scenarios. The three most common infrastructures widely used are Cloud, Fog, and Edge. }

\subsubsection{Cloud}
The essence of cloud computing is to pool resources with virtualization technology. Virtualization technology transforms a physical machine into several virtual machines, greatly changing the mode of application operation and deployment. The term 'cloud' here refers to remote data centers \cite{cloudAIBus2024}. Users usually connect to the servers in the data center through the network to use the computing resources here. The essence of cloud computing is to pool resources, allowing users to purchase resources according to their own requirements and greatly reducing resource waste.

Therefore, through cloud computing, the performance of AI applications has also been significantly improved. We transmit local AI models to remote servers through the network. We can more efficiently utilize some computing resources in the cloud, such as CPUs, GPUs, and memory, and flexibly scale according to our business needs, improving resource utilization and reducing the cost of model training and prediction.

\subsubsection{Fog}

As is well known, fog is closer to the ground than cloud, and the ground here refers to the user's local device. Fog computing technology adopts distributed computing technology, which arranges computing resources on the side closer to the user than cloud computing~\cite{singh2023edge}. It can be said that fog computing has broadened the network computing mode of cloud computing, widening computing capacity from the network center to the network edge, and thus more widely applied to numerous services. Fog computing is more widely distributed geographically and has greater mobility, making it suitable for an increasing number of intelligent devices that do not require extensive computation. For some time delay-sensitive applications such as real-time interaction systems, fog computing also has greater advantages.

Image, video and natural language processing (NLP) are the recently emerging fog computing applications. The placement and processing of images in fog computing is one of the most widely used fields of AI in research and industry. Its goal is to distinguish objects and people from each other and classify and distinguish photos based on image processing algorithms. Using fog computing in image processing-based applications can shorten response delay and improve service quality. In medical applications that require image processing accuracy and fast processing of medical data, deploying effective scheduling algorithms in foggy environments may be beneficial. According to other works, DL algorithms such as Generative Adversarial Networks (GAN) and Convolutional Neural Networks (CNN) can commonly be used in the field of image processing in fog. 

\subsubsection{Edge}

The concept of edge computing is relative to cloud computing. The processing method of cloud computing is to upload all data to the cloud data center or server in the computing resource set for processing. Any request to access this information must be submitted to the cloud for processing \cite{wu2021edge}. Edge computing is a computing model that moves resources and provides edge intelligent services at the network edge near objects or data sources, which improves service quality and information security.
In brief, edge computing analyzes the data produced by the terminal directly in the local device or network near the data generation, without transmitting the data to the cloud data processing center \cite{ahmed2023edge}. Compared with cloud computing, edge computing has shorter network latency, less resource consumption, and higher security of the data to the cloud data processing center. 
        
The so-called edge AI combines two emerging technologies: edge computing and AI. However, the implementation of edge computing is based on the same basic premise, that is, generating, collecting, storing, processing and managing data locally rather than in remote data centers. Edge AI improves this concept to the device level, using ML to imitate human reasoning to reach user interaction points, such as computers, edge servers or IoT devices \cite{misra2023kedge}. Typically, these devices can operate without an Internet connection and make decisions independently. Well-known examples of edge AI technologies include virtual assistants, such as GPT-4o, Apple's Siri, or Amazon Alexa. When the user says ``hey'', these tools will recognize and learn what the user is saying (i.e. ML), interact with cloud-based application programming interfaces (APIs), and store the learned knowledge locally.

\subsection{Application Architecture}
 
Monolithic and microservice, as the two main application architecture patterns~\cite{gos2020comparison}, each has unique advantages and application scenarios that can support AI applications and models.
 
\subsubsection{Monolithic}
Monolithic Architecture, was the traditional model for software development and deployment, used in the past by large companies. And its functionality is encapsulated into one single application. The advantage of this architectural pattern is that it is much easier to develop, deploy, debug and monitor \cite{Errasti-Alcala2013Meta}. Because the interaction between components is directly completed via memory, the performance is better. For some simple and initial AI systems, or applications with small business scales and infrequent changes in requirements, monolithic architecture may be a suitable choice. 

However, with the arising of the business scale and demand, MA has encountered many challenges and shown obvious shortcomings. Due to its inherent tight integration, this design pattern has limitations in scalability, adaptability, and ease of maintenance. Whenever specific modules need to be extended, the entire application needs to be recompiled and deployed, which can lead to low resource efficiency and waste \cite{gill2024modern}. In addition, as the code repository grows, the complexity of development, testing, and deployment also increases, thereby increasing maintenance costs and the risk of errors. Therefore, in large and complex applications, developers are seeking more flexible and scalable architectural patterns, such as microservices architecture, as introduced in the next section.

\subsubsection{Microservice}
{Microservice, as an emerging distributed system architecture model, is gradually changing the landscape of software development. The core idea is, unlike monolithic, to decompose complex large-scale applications into small and independent service units, each focusing on specific business functions or integration domains. Considering the fact that microservices architecture is composed of multiple small components, iterative upgrades of applications are more flexible and efficient \cite{gill2024modern}. Especially for large and complex AI systems, such as e-commerce platforms, social media platforms, etc., microservice architecture is a more suitable choice. These applications typically consist of multiple subsystems, each of which can be independently developed. For example, Taobao has dozens of independent systems, all of which are typical microservice architectures that can support rapid business development and iteration. 

Even so, with the widespread application of microservice architecture, it also faces challenges such as service governance, network transmission efficiency, service expansion, and version iteration. Nevertheless, microservice architecture remains an important development direction for enterprise IT architecture, and its potential and advantages cannot be ignored \cite{Himeur2023Edge}. Future research will require an in-depth exploration of how to overcome these challenges and further promote the development and application of microservice architectures.
}
\subsection{IoT Use Cases}
 
IoT is truly revolutionary, presenting an extensive array of diverse and impactful use cases that are reshaping numerous aspects of our modern world. We can classify IoT use cases into two main categories: static and mobile. As the name suggests, static use cases are like agricultural monitors, which are fixed and usually do not need to be moved, while mobile use cases such as in-car telemetries and wearable devices may frequently move \cite{gill2024ITL}.
 
\subsubsection{Static}
 Static IoT user cases based on edge AI technology are gradually becoming a key driving force for digital transformation. In these cases, data collected by static IoT devices (such as cameras, sensors, etc.) is directly analyzed in real-time on edge devices, and rapid decision-making and response are achieved through edge AI \cite{Himeur2023Edge}. In the field of agricultural ecological environment monitoring, the agricultural IoT predominantly employs high-tech approaches to establish sophisticated agricultural ecological environment monitoring networks, and uses wireless sensor technology, information fusion transmission technology, and intelligent analysis technology to perceive changes in the ecological environment~\cite{iftikhar2023ai}. In 2002, researchers at the University of California, Berkeley conducted a 9-month periodic environmental monitoring of the habitat of the Shanghai Swallow on Duck Island using wireless sensor networks. Regional static MICA sensor nodes were deployed to achieve unmanned and non-destructive monitoring of sensitive wildlife and their habitats. Some countries, including the United States, France, and Japan, have primarily focused on integrating the establishment of agricultural information platforms covering the whole country to achieve automatic surveillance of the agricultural ecological environment and ensure its sustainable development of the agricultural ecological environment. 

\subsubsection{Mobile}
{The emergence of emerging AI such as DL has brought innovations to mobile animal networking. In the field of mobile IoT, edge AI achieves real-time and efficient data processing by deploying the computing power of AI at the edge of devices. For example, in smart health monitoring applications, edge AI enables wearable devices to analyze user physiological data in real-time, providing real-time health feedback to users without uploading data to the "cloud" for processing~\cite {gill2022ai}. These devices harness DL technologies to scrutinize user health metrics and furnish guidance for subsequent lifestyle modifications, yielding substantial benefits, particularly for infants, young children, and the elderly. Due to data not being uploaded to the cloud, it can effectively protect user privacy and security. In addition, in the field of intelligent transportation, AI technology provides timely guidance for vehicle operation by analyzing local data. Similarly, many frameworks utilize DL techniques to predict parking occupancy rates, reduce the time required for vehicles to search for parking spaces, and improve urban traffic management \cite{singh2023edge}. These user cases demonstrate the enormous potential of edge AI technology in improving the performance of mobile IoT applications and enhancing user experience.}
\subsection{Methods}
{In this part, we discuss the dominant methods employed in edge AI which include heuristic algorithm, meta-heuristic algorithm, ML and Deep Reinforcement Learning (DRL). These four types of algorithms have their own characteristics and application scenarios, aiming to optimize the accuracy and performance of AI models and applications.}
\subsubsection{Heuristic}
{
People often refer to methods inspired by the laws of nature or the experiences and rules of specific problems as heuristic algorithms. The current heuristic algorithms are not entirely based on natural laws, but also come from human accumulated work experience. They supply a practical solution for each instance of the combinatorial problem to be solved and maintain an acceptable cost in terms of computation time and space, and the degree of deviation between the practical solution and the optimal solution may not be ascertainable before the implementation in advance \cite{nandhakumar2024edgeaisim}. Heuristic algorithms are a technique that enables the search for the best possible solution within an acceptable computational cost, but may not necessarily guarantee that the obtained solution is the best scheme. In most cases, it is impossible to describe the degree of approximation between the obtained solution and the optimal solution.

In the field of edge AI, heuristic algorithms have been widely used because they conform to the characteristics of human cognitive thinking. Usually, edge devices have limited computing resources~\cite{sharif2023priority}, so reasonable resource scheduling is particularly important. Heuristic algorithms can find satisfactory solutions in a relatively large search space within a short amount of time, so they can help our applications deploy to suitable nodes. In addition, heuristic algorithms can perform efficient data analysis on edge devices, greatly reducing dependence on cloud computing resources and lowering network response latency~\cite{zhuang2022network}.

}
\subsubsection{Meta-Heuristic}
{Meta heuristic is a computational intelligence-based mechanism used to solve complex optimization problems for optimal or satisfactory solutions. The meta heuristic algorithm obtains a sufficiently good solution by searching the space \cite{misra2023kedge}. Meta heuristic algorithms can be seen as an algorithmic framework that can be applied to different optimization problems with slight modifications. In edge AI, the application of meta heuristic mainly focuses on two aspects:

\textbf{Model optimization:} The reasonable deployment of AI applications in edge devices has always been a headache inducing issue. AI applications are computationally intensive services that require high computing resources, while computing resources in edge devices are usually limited \cite{golec2023blockfaas}. The fully heuristic algorithm can help find the optimal parameter configuration for AI models, so that these AI applications can maintain high performance while occupying as few resources as possible.

\textbf{Resource management:} In edge AI systems, multiple tasks may need to run at the same time and they need to share limited resources~\cite{singh2021metaheuristics}. The meta-heuristic algorithm can optimize the allocation of these resources, ensuring that each task receives sufficient resources to run efficiently without depleting the entire system's resources.
}
\subsubsection{Machine Learning}
{As is well known, the ML technique forms the foundation of AI and plays a pivotal role in many applications, such as recommendation systems, text generation, and so on. As edge AI gains prominence, ML is increasingly finding its way onto devices located on the fringes of the network. These edge devices prioritize data processing close to its source, aiming to minimize transmission delays, accelerate response times, and alleviate the reliance on centralized servers \cite{singh2023edge}. Consequently, edge AI systems are often tasked with handling vast volumes of data in real-time scenarios. Following certain ML algorithms, such as sophisticated DL models, these systems can swiftly analyze and process data at the edge, extract crucial insights, and empower a diverse array of real-time applications encompassing autonomous vehicle operations, intelligent manufacturing processes, and robust security systems. In addition, ML techniques can be employed to detect anomalies in real-time on edge devices, such as detecting product quality issues during the manufacturing process, or detecting changes in patient health status in the medical field~\cite{desai2022healthcloud}.
}
\subsubsection{Deep Reinforcement Learning (DRL)}
{Many application problems in AI require algorithms to make decisions and execute actions at every moment. For Go, each step requires determining where to place the pieces on the chessboard in order to defeat the opponent as much as possible; For autonomous driving algorithms, it is necessary to determine the current driving strategy based on road conditions to ensure safe driving to the destination; This type of problem has a common characteristic: to make decisions and actions according to current conditions in order to achieve a certain expected goal \cite{sheng2021deep}. The ML algorithm used to solve such problems is called reinforcement learning (RL). Although traditional reinforcement learning theories have been continuously improved in the past few decades, they are still difficult to solve complex problems in the real world.

DRL is a type of DL with reinforcement learning methods, which enables models to have stronger learning abilities, indicating that machines can autonomously understand and learn the human visual world. Simply put, just like humans, this means inputting visual and other perceptual information, and then directly outputting actions through deep neural networks without the need for manual production. As introduced before, DRL can solve specific problems in edge devices, such as in car systems where the device perceives the surrounding environment and road conditions on its own without the need for human intervention, and selects the appropriate driving route~\cite{kiran2021deep}. In addition, DRL has also driven the development of other fields, such as smart homes.~\cite{iftikhar2023ai}

}
\subsection{Resource Management}
{
Edge computing, which deploys computing, storage, network, and other resources at the network's edge, can drastically minimize data transmission delay, enhance data processing efficiency, and relieve bandwidth demand on the core network \cite{iftikhar2023ai}. However, with the growing number of edge devices and the complexity of applications, how to efficiently manage these resources has become an urgent problem that must be addressed. This section will review how to make resource management in an edge environment from the aspects of resource provisioning, resource allocation, application placement, and workload distribution and prediction.
}
\subsubsection{Provisioning }
{As the name suggests, resource provision is the way of provision of resources by resource suppliers based on the users' pre-established needs and supply strategies. This type of strategy is usually divided into two types: dynamic and static strategies. Static strategies are determined based on user resource needs and constraints, such as QoS and SLAs~\cite{iftikhar2023ai,zhong2022machine,iftikhar2022tesco}. Static strategies are more suitable for stable workloads. For applications with large fluctuations in resource demand, we usually use heuristic algorithms and ML algorithms to predict in advance, which are called dynamic strategies.
}

\subsubsection{Resource Allocation}
{Edge AI is an offline service during model training, different from traditional online services such as microservices, web applications, and API services. Sometimes, it is not sensitive to latency, but usually requires higher resources such as GPU and memory. The computing power of edge devices is often limited by their limited resources, which are often unable to support or complete computationally intensive tasks, such as model training, within an acceptable deadline. This approach at this point is to offload these offline tasks and transfer them to edge servers for completion \cite{10335918}. Edge devices are used for services that need to meet low latency, such as web services and human-computer interaction. Of course, we can also combine some DL techniques to achieve more reasonable resource allocation. This approach aims to minimize resource waste and enhance application performance.}
\subsubsection{Application Placement}
{
Application placement is also an indispensable factor to consider in edge computing resource management, which means formulating an assignment of applications to servers that maximize the QoS for all users in order to optimize the performance for some components that are sensitive to latency such as interactive online games, face recognition, etc. Through reasonable application placement, such as some AI-based approaches~\cite{nayeri2021application}, the computing and storage resources of edge nodes can be fully utilized, improving resource utilization and efficiency.
}
\subsubsection{Workload Distribution and Prediction}
{
Generally, the infrastructure architecture where our application can be described as three separate layers: cloud layer, edge layer, and IoT layer. The workload of the application will have its own characteristics distributed in these three layers ~\cite{carvalho2018iot}. The cloud layer is located at the top level of the architecture and The cloud layer, situated at the top of the architecture, is a robust cluster comprising thousands of virtual machines. Deploying large-scale AI models on the cloud layer is a good choice. The edge layer is located between the cloud layer and the IoT layer composed of a set of nodes that deploy several devices like routers or switches, and it is responsible for load-balancing traffic from the cloud layer and aggregating and analyzing data from the IoT layer. We can deploy some web servers and small-scale AI applications on this layer. The IoT layer is the source of data in this architecture, mainly composed of sensors and wireless devices, such as temperature sensors, cameras, and Bluetooth \cite{sharif2023priority}. These devices need to have high sensitivity and feedback very quickly to user instructions.

Furthermore, accurate predictions can lead to more rational resource management. This is beneficial for both users and service providers. Previous research and technologies such as ARIMA \cite{nguyen2019multivariate} and Holt Winter\cite{winters1960forecasting} were based on linear temporal prediction. However, these technologies often exhibit poor prediction accuracy. With the rise of DL, technologies such as neural networks have been widely applied in data prediction, such as weather, transportation, and finance systems. Especially for recurrent neural networks (RNNs), their inputs not only focus on current data, but also contain information from a period of time in the past. Therefore, it is often used to predict time-series related data. Using RNNs to predict workload in edge computing is a promising approach worthy of consideration.

}
\subsection{ML Model Sizing}
{In the edge computing scenario, the size of the model becomes the key factor to determine whether it can be successfully deployed and applied. Especially in intelligent camera monitoring systems, due to the limitations of edge devices in computing power, storage space, and energy supply, using a full model size DL model is often impractical \cite{10048999}. To overcome these limitations, we usually adopt a strategy of reducing model size. In this section, we analyze and compare the two different strategies of model sizing.}
\subsubsection{Reduced}
{
The training cost and efficiency of AI models are important metrics to assess the quality of the model. Nowadays, many AI giants are progressively increasing the size of model parameters and the volume of training data. The model parameters of GPT-3.5 have reached 175 billion \cite{briouya2024overview}. While this approach significantly enhances model accuracy, it also markedly escalates training costs and hardware requirements, necessitating a trade-off between accuracy and cost. In this way, we need to strike a balance between accuracy and cost. In recent years, there have been numerous studies in this area, such as DenseNet~\cite{huang2017densely}, EfficientNet~\cite{tan2019efficientnet} and EfficientNetV2~\cite{tan2021efficientnetv2}. The goal of these works is to train models to achieve satisfactory accuracy with fewer model parameters. In the field of edge computing, where hardware resources are limited, the reduced-size model will certainly become an important trend of edge AI in the future.
    
Model pruning and model quantization are well-established methodologies for achieving model size reduction. However, nowadays, how to design a lightweight and high-precision neural network has become a focal point of research in the field of AI, such as MobileNet~\cite{howard2017mobilenets} and ShuffleNet~\cite{zhang2018shufflenet}. These models have small parameters and high computational complexity, making them very suitable for running on edge devices.
}
\subsubsection{Full}
{
Unlike MobileNet and ShuffleNet, GPT-3 is a full model with 175 billion parameters, which is hundreds of times the number of GPT-2 parameters (3 billion). Tom Brown~\cite{brown2020language} demonstrated that GPT-3 has completed various NLP tasks, such as translation, question answering, etc., with minimal sample training. Due to its outstanding performance in the domain of NLP, this model has greatly promoted the development of large language models. Currently, many edge computing frameworks, such as KubeEdge~\cite{xiong2018extend}, have integrated plugins that support the deployment of these extensive language models, thereby extending their applicability and utility in edge environments.
}
\subsection{Heterogeneity}
{Heterogenous environments in edge devices are employed to run various IoT applications. Their diversities are embodied in three aspects: computational heterogeneity, hardware heterogeneity and platform heterogeneity.
}
\subsubsection{Computational Heterogeneity}
{
Computational heterogeneity in edge computing emphasizes the variability in application behavior during computational operations. For AI applications, there is a large amount of vector operation logic in the model code, which determines that such applications are suitable for parallel computing rather than serial computing. For web services, universal computing is the main approach \cite{singh2022machine}.  This distinction manifests in hardware requirements, where AI applications rely on GPU acceleration, whereas web services operate efficiently with CPU resources alone.

For many microservices, such as the web services mentioned earlier, their performance bottleneck often is not in CPU but in disk read and write speed, as most of the time is spent accessing databases. In other words, they are IO-intensive services rather than computationally intensive. In order to reduce network latency, microservices are usually deployed in edge nodes. How to reduce the performance loss caused by slow disk read and write speeds is a problem we need to consider.

}
\subsubsection{Hardware Heterogeneity}
{
Edge devices have many differences in processor and hardware architecture due to the computing characteristics of the applications deployed on them. The instruction set of CPU can be divided into two categories: ARM and AMD, and software running on different instruction sets may have differences in performance. Some infrastructure deployed on edge nodes, such as routers and switches, are responsible for tasks such as data forwarding and protocol conversion, and therefore require CPU support. The application of image generation and virtual reality requires high-performance graphics rendering and environment recognition. In addition to processing CPUs, GPUs and other chips are also necessary. Nowadays, many container frameworks have support for CPU and GPU hardware resource isolation, such as Docker and Container, which can eliminate the impact of service instability caused by hardware resource competition.

In 2018, Google launched Edge TPUs~\cite{you2019fast}, specially designed for inference and training of neural networks on edge devices with limited resources. Edge TPUs demonstrate strong capabilities in computer vision~\cite{sun2021deep}. Some IoT applications for autonomous driving and facial recognition can benefit greatly
}
\subsubsection{Platform Heterogeneity}
{
Due to the rise of edge computing, the world's major technology giants have also launched their own edge computing platforms. For example, Amazon's AWS IoT Greengrass, Microsoft's Azure IoT Edge, and Google's Cloud IoT Edge. They all support the effective operation of AI models on edge devices, providing service management and data analysis capabilities. Other open-source platforms also deserve attention, such as KubeEdge and OpenYurt, which are extensions of Kubernetes in the field of edge computing and provide container management, automatic operation and maintenance and other functions.

}
\subsection{Security}
{
    With the rapid development of edge computing techniques, an increasing number of enterprises and organizations are deploying edge computing solutions to meet high demands for real-time capabilities, security, and privacy protection. However, simultaneously, the edge computing environment also faces numerous security challenges~\cite{singh2023edge,casalicchio2020state}. To ensure the stable operation of edge computing systems and data security, we need to consider and ensure security comprehensively from three aspects: Platform, Host, and Network.
}
\subsubsection{Platform}
{Blockchain is a distributed, decentralized, and tamper proof database. It is often used to build a secure and trusted intelligent platform, which can solve the security problems in edge computing. Zhang \textit{et al.}~\cite{zhang2021blockchain} utilized blockchain technology to construct a highly secure trusted edge platform, providing a secure environment for AI applications on edge nodes. Wang \textit{et al.} ~\cite{wang2018secure} proposed an integrated trust evaluation mechanism based on cloud and edge computing, along with a new architecture of service templates and balanced dynamics, to address security challenges.  In this architecture, the design of edge networks and edge platforms is aimed at reducing resource consumption and ensuring the scalability of trust evaluation mechanisms, respectively. Other security technologies such as Role Based Access Control (RBAC) have also been widely applied to some distributed platforms, such as Kubernetes.}
\subsubsection{Host}
{Host security is defined as the security of all hardware and software deployed on a single edge server or device. Due to the proximity of edge devices to the human body, such as healthcare systems and intelligent driving systems. Imagine that if a car is using intelligent driving and its intelligent driving system is hacked, it will pose a serious threat to the safety of passengers and other vehicles on the road~\cite{gharaibeh2017smart}.

We can take many measures to defend against external attacks on the host. Firewall rules can be configured to block access from unauthorized IP addresses. Moreover, by installing antivirus software and regularly updating patches, the security factor can also be improved.
}
\subsubsection{Network}
{Distributed Denial of Service (DDoS) attack, which causes significant economic losses to society every year, is one of the most common attack methods in computer networks, and it also has strong destructive power on IoT devices. From the time of the 2016 botnet Mirai attack on KrebsOnSecurity~\cite{krebs2016} and Dyn~\cite{dyn2016}, it can be seen that DDoS attacks are seriously threatening the security of IoT applications. With the development of edge computing, the threat of such attacks on large-scale IoT devices is growing, which may lead to incalculable economic losses. For example, in the field of automation, AI technology is widely used to make decisions and adjust plans. If edge AI is subjected to network attacks, it can lead to AI models making incorrect decisions, resulting in product quality issues.

Although edge nodes exhibit the potential to isolate most of the IoT data at the network edge and detect and intercept attacks near the source in the first place, they encounter significant challenges in practical applications. The main reason is that edge nodes are unable to capture the aggregated network traffic required for IoT DDoS detection, nor can they scale and provide the necessary resources like elastic clouds~\cite{bhardwaj2018towards}. Therefore, directly deploying existing cloud-based defense solutions on edge nodes is far from achieving ideal results. We need to redesign the DDoS defense scheme based on edge computing to solve the special and severe security problems in the edge environment.
 }
\subsection{Scheduling}
{Resource scheduling is the process of efficiently allocating and managing system resources, ensuring optimal utilization of resources according to demand and priority. In edge environments, resource scheduling is exceedingly crucial for achieving real-time, low-latency services. Especially in AI scenarios that have high demands for computing and network resources, only by allocating edge device resources reasonably can we support the rapid response and efficient operation of AI applications, and improve overall system performance and user experience. Resource scheduling is also a hot research direction, and there has been a lot of work in this area before~\cite{iftikhar2023ai,casalicchio2020state,singh2023edge}. In this section, we discuss scheduling from the following four granularities, because the four constitute the core unit for application deployment and management in container orchestration systems such as Kubernetes and KubeEdge.
}
\subsubsection{Container}
{Containers are a software virtualization technology that laid the foundation for the development of microservices. Nowadays, containers are also widely used in the field of edge computing. There is also much research on containers in edge computing, which stems from the growing demand of users for millisecond delay computing. In ~\cite{oleghe2021container}, the authors elucidate the concepts of container placement and migration between edge servers, and propose a container scheduling framework grounded in multi-objective optimization models or graph network models. 

In addition, some open-source container orchestration and scheduling frameworks are worth paying attention to, such as KubeEdge. KubeEdge can extend its powerful cloud computing capabilities to edge devices. Especially suitable for some AI applications, model training can be completed in the cloud and then deployed to edge devices. In addition, KubeEdge can optimize scheduling performance based on different AI application business scenarios by configuring the algorithm and parameters of the kube-scheduler.}
\subsubsection{Task}
{In the scheduling task of edge computing, there are usually two problems to be solved: scheduling time and resource allocation. Most previous research~\cite{li2019credit,zhang2019resource,tran2018joint} on task scheduling has focused on these two aspects. With the advancement of AI technology, ML technology has shown unique advantages in task scheduling. Markov Decision Process (MDP) is a robust and effective method for modeling temporal data and providing high-precision predictions. The problem of resource allocation in edge devices can be described as MDP, and the deep Q network (DQN) algorithm uses multiple replay memories to minimize the total delay and resource utilization. the study in~\cite{wei2018drl} addresses the intricate issue of joint task offloading and resource allocation problems for computationally intensive tasks in fog computing. This intricate problem is formulated as a partially observable MDP, and the Deep Recursive Q-Network (DRQN) algorithm is adopted to approximate the optimal value function.
}
\subsubsection{Pod}
{In container orchestration systems such as Kubernetes, Pod is the smallest unit of work composed of several containers. Pod scheduling is the process of assigning Pods to a node based on a certain algorithm strategy, which is of great significance for ensuring high availability, resource utilization, and performance of the system. Pod scheduling is mainly controlled by kube-scheduler, and its process includes two stages: screening and scoring~\cite{carrion2022kubernetes}. During the filtering phase, the scheduler checks all nodes to determine which ones have the resources (such as CPU and memory) and other requirements (such as node selector labels) needed to run Pod. Then, the selected node will enter the scoring stage, and the scheduler will rate each node based on a series of criteria such as node affinity, resource utilization, etc. The node scoring the highest will be designated as the running location for Pod. kube-scheduler supports custom scheduling plugins, and users can develop some extension plugins based on the business characteristics of the enterprise. 
}
\subsubsection{Service}
{Service refers to software or system components deployed at the edge of a network that provides specific functions or resources to meet the real-time, low latency, and high bandwidth needs of users or devices. Service can be a computing service, data processing service, storage service, or any form of network service that optimizes resource utilization and reduces data transmission latency, bringing better service quality and experience to users.

Service scheduling is the deployment, allocation, and scheduling process for these services \cite{ahmed2023vehicular}. In the edge computing environment, Service Scheduling is responsible for arranging and scheduling the execution sequence and location of services reasonably based on application requirements, resource conditions, and network conditions. Effective service scheduling can ensure that the service can efficiently use limited edge computing resources, achieve load balancing, reduce service latency, and improve the performance and reliability of the entire system. 
}
\subsection{Container Migration}
{In distributed and cloud computing environments, containers need to be migrated from one node to another due to node failures, load imbalance, resource upgrades, and other reasons. At this point, container migration technology is needed to achieve rapid migration and recovery of containers.
}
\subsubsection{Stateful vs Stateless containers }
{Stateful containers and stateless containers are two major classifications of containers, which are important criteria for container expansion, contraction, and migration.

\textbf{(i) Stateful containers:} The so-called state essentially refers to the data in the running container. When migrating such containers, it is usually necessary to migrate their data together, such as a database. Due to the involvement of data replication, such containers need to consider issues such as data loss and data integrity. Specific migration tools or strategies may be needed to ensure accurate migration and recovery of data~\cite{junior2020stateful}. All containers managed by a StatefulSet controller in Kubernetes are considered stateful. 

\textbf{(ii) Stateless containers:} These containers are containers that do not save any state during runtime. For example, a web server that provides services for static pages, treats each request as independent, and the container does not need to remember previous interactions. The migration process is very simple, just pull up the container on other nodes and delete the container from the original node. All Pods under a Deployment in Kubernetes are stateless.
}
\subsubsection{Inter versus Intra cluster migrations}
Inter-cluster and intra-cluster migration are discussed briefly below:\\
\textbf{(i) Inter-cluster migration:} When a company or organization needs to migrate its data center from one geographic location to another, inter-cluster migration is an indispensable step. Inter-cluster migrations involve node migration between different clusters, typically requiring consideration of cross-cluster communication factors such as network latency and bandwidth limitations \cite{10440418}. Due to the collaborative work of multiple clusters and nodes involved in cross-cluster migration, the migration process is relatively complex and requires ensuring data consistency and service continuity.

\textbf{(ii) Intra-cluster migration:} In a cluster, migration within the cluster can take effect when a node experiences performance degradation or longer response time due to excessive workload. Administrators or automation tools can migrate a portion of the workload (such as containers, virtual machines, or services) on that node to other nodes in the cluster to balance the load and optimize performance \cite{atan2022ai}. Compared to inter-cluster migrations, The complexity of intra-cluster migrations is relatively low because it only involves nodes and data migration within the same cluster.

\subsubsection{Migrations at cloud/edge/fog}
{Migration at cloud is the process of migrating applications, data, and other business processes from traditional local devices or servers to cloud platforms, including the migration to IaaS, PaaS, and SaaS~\cite{amin2021opportunities}. IaaS migration is the most ideal and applicable cloud migration solution. Because we can entrust all programs and data to cloud vendors such as Alibaba Cloud and AWS. Users do not need to consider all operational and deployment issues. 

Edge computing has become an important technical support for the development of IoT~\cite{chiang2016fog}. A thorny problem in edge computing is service migration, especially in the mobile IoT device environment. Due to the limited coverage of a single-edge server network, the migration of mobile services between servers is likely to reduce the QoS of the services. State preservation of services (such as stateful services), data loss, and cost control have become challenges in the migration of edge computing services.

Migration at fog is the process of migrating applications, services, or data from traditional centralized data centers or cloud environments to fog computing environments. The purpose of this migration is to achieve low latency, bandwidth optimization, enhanced security, improved scalability, and fault tolerance. Fog migration involves redesigning applications to adapt to the distributed architecture of fog computing, including modular design and the ability to handle network dynamics \cite{verma2018fog,iftikhar2022fogdlearner}. Fog migration can provide more effective support for IoT devices, mobile devices and other applications that need rapid response, and achieve the goal of intelligent edge computing.
}
\subsubsection{Simulations versus real-world testbed migrations}
{When discussing container migration, two different testing and validation methods are usually involved: simulations and real-world testbed migrations. Here is a comparison between these two methods:

\textbf{(i) Simulations:} It uses models to replace actual or conceptual systems for training, analysis, argumentation, experimentation, experimentation, and planning methods, techniques, and activities~\cite{iftikhar2023ai}. Simulations can predict system performance and efficiency, validate and iterate modeling and simulation through real experimental data, support, optimize and expand experimental identification, accelerate development and reduce risk costs \cite{nandhakumar2024edgeaisim}.

\textbf{(ii) Real-world testbed migrations: }Its definition is the process of testing and validating a system or application in a real physical environment, involving the migration of the system or application from one environment to another. Since it is conducted in a real-world environment, it can directly evaluate the performance, reliability, and safety of the system or application under actual operating conditions to ensure that it meets practical needs \cite{10335918}.

When conducting container migration testing and validation, simulations and real-world testing platforms are usually combined. Simulation can quickly validate concepts and strategies in the early stages, while real-world testing platforms are used to test and optimize migration strategies under conditions close to actual operational environments. This combination of methods can balance the cost, time, and accuracy of results, providing a comprehensive evaluation for fog migration.

}

\subsection{Container Scaling}
{With the continuous development of cloud computing and container technology, container scaling has become an important means to ensure application performance, high availability, and resource optimization. This section will explore the strategies and practices of container scaling from two key perspectives: firstly, the scaling decisions of proactive and reactive, which exhibit different characteristics and advantages in dealing with load changes; next horizontal vertical and hybrid scaling strategies represent how effectively container resources are in different scenarios.  }
\subsubsection{Proactive versus reactive scaling decisions }
{The scaling decisions of Proactive and Reactive reflect two different strategies, which have a significant impact on the performance and resource allocation of container applications. The following are specific explanations of these two strategies:

\textbf{(i) Proactive scaling decision:} This method will use historical data of container load to train a specific AI model, through which future changes in container resource load can be perceived and predicted in advance~\cite{imdoukh2020machine}. It allows administrators or systems to automatically adjust resources to maintain optimal performance and efficiency. For example, this strategy can predict based on historical data that as long as it reaches 7 pm or 8 pm, the QPS of AI applications will significantly increase because everyone is off work, which is the entertainment time at night.

\textbf{(ii) Reactive scaling decision}: This is a strategy that utilizes third-party resource monitoring tools, such as Promethues~\cite{prometheus}, to make real-time decisions on the number of replicas and resource allocation in containers. The container orchestration tool determines whether to expand or dissolve based on the resource change data of the relevant containers in the monitoring tool. When the load increases, reactive scaling will start adding containers; When the load decreases, it will decrease the number of containers~\cite{klinaku2018caus}. The decision-making of reactive scaling is based on real-time load data.  When training the model, the utilization of GPU and GPU memory inside the container may reach 80\%-90\%. At this time, the system will immediately detect the high utilization rate and scale up the capacity promptly.}

\subsubsection{Horizontal, Vertical and Hybrid scaling} Three types of scaling techniques are described below:

\textbf{(i) Horizontal scaling:} It is a way to cope with load changes by increasing or decreasing the number of container instances~\cite{ali2019survey,nguyen2020horizontal} (such as Pods, container groups, etc.). It can respond very quickly to load changes and adjust overall processing power by adding or removing container instances. Each container instance is independent and has good fault isolation, a fault in one instance will not affect other instances. Horizontal scaling is very suitable for scenarios with stateless services and the need to handle a large number of concurrent requests.

When AI applications need to handle a large number of concurrent requests and each request has a relatively short processing time, horizontal scaling is a good choice. For example, online recommendation systems, real-time advertising delivery systems, etc. In some application scenarios that require a large amount of computing resources (such as CPU, GPU, memory, etc.), horizontal scaling can provide sufficient resources by adding more machines. For example, DL model training, large-scale image recognition, etc.

\textbf{(ii) Vertical scaling:} It is adjusting the processing power of a single container instance by increasing or decreasing its resource allocation (such as CPU, memory, storage, etc.). It does not require managing multiple container instances, only adjusting the resource allocation of a single instance and can accurately adjust resource allocation based on actual load conditions, avoiding resource waste. Vertical scaling is suitable for stateful services\cite{rossi2019horizontal}. However, this scaling strategy also has its drawbacks, as it poses a challenge to the computing and storage capabilities of individual machines.

When AI applications encounter performance bottlenecks stemming from the capabilities of individual nodes, vertical scaling offers an effective solution by enhancing hardware capabilities, such as deploying faster CPUs, increasing memory capacity, or leveraging more efficient GPUs. For AI applications that do not necessitate extensive concurrent processing or substantial computing resources, vertical scaling can serve as a more cost-effective and straightforward approach, minimizing complexity while maximizing performance within the confines of a single node. 

\textbf{(iii) Hybrid scaling:} Hybrid scaling is a scaling strategy that combines both horizontal scaling and vertical scaling~\cite{zhong2022machine}. Based on the load characteristics and requirements of the application, use both horizontal and vertical scaling methods to optimize resource allocation and performance. Being able to flexibly choose scaling methods based on different scenarios and needs, and combining horizontal and vertical methods can more effectively utilize resources, and improve application performance and stability.

For AI applications where demand often changes or is difficult to predict, hybrid expansion can dynamically adjust the ratio of horizontal and vertical expansion based on actual demand.

\section{Comparisons of Existing Edge AI approaches based on Taxonomy}\label{sec:Comparisons} 
In this section, we compare the existing edge AI approaches based on the proposed taxonomy.

\subsection{Infrastructure}
Cloud computing, fog computing, and edge computing play different roles in realizing offline, low-latency, privacy-preserving AI services \cite{Hu2020Edge}. Among them, cloud computing provides powerful computing and storage resources for training large-scale DL or other algorithm models, and processing massive amounts of data, which are usually used in Edge AI to handle time-insensitive tasks, such as large model training, multi-data analysis and model optimization, and finally, cloud computing distributes well-trained models to various user devices;

Fog computing moves computing resources to the edge of the network, reduces data transmission latency, improves response speed, and is typically used in Edge AI to handle tasks that require high real-time responses, such as speech recognition.

Edge computing deploys computing resources directly near terminal devices to further reduce data transmission latency, which is usually used for real-time reasoning and decision-making in Edge AI, such as intelligent monitoring, smart home, etc., edge computing can realize real-time processing of data on user devices, maximize the protection of users' data privacy, and at the same time reduce the dependence on network bandwidth and reduce the pressure on the core network.

In summary, cloud computing, fog computing, and edge computing have their own focus on Edge AI, and these three together build a complete edge intelligence ecosystem. Cloud computing provides powerful computing and storage resources, fog computing emphasizes real-time and low latency at the edge of the network, and edge computing enables real-time data processing and decision-making closest to the end device. These three work together to provide comprehensive support for the development of AI at the edge.

As shown in Fig.~\ref{fig_model_feature}, we compare the emphasis, advantages and disadvantages of cloud computing, fog computing, and edge computing under different indicators:
\begin{figure*}[!ht]
\centering
\includegraphics[width=6.3in]{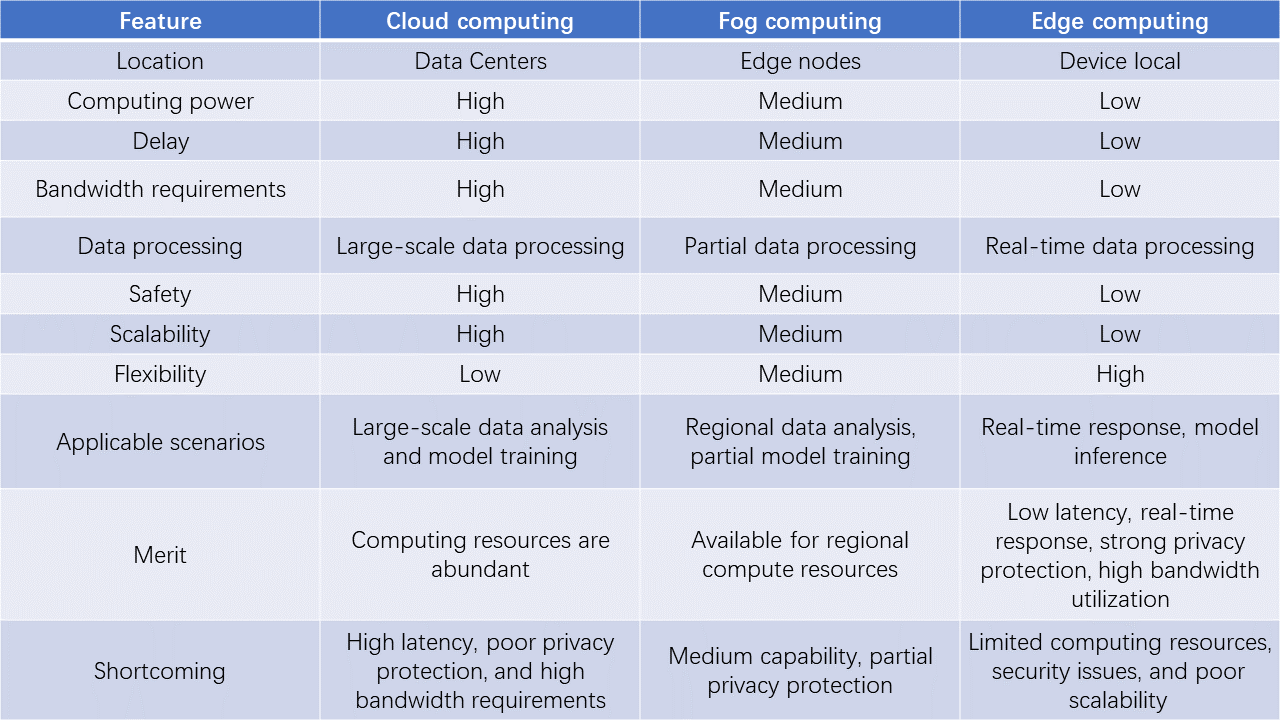}
\caption{Advantages, disadvantages and emphases of the three computational paradigms}
\label{fig_model_feature}
\end{figure*}
\subsection{Application}
Monolithic and microservices have their own advantages and disadvantages in edge AI, and we will compare the advantages and disadvantages of these two in terms of flexibility, performance and resource utilization, as well as deployment and scalability.

{
\textbf{(i) Flexibility:} Since all functional modules of the monolithic architecture run inside the same application, if we need to modify a module, we may have to recompile and deploy the entire application, so the monolithic architecture is not conducive to modularity and independent development, that is, it is less flexible; Each microservice in the corresponding microservices architecture can be deployed, scaled, and updated independently, making it easy to develop and maintain independently, thus increasing the flexibility of the overall application.

\textbf{(ii) Performance and resource utilization:} Monolithic architectures may have resource contention and performance bottlenecks because all modules share the same process and resources, but from a resource utilization perspective, monolithic architectures may make more efficient use of resources because they do not require additional communication and management overhead; The microservices architecture, on the other hand, can independently deploy and scale the corresponding microservices according to the needs, thereby improving the performance of the application. From the perspective of resource utilization, services in a microservice fabric need to communicate with each other, which may increase the latency and bandwidth consumption of the system, and the microservice system may require more resources to manage and run due to the need to maintain multiple services.

\textbf{(iii) Deployment and scalability:} Monolithic architectures are typically simple and easy to implement and deploy, but they often lack scalability to cope with frequently changing requirements; The corresponding microservices architecture, while more complex to deploy and often requires additional development and management efforts, scales flexibly and allows services to be added or removed quickly as needed.
}

In summary, the monolithic architecture focuses on simple deployment and performance optimization, which is suitable for simple, relatively fixed edge AI scenarios, while the microservice architecture focuses on flexible scaling and maintainability, and is suitable for complex edge AI application scenarios that need to be dynamically adjusted.

\subsection{IoT Use Cases}
\begin{figure*}[!ht]
\centering
\includegraphics[width=6.4in]{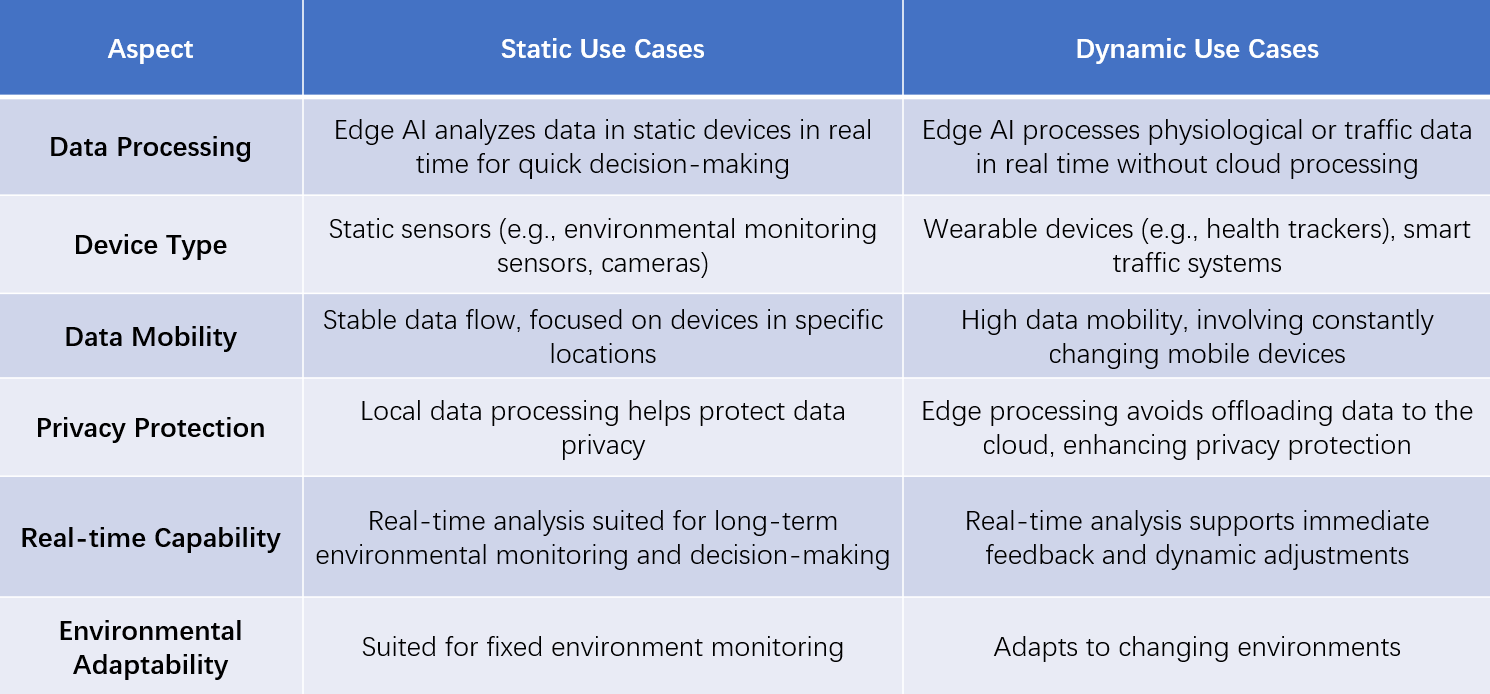}
\caption{Comparison of different IoT Use Cases}
\label{use_cases}
\end{figure*}

IoT use cases can be divided into static and dynamic in terms of user mobility. As shown in Fig.~\ref{use_cases} 2, we will show the role of edge AI in two different IoT use cases from different perspectives.

\begin{figure*}[!ht]
\centering
\includegraphics[width=6.4in]{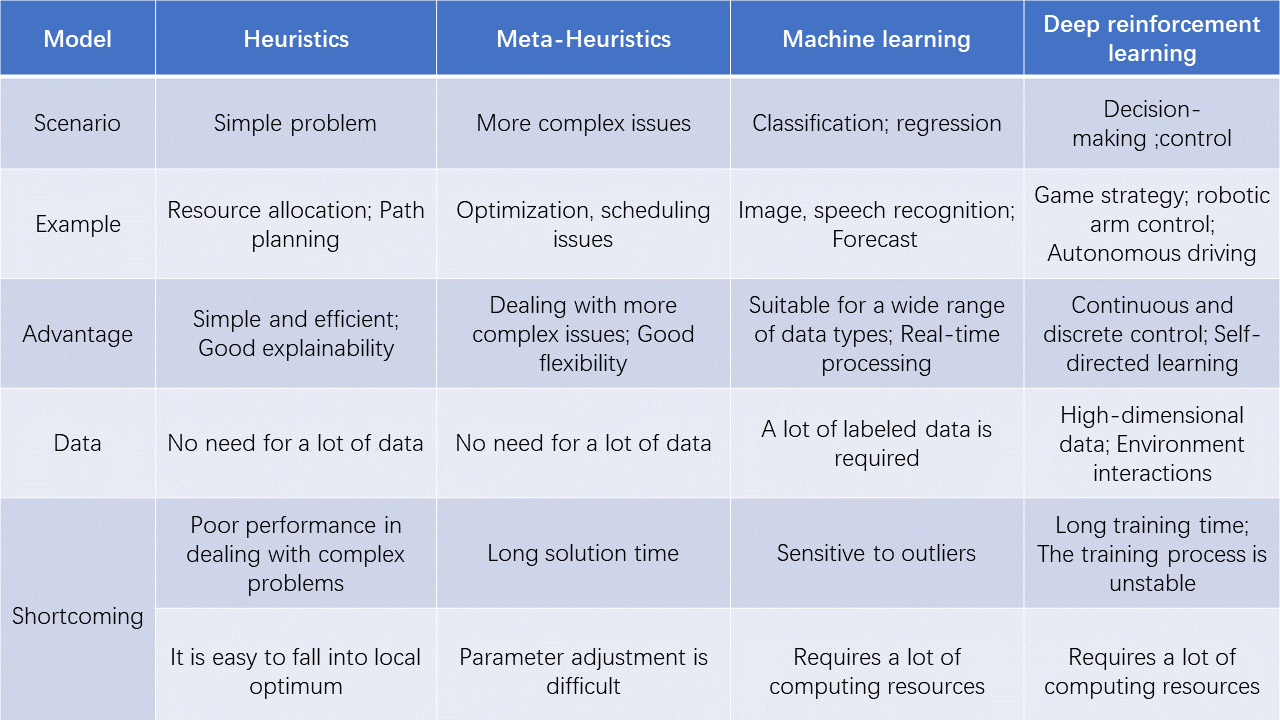}
\caption{Comparison of different models}
\label{fig_model_com}
\end{figure*}

\subsection{Methods}
For the four main AI methods, heuristics\cite{Dai2019Heuristic}, meta-heuristics\cite{Errasti-Alcala2013Meta}, machine learning\cite{Mattmann2020Machine}, and deep reinforcement learning\cite{Kaloev2021Experiments}, we will compare them from the perspectives of applicable scenarios and problem complexity, data scale and training cost, real-time requirements and resource consumption, and generalization. Fig.~\ref{fig_model_com} shows the specific comparison. 

In summary, choosing the right algorithm depends on the specific edge AI application scenario, data scale, data type, real-time requirements, and resource consumption. Heuristics and meta-heuristics are generally suitable for simple to medium-complexity problems, and the requirements for data and resources are generally not very high. ML and DRL are more suitable for dealing with some complex and nonlinear problems, and have high requirements on data volume, data quality, and computing resources.

\subsection{Resource Management}
\begin{figure*}[!ht]
\centering
\includegraphics[width=6.4in]{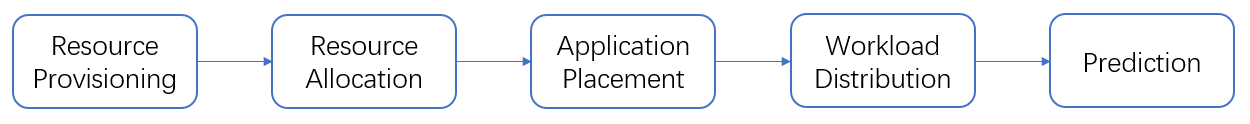}
\caption{Process of Resource Management}
\label{Resource Management}
\end{figure*}

With respect to the methods of resource provisioning, resource allocation, application placement, and workload distribution and prediction in edge AI resource management\cite{hong2019resource}, we will further describe the relationship between these methods in detail, as shown in Fig.~\ref{Resource Management}. 

As shown in Fig.~\ref{Resource Management}, when deploying edge AI, we first need to ensure that the edge devices have sufficient computing, storage, and network resources, and once they have sufficient resource provision, we also need to allocate these resources to different applications or computing tasks. Resource allocation ensures that each application or compute task gets the resources it needs to meet its performance requirements. Once the resources are allocated, the application needs to be placed on the appropriate edge device. Application placement needs to take into account the characteristics and needs of each application, as well as the state information of the edge device to achieve the best placement strategy. Once the application is placed, we can send compute tasks and compute workloads to various edge devices, and the workload distribution enables parallel processing tasks and load balancing, thereby improving the performance and efficiency of the system. Finally, we can predict future resource demand and workload changes through ML models, for example, so that we can make adjustments and optimizations in advance.

\subsection{ML Model Sizing}
Regarding the deployment of AI models on edge devices, we generally have two deployment methods: reduced model \cite{ebrahimi2022combining, kang2017neurosurgeon} and full model\cite{liang2021super}, and we compare these two methods from five aspects: model size, inference speed, accuracy, training and deployment cost, and application scenario, as shown in Table \ref{tab:model_size}.
\begin{table*}
\caption{Comparison of different deployment methods}
\centering
\begin{tabular}{|c|c|c|}
\hline
Model & Full model & Reduced model\\
\hline
Model size & Large & Small\\
\hline
Inference speed & Slow & Fast\\
\hline
Accuracy & High & Slightly lower\\
\hline
Training and deployment cost & High & Low\\
\hline
Application scenario & Resource-rich equipment & Resource-constrained devices\\
\hline
\end{tabular}
\label{tab:model_size}
\end{table*}
\subsection{Heterogeneity}
The different types of heterogeneity\cite{sanaei2013heterogeneity} involved in edge AI deployment mainly include computing heterogeneity, hardware heterogeneity, and platform heterogeneity, as shown in Table \ref{tab:Heterogeneity}.
\begin{table*}
\caption{Comparison of different types of heterogeneity}
\centering
\begin{tabular}{|c|c|c|c|}
\hline
Heterogeneity & Computing heterogeneity & Hardware heterogeneity & Platform heterogeneity\\
\hline
Definition & Different types of computing tasks & Different kinds of hardware & Devices with different functions\\
\hline
Example & Image recognition; NLP & CPU, GPU, FPGA, ASIC & Cloud server; Edge device\\
\hline
Difference & Differences in demand & Diversification of hardware & Differences between devices\\
\hline
\end{tabular}
\label{tab:Heterogeneity}
\end{table*}
\subsection{Security}
\begin{figure*}[!ht]
\centering
\includegraphics[width=5in]{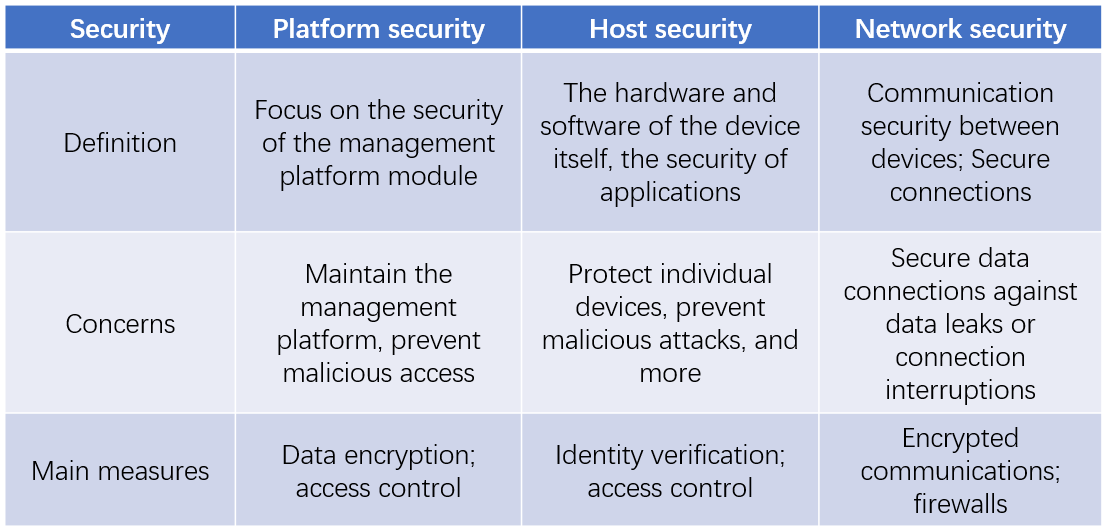}
\caption{Comparison of different types of security}
\label{Security}
\end{figure*}

Starting from the security of deploying edge devices, we mainly consider platform security, host security, and network security\cite{Marin2005Network}. As shown in Fig.~\ref{Security}, we will compare the roles of the three in detail. 
\subsection{Scheduling}
\begin{figure*}[!ht]
\centering
\includegraphics[width=6.4in]{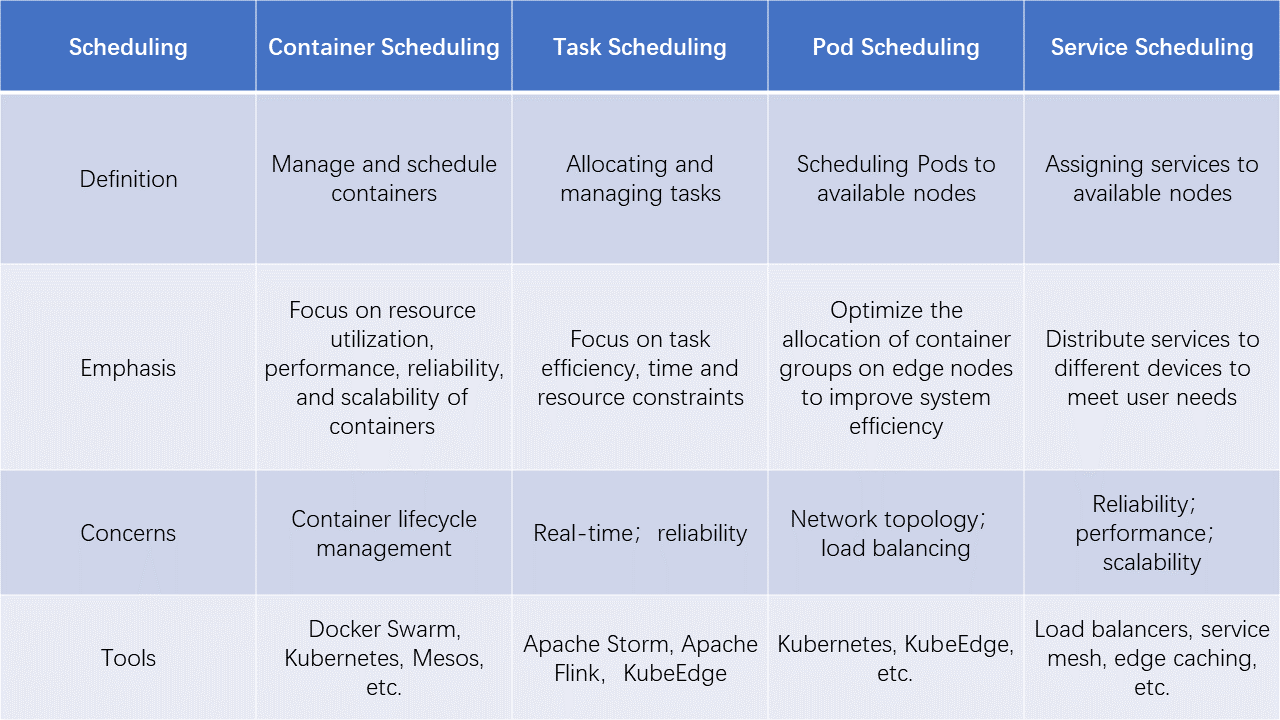}
\caption{Different types of resource scheduling methods}
\label{scheduling}
\end{figure*}

For the resource scheduling categories in edge AI, there are mainly container scheduling, task scheduling, pod scheduling, and service scheduling \cite{zhang2019resource}, and we will compare these four different scheduling types from the aspects of emphasis, scheduling measures, and scheduling tools, as shown in Fig.~\ref{scheduling}.
\subsection{Container Migration}
\begin{figure*}[!ht]
\centering
\includegraphics[width=6.4in]{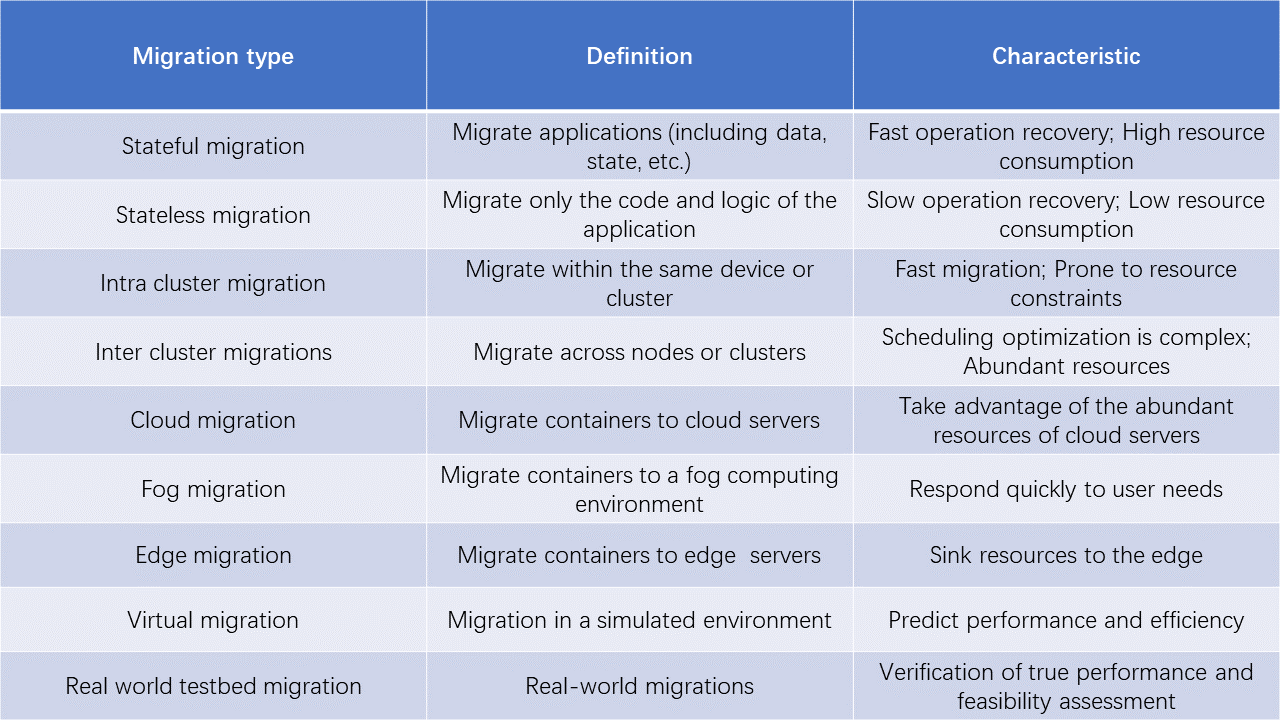}
\caption{Different types of migrations}
\label{Container Migration}
\end{figure*}

There are four main types of container migration\cite{puliafito2019container, singh2021taxonomy} in edge AI: stateful migration and stateless migration, intra-cluster migration and inter-cluster migration, cloud/fog/edge migration, virtual migration and real-world testbed migration. We have made a detailed comparison of the different migration methods, as shown in Fig.~\ref{Container Migration}.
\subsection{Container Scaling}
\begin{figure*}[!ht]
\centering
\includegraphics[width=5in]{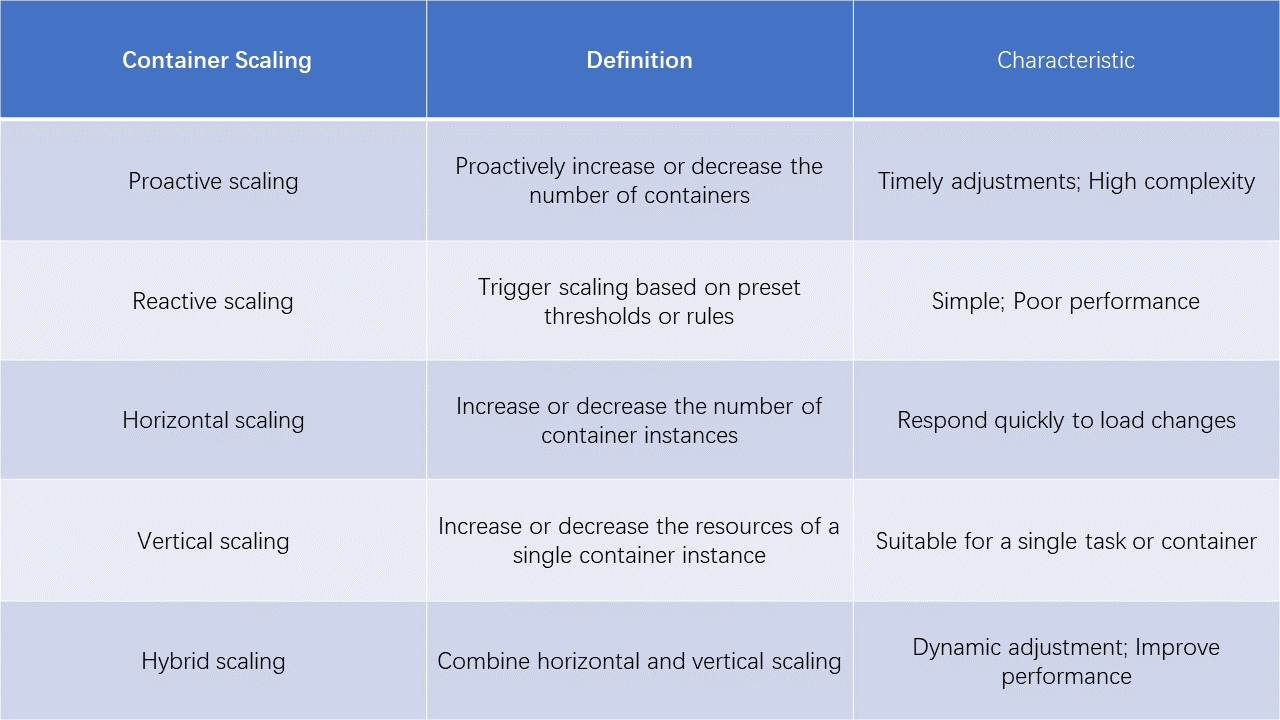}
\caption{Different types of container scaling}
\label{container scaling}
\end{figure*}

Regarding container scaling\cite{rossi2019horizontal} in edge AI, there are two main ways to actively scale and passively scale from the perspective of system response. From the expansion mode, there are mainly horizontal expansion, vertical expansion and hybrid expansion. As shown in Fig.~\ref{container scaling}, we make a detailed comparison of these two categories.

\begin{figure}[h]
    \centering
    \includegraphics[width=3.5in]{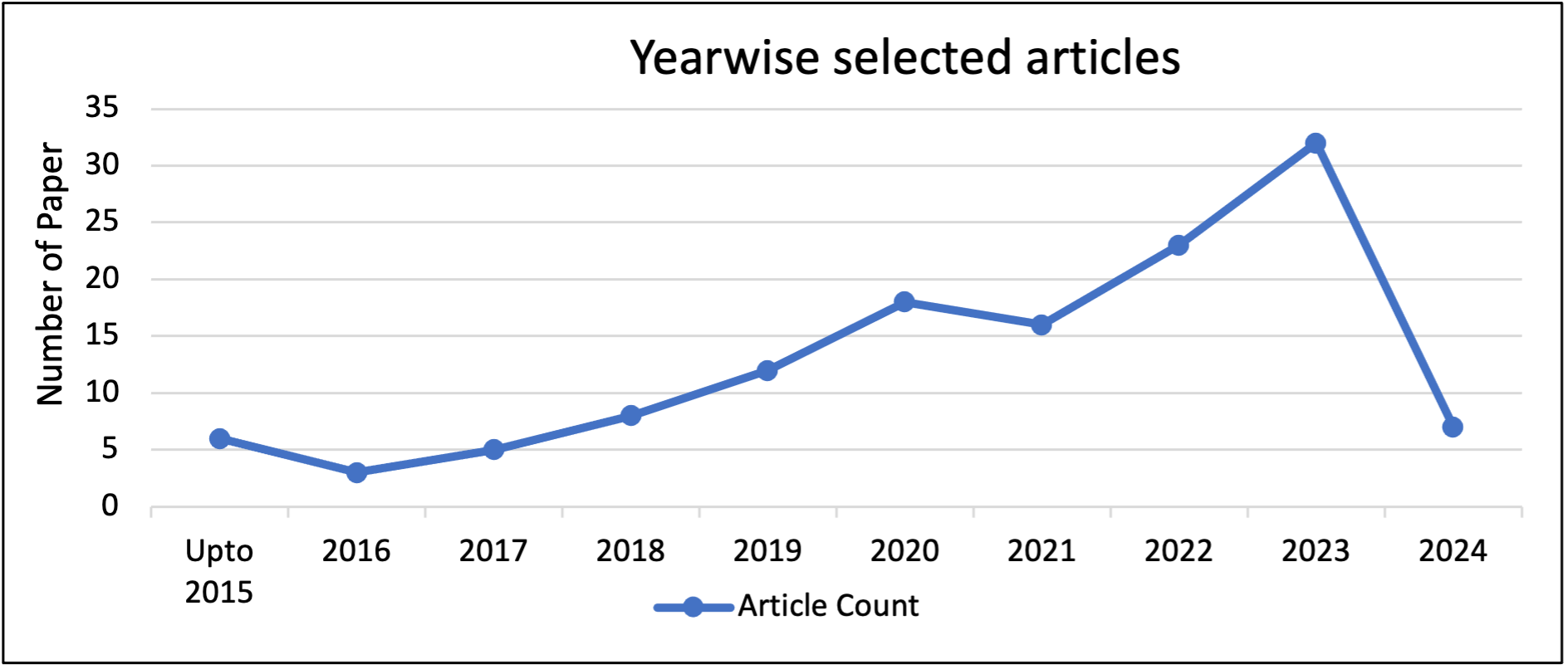}
    \caption{Year wise Publication}
    \label{fig:yearwisepublication}
\end{figure}

\section{Analysis and Result Outcomes}\label{sec:analysis}
The survey enriches various prospects associated with Edge-empowered AI such as infrastructure support, IoT use cases, resource management strategies, security concerns and many more in the form of various state-of-the-art studies. The authors have systematically reviewed numerous articles in order to understand the prevailing status of Edge AI in distinctive domains along with intelligent paradigms like ML and DL. There are lots of works going in this direction to improve the lifestyle of people and solve real-time problems. Hence, this section signifies the importance of our work referred to in the form of year-wise papers, publication count, type of implementation (Simulation or experimental-based) and nevertheless QoS parameters addressed. Fig. \ref{fig:yearwisepublication} presents the year-wise analysis of related work carried out in the form of a number of papers referred from each year. The taxonomy of our study has been proposed with reference to articles from year 2015 to 2024. As depicted in Fig. \ref{fig:yearwisepublication}, it is concluded that a major chunk of the referred articles is recent and are from the year 2023. This clearly illustrates the fact that our survey includes the latest work done by the researchers.

\begin{figure}[h]
    \centering
    \includegraphics[width=3.5in]{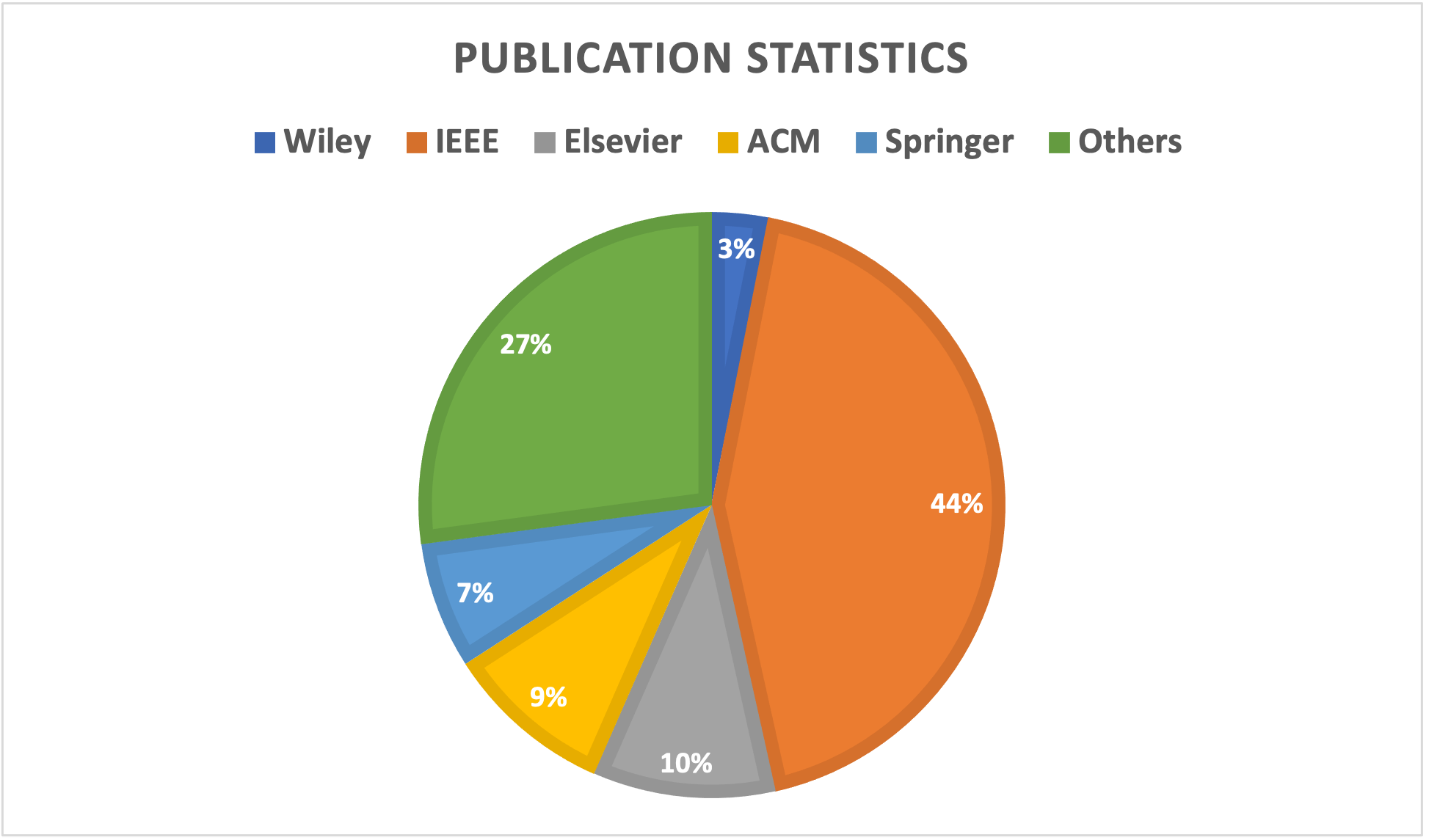}
    \caption{Publication Statistics}
    \label{fig:publicationStats}
\end{figure}

\begin{figure}[h]
    \centering
    \includegraphics[width=3.5in]{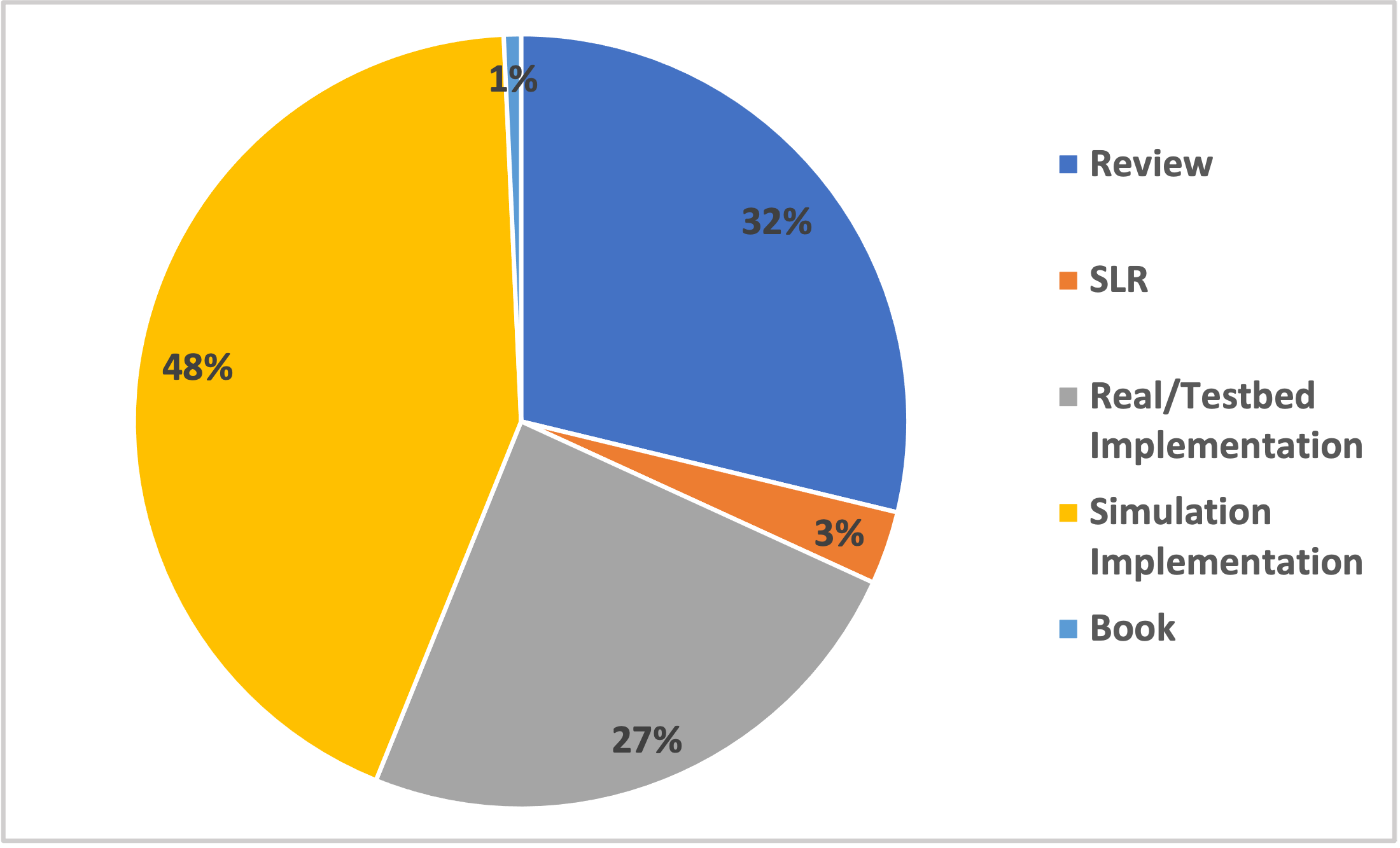}
    \caption{Categorization of Articles}
    \label{fig:ArticleCategorization}
\end{figure}

\begin{figure*}[h]
    \centering
    \includegraphics[width=7in, height=3.7in]{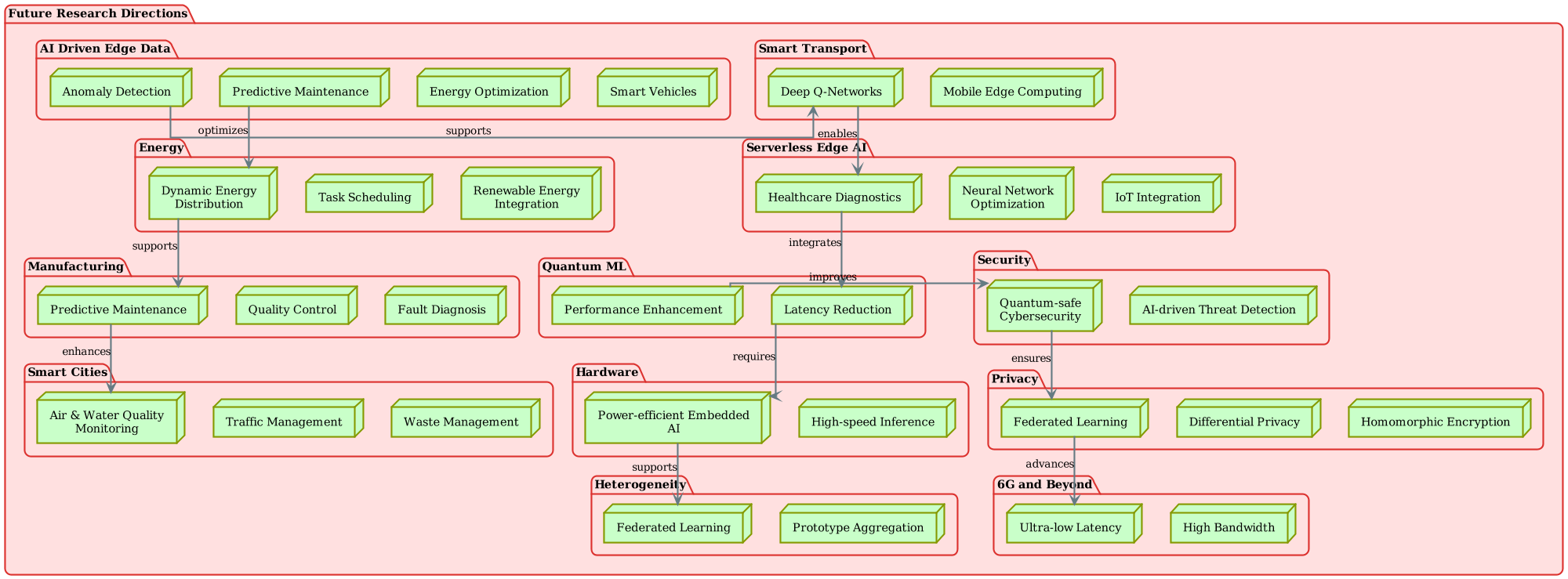}
    \caption{Summary of Future Research Directions}
    \label{fig:futuredirections}
\end{figure*}

Apart from that, we rigorously reviewed the publication-based statistics for the extensive study conducted highlighting its importance in real-time data processing. In total 1253 articles were collected during the data collection phase from various sources such as IEEE, ACM, Wiley, Science Direct, Taylor \& Francis and Springer. Afterwards, the filtering stage excluded collected articles based on redundancy and inclusion and Exclusion criteria. The final stage comprises articles that the authors believe contributed the most towards shaping up the survey as depicted in Fig. \ref{fig:publicationStats}. Furthermore, the articles have been thoroughly reviewed and divided into 4 categories: review, Systematic Literature Review (SLR), implementation (simulation-based), implementation (Real/Testbed-based) and book. Fig. \ref{fig:ArticleCategorization} illustrates that the major portion of the articles referred, based on implementation (simulation-based), which signifies the fact that the implication of edge-empowered AI is yet to be tested on real-life IoT-based use cases. Several real-time works have been proposed by the researchers in recent years to improve the IoT applications-based architecture using intelligent paradigms like ML, DL and reinforcement learning.  This highlights a potential research direction for future studies to explore and validate edge-empowered AI in practical, real-world IoT environments. In addition, this article will motivate the researchers to propose novel solutions to improve society 5.0.

\section{Future Research Directions}\label{sec:future}

Edge AI is continuously evolving and showing potential across various domains, offering numerous opportunities for innovation and improvement. This section examines critical future research directions that promise to enhance the capabilities and applications of Edge AI as summarized in Figure \ref{fig:futuredirections}. These directions include optimizing energy use, strengthening security, and integrating with next-generation networks like 6G, highlighting the transformative impact of Edge AI across multiple sectors.

\subsection{AI-driven Edge Data} 

AI-equipped edge devices process data locally, enabling real-time decision-making without the latency caused by cloud transmission. This is \textcolor{black}{essential for latency-sensitive applications such as predictive maintenance, real-time video analytics, and autonomous systems. In predictive maintenance, for instance, AI models deployed at the edge process sensor data from machines to predict failures and schedule repairs before breakdowns occur. These systems often rely on time-series forecasting models, such as Long Short-Term Memory (LSTM) networks, which are optimized for real-time edge inference through techniques like pruning and quantization \cite{Kristiani2020Optimization}.}

Using video, audio, and sensor data for anomaly detection, edge AI enhances security in smart cities by identifying threats in real-time \cite{Saha2023AI-Driven}. \textcolor{black}{Anomaly detection often employs CNNs or RNNs to detect irregularities in sensor streams. CNNs, such as MobileNet, use depthwise separable convolutions to reduce the number of operations while preserving spatial hierarchies, making them suitable for resource-constrained edge devices. To further optimize performance, techniques like model compression and pruning are used to reduce the model size without significantly compromising accuracy \cite{Kristiani2020Optimization}. RNN-based architectures, including Gated Recurrent Units (GRUs), are also deployed at the edge for detecting anomalies in sequential data, such as traffic patterns or environmental sensors.}

\textcolor{black}{Reinforcement learning algorithms are widely used to manage energy consumption at the edge, particularly for dynamic resource allocation. In HVAC systems, RL models learn optimal policies to balance energy consumption and occupant comfort by interacting with the environment and receiving feedback in the form of rewards \cite{Liang2022ModeldrivenCR}. In these systems, deep Q-learning (DQN) is used to handle large state-action spaces efficiently. Edge devices also deploy on-policy methods like Proximal Policy Optimization (PPO), which allow continuous adjustments based on real-time data. These models are further optimized through techniques like experience replay, which reduces memory usage and computation load, critical for edge deployments.}

Edge AI plays a pivotal role in smart vehicles and drones, where real-time sensor data processing is crucial for navigation and decision-making \cite{Shi2020Communication-Efficient}. \textcolor{black}{Autonomous vehicles rely on edge AI to process data from LIDAR, RADAR, and cameras using sensor fusion techniques, which combine multiple sensor inputs to improve accuracy and robustness \cite{Zou2023Robust}. For instance, Kalman filtering is employed to integrate noisy sensor measurements, while CNNs perform object detection and classification. Edge AI uses these techniques to make split-second decisions, such as obstacle avoidance, without cloud dependencies. Privacy-preserving models like federated learning are also crucial in this context, enabling local data processing on vehicles while sharing only model updates, ensuring that sensitive location data remains private. Advanced techniques such as differential privacy and homomorphic encryption are integrated into FL to protect against data leakage during model updates.}

\textcolor{black}{However, deploying AI at the edge presents significant challenges, particularly in terms of computational limitations and energy constraints. Techniques like model quantization, where neural network weights are reduced from 32-bit floating point to 8-bit integers, help decrease the model size and improve inference speed \cite{Kristiani2020Optimization}. Moreover, hardware-specific optimizations, such as leveraging the parallelism of Tensor Processing Units (TPUs) or Graphics Processing Units (GPUs), play a crucial role. For example, Google’s Edge TPU accelerates inferencing tasks with high energy efficiency, while Nvidia's Jetson platform provides scalable computing power for more complex tasks.}

In edge computing, the communication paradigm plays a critical role in system performance, especially in distributed learning models like Federated Learning \cite{Shi2020Communication-Efficient}. \textcolor{black}{FL enables devices to compute local updates on their own datasets and only transmit model gradients, thereby reducing communication overhead. Challenges such as non-IID (non-independent, identically distributed) data among edge nodes, which can lead to model biases, are addressed through methods like FedProx, which adds a regularization term to prevent drastic divergence from the global model. Additionally, communication-efficient techniques such as gradient sparsification, where only significant gradients are transmitted, and asynchronous updates ensure that edge nodes can update models independently without waiting for synchronization, thus improving efficiency in bandwidth-constrained environments.}

The convergence of edge AI and advanced hardware solutions continues to drive innovations in AI deployments. \textcolor{black}{Custom AI accelerators, such as Google's Edge TPU, are optimized for low-power inference tasks, supporting applications that require high-speed processing, such as object detection in real-time video feeds \cite{Zou2023Robust}. NVIDIA's Jetson platform, on the other hand, provides a scalable solution for more compute-intensive tasks, such as deep learning-based robotics or autonomous navigation. These platforms support parallelized inference operations, maximizing throughput while minimizing energy consumption, making them ideal for edge environments. Future research directions include developing neuromorphic hardware, which mimics biological neural networks, offering significant reductions in energy consumption while maintaining high processing speeds.}

\subsection{Energy}

Optimizing energy use in AIoT systems through intelligent edge computing requires focusing on sophisticated algorithms for dynamic energy distribution, task scheduling, ML for workload management, and the design of low-power hardware \cite{du2023computation}. \textcolor{black}{For example, dynamic energy distribution can be handled using RL-based approaches where the system continuously learns and adapts to energy usage patterns. Techniques such as DQN and Multi-agent Reinforcement Learning (MARL) are effective for decentralized energy optimization, allowing edge devices to work autonomously while coordinating with other nodes in the network to maximize efficiency. MARL models allow each device to act as an independent agent, optimizing energy at both local and global levels by sharing information across the network \cite{Johnson2022Multi-Agent}. Furthermore, task scheduling in AIoT systems can benefit from heuristic algorithms like Genetic Algorithms (GA) and Particle Swarm Optimization (PSO), which provide near-optimal solutions for task allocation in energy-constrained environments. These methods are computationally efficient, making them well-suited for edge devices with limited resources \cite{Dong2020Task,Chhabra2020Multi-criteria}.}

Improvements in communication protocols and integration of renewable energy sources will further enhance system efficiency and scalability. \textcolor{black}{Protocols such as Low-Power Wide Area Networks (LPWANs), including LoRaWAN and NB-IoT, are particularly useful for AIoT systems as they allow for long-range communication with minimal power consumption. These protocols, when combined with AI models running at the edge, enable real-time data exchange between devices while conserving energy. Additionally, optimizing the scheduling of data transmissions based on energy availability or demand-response signals can significantly reduce communication overhead in energy-constrained environments \cite{Roeder2021Energy-aware}.}

Literature \cite{Yousef2023Artificial,chowdhury2023covidetector} reported that AI techniques for managing renewable energy sources have been investigated, emphasizing advanced ML models for accurate forecasting and optimizing energy storage and grid integration. \textcolor{black}{For instance, LSTM networks and Autoregressive Integrated Moving Average (ARIMA) models have been widely applied in energy forecasting tasks. LSTMs are especially effective in capturing the temporal dependencies in renewable energy generation data (e.g., solar and wind), allowing for more accurate predictions of energy availability. These forecasts can then be used to optimize the allocation of tasks across the grid, improving the balance between energy demand and supply. In energy storage, AI models like Gradient Boosting Machines (GBM) and Support Vector Machines (SVM) have been applied to optimize the charge-discharge cycles of batteries, maximizing the longevity and efficiency of storage systems \cite{Wang2023Task}. Integrating these AI models with grid control systems enables real-time decision-making to manage fluctuations in energy generation and consumption effectively.}

Implementing edge AI will reduce latency and enable real-time decision-making, increasing system responsiveness and resilience \cite{energyedge1}. \textcolor{black}{In edge computing environments, AI models can be used to make decisions locally without needing to transmit large amounts of data to a central cloud, which reduces the overall latency and bandwidth usage. This is especially beneficial in scenarios where immediate action is required, such as energy demand-response events or real-time fault detection in smart grids. Edge AI, combined with lightweight models such as MobileNet or TinyML, can process sensor data in real time, detecting anomalies or optimizing energy usage without draining significant computational resources. This local processing not only reduces decision latency but also enhances the system's resilience to network disruptions.}

It has been identified that incorporating edge AI in the Internet of Energy (IoE) presents unique opportunities, particularly in areas like secure edge computing, blockchain for data security, lightweight AI algorithms, standardization for interoperability, and 5G networks for low-latency communication \cite{Himeur2023Edge,doyle2022blockchainbus,golec2022aiblock}. \textcolor{black}{Blockchain technology offers decentralized solutions for secure energy trading and data management within IoE networks. The integration of blockchain with edge AI can ensure the transparency and immutability of energy transaction data while minimizing the energy overhead typically associated with blockchain’s consensus mechanisms. By using energy-efficient consensus protocols like Proof of Stake (PoS) or Proof of Authority (PoA), IoE systems can maintain security without incurring significant computational costs. Additionally, 5G and the upcoming 6G networks offer ultra-low latency and high-speed connectivity, enabling real-time energy optimization by allowing AI models to dynamically offload tasks between the edge and cloud based on energy constraints. Network slicing within 5G/6G networks ensures that energy-critical tasks receive prioritized resources, improving the overall performance and responsiveness of the IoE system.}

Advances in federated edge AI and DRL will optimize energy distribution, enhancing the efficiency and resilience of IoE systems \cite{singh2023next}. \textcolor{black}{Federated Learning allows edge devices to collaboratively train a global model without exchanging raw data, thereby preserving privacy and reducing communication overhead. In energy management, FL can be adapted to enable energy-aware federated updates, where devices with limited energy resources reduce their participation in the model training process\cite{Wang2023Task}. This approach ensures that the overall system remains efficient, even when certain devices operate under low-power conditions. DRL further improves energy optimization by learning policies for real-time control of energy resources, such as dynamically adjusting the operation of energy storage systems or scheduling energy-consuming tasks during periods of high renewable energy generation. These algorithms, when combined with edge AI, enable IoE systems to respond to changes in energy supply and demand with minimal delay.}

\subsection{Manufacturing}

AI meets rising customer expectations for customization and high-value production by integrating capabilities at the network edges. \textcolor{black}{The integration of edge AI in manufacturing introduces opportunities for decentralized, low-latency processing, which is crucial for real-time operations in smart factories. Edge AI enables data-driven decision-making directly on the factory floor, minimizing delays due to communication with centralized cloud servers \cite{Vermesan2022An}. This shift towards distributed intelligence allows manufacturers to rapidly adapt to changing production demands, offering more customized products and services. Furthermore, AI-driven automation helps reduce manual labor, improving both productivity and safety in industrial environments.}

It has been identified that integrating AI in manufacturing enhances collaboration with experts through tools like Google VisionAI for data science applications \cite{kovalenko2023opportunities}. \textcolor{black}{Vision-based AI tools, such as Google VisionAI, enable real-time monitoring and analysis of production lines by processing images and video streams at the edge. These tools can be integrated with machine vision systems to automatically inspect product quality, detect defects, and identify anomalies in manufacturing processes. By deploying AI models at the edge, manufacturers benefit from low-latency responses, which are critical for high-speed production environments where defects need to be detected and corrected in real time. This significantly reduces the waste associated with defective products, leading to increased production efficiency and cost savings \cite{Cinar2020Machine}.}

Key research areas include manufacturing scheduling and planning due to abundant data and productivity improvement opportunities. \textcolor{black}{In the context of scheduling and planning, AI can optimize the allocation of resources and machinery to minimize downtime and maximize throughput. Traditional scheduling algorithms, such as job shop scheduling, have limitations in handling the complexity and variability of modern manufacturing environments. AI techniques, such as RL and constraint satisfaction algorithms, are being applied to dynamically adjust production schedules in response to real-time data \cite{Chien2020Artificial}. For example, RL can be used to model complex environments with multiple variables, learning optimal policies to allocate resources efficiently while adapting to unforeseen disruptions, such as machine failures or supply chain delays. Additionally, advanced optimization algorithms like GA and PSO can be integrated into manufacturing planning systems to solve multi-objective optimization problems, balancing factors like energy consumption, production speed, and quality \cite{Ying2021Edge-enabled}.}

There is a need to investigate edge AI's real-time analysis for predictive maintenance, quality control, and fault diagnosis in manufacturing, improving efficiency, reducing waste, and optimizing resources \cite{NAIN2022588}. \textcolor{black}{Predictive maintenance is a key area where edge AI can significantly enhance manufacturing operations. Using AI models like RNNs and LSTM networks, edge devices can analyze data from sensors embedded in machinery to predict when a machine is likely to fail. This allows manufacturers to perform maintenance only when necessary, reducing downtime and extending the life of equipment. In addition, real-time fault detection models deployed at the edge can immediately identify deviations from normal operating conditions, allowing for immediate corrective actions. Techniques such as CNNs can be employed for real-time image and video analysis in quality control, detecting surface defects, dimensional inaccuracies, and assembly errors with high accuracy \cite{Ringler2023Machine}. Implementing AI for fault diagnosis can also leverage unsupervised learning methods, such as autoencoders and clustering algorithms, to identify abnormal patterns in sensor data without needing labeled fault data.}

Implementing ML models at the edge allows continuous monitoring, early fault detection, and immediate corrective actions, enhancing intelligent manufacturing. \textcolor{black}{The real-time processing capabilities of edge AI offer a significant advantage for continuous monitoring in manufacturing environments. By running ML models at the edge, data from production lines can be processed in real time, enabling early detection of equipment malfunctions and process anomalies. For example, edge devices can run anomaly detection algorithms using principal component analysis (PCA) or one-class SVMs to flag deviations from normal production patterns. This early detection helps prevent costly downtime and reduces the risk of producing defective products, enhancing overall manufacturing efficiency\cite{Pule2022Application}.}

Additionally, edge devices can optimize power generation and consumption by analyzing real-time data, promoting renewable energy use and cost savings. \textcolor{black}{Energy efficiency is becoming a critical aspect of modern manufacturing systems, especially as manufacturers strive to reduce their carbon footprint. Edge AI can be integrated with energy management systems to monitor power usage in real time, allowing for dynamic optimization of energy consumption across various production processes. Machine learning algorithms, such as reinforcement learning and Bayesian optimization, can be employed to balance power consumption with production goals. For example, RL-based systems can learn optimal policies for turning machines on or off based on current energy prices, renewable energy availability, and production schedules. This integration of edge AI with energy optimization not only reduces operational costs but also aligns with sustainable manufacturing practices \cite{Yu2020A}.}

\subsection{Smart Cities}

The future of Edge AI in smart cities holds the promise of several significant advancements \cite{9591732}. \textcolor{black}{Edge AI enables real-time data processing, allowing cities to respond to dynamic situations instantly without relying on cloud processing. By deploying AI models at the edge, data from sensors (e.g., air quality monitors, traffic cameras, and IoT waste bins) can be analyzed locally, reducing the need for constant data transmission to centralized servers. This not only lowers the load on cloud infrastructure but also enhances the responsiveness of city services. For example, in traffic management, edge AI can detect congestion or accidents in real time using computer vision models such as CNNs applied to video feeds. These systems can dynamically adjust traffic signals or reroute vehicles based on real-time conditions, reducing congestion and improving urban mobility.\cite{Pule2022Application}}

First, enhancing data processing capabilities at the edge reduces cloud load and latency, enabling real-time decision-making for applications like air and water quality monitoring, traffic management, and waste management. \textcolor{black}{In air and water quality monitoring, edge AI can analyze sensor data from distributed nodes throughout a city to detect harmful pollutants or water contamination. Machine learning models such as random forests, SVMs, or deep learning models like LSTMs can be employed at the edge to predict air quality trends based on historical data and real-time inputs. These models can continuously adjust ventilation systems in buildings or trigger alarms in high-risk areas, ensuring a more responsive and automated environmental management system. Similarly, edge devices in smart waste management systems can monitor bin levels using IoT sensors, and AI models can optimize waste collection routes by predicting when and where waste accumulation is likely to occur, thus reducing operational costs and environmental impact\cite{Cojbasic2023Application}.}

Integration with AI algorithms will facilitate more intelligent decision-making by analyzing sensor data from various city domains. \textcolor{black}{For example, AI-driven edge systems can aggregate and process data from traffic lights, parking sensors, public transport, and emergency services to optimize city-wide mobility. RL models can be used to manage traffic lights dynamically, learning from past traffic flow patterns to minimize delays and congestion.\cite{Wang2022Deep} RL algorithms such as DQN or PPO can be applied in such environments, where the model learns to optimize traffic signals based on real-time sensor data, improving traffic flow efficiency over time. Additionally, AI-based predictive models like LSTMs can forecast urban energy demand by analyzing sensor data from smart meters across the city. This allows utility companies to balance energy generation and distribution in real time, reducing the risk of blackouts and improving energy sustainability\cite{Wang2023A}.}

Innovations in low-power hardware and efficient communication protocols are crucial to ensure scalability and energy efficiency. \textcolor{black}{Edge AI systems in smart cities must be designed with energy efficiency in mind, particularly given the large number of distributed edge devices required. Specialized hardware accelerators, such as Google’s Edge TPU or Nvidia’s Jetson Nano \cite{Ferraz2023Benchmarking}, are designed to perform AI inference with minimal energy consumption, making them ideal for edge deployments in smart cities. These hardware solutions are typically paired with lightweight AI models like MobileNet or TinyML to further reduce power consumption while maintaining high accuracy. Additionally, communication protocols such as \textbf{LoRaWAN and Narrowband IoT (NB-IoT)\cite{Marini2022Low-Power}} are optimized for low-power, wide-area networks, ensuring that edge devices can communicate over long distances without significant energy overhead. Integrating these protocols with AI-driven edge systems ensures that smart cities can scale without overwhelming energy resources.}

Developing secure edge computing and blockchain technologies will also address data security and privacy concerns \cite{gharaibeh2017smart}. \textcolor{black}{One of the key challenges in deploying AI at the edge is ensuring data security and privacy. Edge devices often handle sensitive information, such as personal location data or surveillance footage, making them attractive targets for cyberattacks. To address this, blockchain technology can be integrated with edge AI to create a decentralized and secure method of data management\cite{Nguyen2021Federated}. Smart contracts can automate the verification and exchange of data between devices, ensuring that only authorized entities can access or modify the data. Furthermore, blockchain-based systems can offer tamper-proof audit trails, ensuring transparency and accountability in data usage. Privacy-preserving AI techniques, such as federated learning and differential privacy\cite{Adiwijaya2023Federated}, can also be implemented at the edge to allow AI models to be trained on decentralized data without exposing sensitive information. Federated learning allows edge devices to collaboratively train a global model without sharing raw data, while differential privacy ensures that individual data points cannot be reverse-engineered from model outputs.}

Finally, advancements in 5G networks will provide the necessary infrastructure for high-speed, low-latency communication, further enhancing the responsiveness and resilience of intelligent city systems. \textcolor{black}{The deployment of 5G technology in smart cities enables real-time, high-bandwidth communication between millions of connected devices. 5G’s ultra-reliable low-latency communication (URLLC)\cite{Mutalemwa2020A} can significantly enhance applications that require instantaneous responses, such as autonomous vehicles, real-time traffic management, and emergency response systems. Additionally, network slicing\cite{Antevski2021A} within 5G allows different services to have dedicated virtual networks with tailored resources, ensuring that critical services like emergency response or traffic control always receive the necessary bandwidth and priority. As smart cities evolve, the combination of edge AI and 5G will be crucial for enabling real-time decision-making, where AI models deployed at the edge can interact with central cloud systems when needed, without suffering from communication delays. Future advancements in 6G networks\cite{Adhikari20226G-Enabled} will likely take these capabilities further by offering even higher data transfer speeds and supporting more complex AI models in real-time applications.}

\subsection{Smart Transport}

Applying DRL, specifically DQN, to mobile edge computing in smart transportation systems plays a crucial role in balancing computing capability and traffic state. \textcolor{black}{DQN, as a value-based reinforcement learning algorithm, operates by learning a Q-function that maps states (such as traffic conditions) to the expected rewards of taking certain actions (like adjusting traffic lights or rerouting vehicles). In smart transport, DQN-based models~\cite{Guo2023DQN} are deployed at the edge to optimize local decision-making in real time, minimizing the delays caused by communication with centralized cloud servers. For instance, in an urban environment, a DQN-based system can learn to adjust traffic signals based on current traffic flow, historical patterns, and predicted congestion, thus improving traffic efficiency \cite{Liu2023Deep} and reducing fuel consumption. The use of edge computing here ensures low-latency decision-making, which is critical in dynamically evolving traffic situations where even minor delays can significantly impact traffic flow.}

This approach highlights the need for further research on trade-offs and optimization techniques to enhance efficiency and performance in edge AI applications \cite{10207661}. \textcolor{black}{In the context of smart transport, there are several trade-offs to consider, particularly in the computational complexity of DRL models versus the energy and processing limitations of edge devices. DRL models like DQN, while effective, can be computationally intensive due to the large state-action space that must be explored. To address this, techniques such as \textbf{Double DQN} and \textbf{Dueling DQN}\cite{Cheng2022Task} have been introduced to improve the stability and efficiency of Q-learning by reducing overestimation biases and learning more granular value functions. These variants reduce the number of updates required for the Q-function to converge, which is critical in resource-constrained edge environments.}

\textcolor{black}{In addition to improving algorithmic efficiency, optimizing resource allocation in mobile edge computing environments is an ongoing area of research. Traffic state optimization is a multi-objective problem that involves balancing computational load, energy consumption, and communication latency. Techniques such as \textbf{multi-agent DRL (MADRL)\cite{Seid2021Multi-Agent}} can be applied, where each vehicle or edge device is treated as an independent agent learning to optimize its local performance while contributing to the global traffic management system. MADRL\cite{Peng2023Joint} allows for decentralized decision-making, where agents can communicate with each other or a central node to share state information (such as traffic density or road conditions), thus improving the coordination of traffic signals and vehicle routing across a city. The challenge lies in managing the communication overhead, which increases with the number of agents, while maintaining real-time performance.}

\textcolor{black}{Further research is also needed to explore hierarchical reinforcement learning (HRL)\cite{Gao2022Cola-HRL:}, which can decompose complex traffic control tasks into a hierarchy of simpler sub-tasks. HRL enables smart transport systems to break down large-scale optimization problems (e.g., optimizing city-wide traffic flow) into smaller, manageable sub-problems (e.g., optimizing traffic flow at individual intersections). This reduces computational overhead and makes the learning process more scalable, particularly when applied at the edge. Additionally, \textbf{PPO}\cite{Dai2020Queueing}, a popular policy-based DRL algorithm, could be explored for continuous control in smart transport applications. PPO is known for its robustness in high-dimensional environments and could improve the real-time adaptability of transport systems to unpredictable traffic conditions or sudden changes in road infrastructure.}

\textcolor{black}{To further optimize the integration of DRL in smart transportation, future research should focus on energy-efficient hardware accelerators such as Google’s Edge TPU or Nvidia’s Jetson\cite{Baller2021DeepEdgeBench:}, which are designed to handle AI workloads with minimal power consumption. These devices can run DRL models locally, allowing edge nodes (e.g., traffic lights or smart vehicles) to process large volumes of data in real time without overburdening the power infrastructure. Furthermore, low-power communication protocols, such as \textbf{Vehicular Ad-hoc Networks (VANETs)\cite{Pan2022Artificial}}, can be integrated with edge AI systems to ensure efficient data sharing among vehicles and infrastructure while minimizing energy usage. VANETs allow vehicles to communicate with roadside units (RSUs) and other vehicles in real time, enabling cooperative decision-making for optimized traffic control.}

\textcolor{black}{Finally, with the advancement of 5G networks, mobile edge computing in smart transportation will benefit from ultra-low-latency communication, enabling more efficient interaction between vehicles, infrastructure, and edge devices. The use of \textbf{network slicing} in 5G networks can provide dedicated virtual network resources for traffic management, ensuring that time-sensitive tasks, such as emergency vehicle routing or accident detection, are handled with priority. This integration of 5G\cite{Antevski2021A} with edge AI will allow transport systems to scale effectively, handling large volumes of data with minimal delay while maintaining energy efficiency. Future research may also explore the potential of \textbf{6G networks}, which are expected to provide even faster data transfer speeds and greater network capacity, allowing for the deployment of more sophisticated AI models at the edge in smart transport systems\cite{Adhikari20226G-Enabled}.}

Applying DRL to mobile edge computing in smart transport offers numerous opportunities for optimizing traffic management and improving transportation efficiency. However, ongoing research is required to address the trade-offs between model complexity, computational resources, and energy consumption in edge devices, while leveraging the latest advancements in communication protocols and network technologies.

\subsection{Serverless Edge AI}

Leveraging the flexibility and scalability of serverless architectures will significantly enhance the deployment of ML models in healthcare \cite{10440418,golec2023cold}. \textcolor{black}{Serverless computing offers a key advantage by abstracting the underlying infrastructure, allowing developers to focus solely on deploying and scaling machine learning (ML) models without needing to manage servers. This architecture also allows for automatic scaling based on demand, meaning ML models can be deployed in a cost-efficient manner. Specifically in healthcare, real-time diagnostics and monitoring are vital for patient care, and serverless edge AI\cite{Eapen2020Serverless} enables low-latency data processing at the edge without needing continuous cloud connectivity. For instance, ML models can be used to process patient data directly from IoT devices such as wearable sensors, providing immediate alerts for abnormal health conditions like arrhythmias or glucose fluctuations.} 

This approach will enable real-time, cost-effective diagnostics without managing backend infrastructure. \textcolor{black}{One of the key challenges in serverless environments, however, is managing the "cold start" problem \cite{Solaiman2020WLEC:}, where there is a delay in invoking a function due to the time it takes to spin up resources when a function is first triggered. In healthcare, where every second matters, reducing cold start latency is critical. Future research could explore techniques such as \textbf{pre-warming} serverless functions by maintaining a pool of ready-to-go containers or employing \textbf{just-in-time compilation (JIT)} to reduce function invocation time. Additionally, lightweight neural network architectures such as \textbf{TinyML} or \textbf{MobileNet} can be optimized for serverless environments, minimizing the computation load while maintaining high accuracy in diagnostics tasks.} 

Future research will optimize neural network models for serverless environments, reduce cold start latencies, and enhance model performance through adaptive learning techniques \cite{senjab2023survey}. \textcolor{black}{Adaptive learning techniques can further improve the efficiency of ML models deployed in serverless architectures. For instance, \textbf{on-demand model loading} could allow serverless functions to dynamically load specific portions of a neural network based on the current task, thus reducing memory and processing requirements. This concept, known as \textbf{model partitioning}, divides a model into smaller, callable sections, enabling efficient utilization of serverless resources. Furthermore, the use of \textbf{quantization} and \textbf{model pruning} techniques can reduce the model size, ensuring faster execution times and lower computational overhead, which is crucial for real-time diagnostics in healthcare settings. These optimizations not only reduce latency but also lower the costs associated with serverless computing by minimizing the resources consumed during function execution.}

Integrating serverless edge AI with IoT frameworks will facilitate continuous monitoring and rapid response in medical applications \cite{10517641}. \textcolor{black}{In healthcare, IoT devices such as smartwatches, biosensors, and connected medical devices generate a constant stream of health data that requires real-time analysis. By integrating \textbf{serverless edge AI} with these IoT frameworks, the data can be processed locally on the device or at nearby edge servers, reducing latency and ensuring real-time feedback to both patients and healthcare providers. This continuous monitoring can be critical in managing chronic conditions like diabetes or cardiovascular disease, where immediate response to changes in vital signs is essential. In the case of \textbf{federated learning}, ML models can be collaboratively trained across multiple edge devices without sharing sensitive health data, further enhancing patient privacy and complying with healthcare regulations like HIPAA, GDPR, etc.}

Additionally, improving interoperability and security protocols will ensure the safe and efficient handling of sensitive healthcare data in serverless architectures, paving the way for the next generation of intelligent, responsive, and secure healthcare systems. \textcolor{black}{One of the major challenges in serverless edge AI for healthcare is ensuring the security and privacy of sensitive patient data. The distributed nature of serverless architectures, combined with the use of IoT devices, increases the attack surface for cyber threats. Future research should focus on developing secure execution environments such as \textbf{trusted execution environments (TEEs)}, which provide hardware-based security features that isolate sensitive data during function execution. TEEs can be integrated with \textbf{differential privacy} and \textbf{homomorphic encryption}\cite{Wei2023THE-V:} to allow secure data processing and model training without exposing raw data. Moreover, ensuring \textbf{interoperability} across different healthcare systems and IoT devices is crucial to enable seamless data sharing and processing. \textbf{Standardization efforts} in APIs, data formats (such as HL7 or FHIR for healthcare data)\cite{Jaffe2023Implementing}, and communication protocols will be vital to ensuring compatibility across platforms, reducing the complexity of integrating serverless edge AI into existing healthcare infrastructure. These advancements will make it possible to deploy scalable, secure, and efficient healthcare systems that leverage the full potential of serverless computing.}

\subsection{Quantum Machine Learning (QML)}

Integrating Quantum Machine Learning (QML) with edge AI technologies, such as large, intelligent surfaces and visible light communications, will significantly reduce latency and enhance performance \cite{gill2024quantum}. \textcolor{black}{QML leverages the principles of quantum computing, such as quantum superposition and entanglement\cite{Dupont2022Entanglement}, to process data more efficiently than classical computing methods. By integrating QML with edge AI, it becomes possible to execute complex machine learning algorithms in parallel, drastically reducing computation time. For example, quantum algorithms such as the \textbf{Quantum Approximate Optimization Algorithm (QAOA)\cite{Zhang2020QED}} and \textbf{Variational Quantum Eigensolver (VQE)}\cite{Kirby2020Variational} can solve optimization problems faster than their classical counterparts, enabling real-time decision-making in edge AI systems. These quantum algorithms are particularly well-suited for tasks like traffic flow optimization, supply chain management, and resource allocation in smart cities, where traditional algorithms struggle with computational complexity.}

This combination enables efficient processing and decision-making at the network edge, which is crucial for managing the vast data generated by IoT devices and other edge sources \cite{golec2024quantum}. \textcolor{black}{As the number of connected IoT devices grows exponentially, classical edge AI systems face significant challenges in processing large volumes of data in real time. By incorporating quantum-enhanced models, edge AI can offload more complex computations to quantum processors, allowing for faster and more efficient data analysis. For instance, \textbf{Quantum SVMs (QSVMs)} \cite{Du2020On} can be used to classify high-dimensional data generated by IoT sensors more efficiently than classical SVMs, leading to faster and more accurate decision-making at the edge. Additionally, \textbf{Quantum Neural Networks (QNNs)}\cite{Zhao2021QDNN:} have the potential to significantly reduce the training time for deep learning models, allowing edge AI systems to adapt more quickly to changing environments. This is especially important in autonomous systems, such as self-driving cars or drones, where real-time responsiveness is critical.}

Leveraging quantum computing capabilities in edge AI applications will achieve unprecedented network performance, leading to more responsive and adaptive AI systems \cite{ssgillquantum}. \textcolor{black}{Quantum edge AI systems can utilize \textbf{quantum parallelism} to explore multiple solutions simultaneously, which significantly enhances the performance of optimization and search tasks. For example, in a smart transportation system, quantum-enhanced algorithms could simultaneously evaluate multiple traffic routes to find the most efficient path, significantly reducing computation time compared to classical algorithms. Furthermore, \textbf{quantum annealing} can be used to solve combinatorial optimization problems in real time, which is highly beneficial for applications like dynamic resource allocation in 5G networks.} 

The convergence of QML and edge AI in 6G networks will drive innovative solutions for real-time analytics and intelligent automation, meeting the increasing demands for low-latency and high-efficiency edge computing environments \cite{8681450}. \textcolor{black}{6G networks are expected to provide ultra-reliable low-latency communication (URLLC), which is essential for supporting the high-speed data exchange required by quantum-enhanced edge AI systems. By integrating QML with 6G networks, edge devices can harness the power of \textbf{quantum communication protocols}, such as \textbf{quantum key distribution (QKD)}, to ensure secure data transmission between devices and the cloud. QKD offers provably secure communication, which is particularly important for applications like autonomous vehicles and smart grid management, where data integrity is critical. Additionally, the high bandwidth offered by 6G will enable edge devices to offload quantum computations to nearby quantum processors with minimal delay, further enhancing the system's overall responsiveness \cite{ssgillquantum}.}

\textcolor{black}{Another area of research is the use of \textbf{quantum machine learning for federated learning (QFL)} in edge AI environments. In traditional federated learning, edge devices collaboratively train a global model without sharing their local data, which minimizes privacy risks. However, with the increasing complexity of AI models, the communication overhead in FL can become a bottleneck. By using QFL, edge devices can leverage quantum algorithms to reduce the amount of data that needs to be shared, as quantum communication enables the transmission of compressed information with higher efficiency. This can lead to faster convergence times in federated learning models while maintaining the privacy of sensitive data, which is crucial in sectors like healthcare and finance.}

\textcolor{black}{Moreover, future research should focus on developing \textbf{quantum-classical hybrid algorithms} \cite{Bravyi2020Hybrid} that can efficiently combine the strengths of both quantum and classical computing. In scenarios where quantum hardware is limited, hybrid approaches can offload specific sub-tasks (such as optimization or matrix inversion) to quantum processors while performing other parts of the computation on classical hardware. These hybrid systems are particularly well-suited for edge AI, where devices may not have direct access to quantum hardware but can offload tasks to quantum cloud services. This will enable a more scalable and flexible approach to integrating QML in edge AI applications.}

As 6G networks and quantum hardware continue to evolve, the convergence of these technologies will drive new innovations in areas like intelligent automation, secure communications, and resource optimization across smart cities and autonomous systems. \textcolor{black}{The combination of quantum computing and AI at the edge will address critical challenges in real-time processing, allowing for the handling of vast data streams with minimal latency and maximizing the use of network and computational resources \cite{ssgillquantum}.}

\subsection{Hardware}

The physical boundaries for AI systems are set by the hardware of edge nodes, which drives significant efforts in designing specialized edge AI hardware. \textcolor{black}{Edge AI hardware must balance several constraints, including computational performance, power efficiency, size, and cost. This has led to the development of specialized hardware like \textbf{Nvidia’s Jetson TX2}, which is designed for power-efficient embedded AI computing, and \textbf{Google’s Edge TPU}, optimized for high-speed inference at the edge \cite{gill2024modern}. The \textbf{Jetson TX2} integrates GPU-based architectures with AI acceleration cores to handle tasks requiring significant parallel processing power, such as deep learning model inference. It is particularly well-suited for applications like autonomous robots, drones, and intelligent surveillance systems, where real-time processing of sensor data is critical. On the other hand, \textbf{Google’s Edge TPU} focuses on accelerating machine learning models in a cost-effective and energy-efficient manner, often used in IoT devices, smart cameras, and wearable technology for lightweight, high-speed AI tasks. These hardware platforms allow AI models to run at the edge with reduced latency and power consumption compared to traditional cloud-based AI systems.}

However, these devices mainly concentrate on handling entire tasks, especially local edge inference. \textcolor{black}{Moving forward, edge AI hardware design will evolve to address the diverse requirements of various AI workloads. For example, high-performance tasks such as 3D object recognition, complex signal processing, and multi-modal data fusion require more powerful hardware accelerators that can manage large-scale computations at the edge. This is where \textbf{Field Programmable Gate Arrays (FPGAs)} \cite{Kalapothas2022Efficient} and \textbf{Application-Specific Integrated Circuits (ASICs)} come into play. \textbf{FPGAs} offer reconfigurable hardware that can be customized for specific AI tasks, such as accelerating CNNs or RNNs. The flexibility of FPGAs makes them ideal for environments where the AI workload can change dynamically, such as in autonomous vehicles or industrial automation. \textbf{ASICs} \cite{Hu2022A}, on the other hand, provide dedicated hardware that is optimized for specific AI algorithms, offering superior performance and energy efficiency for large-scale deployment in fixed-function systems like smart grids and edge data centers.}

\textcolor{black}{To further enhance energy efficiency, \textbf{neuromorphic processors} \cite{Vitale2022Neuromorphic} are being explored as a promising solution. These processors mimic the structure and operation of the human brain, using spiking neural networks (SNNs) that only consume power when active. This event-driven computation model is highly advantageous for edge applications such as continuous sensor monitoring, where devices need to remain energy-efficient while processing intermittent data streams. For example, neuromorphic chips can be integrated into wearable devices for health monitoring or in environmental sensors for smart city applications, reducing overall power consumption without compromising performance. Research in neuromorphic computing focuses on increasing these processors' scalability and accuracy for more complex AI tasks at the edge.}

Moving forward, we'll see a variety of edge AI hardware designed specifically for different AI system architectures and applications \cite{Shi2020Communication-Efficient}. \textcolor{black}{Future advancements in edge AI hardware will likely focus on co-designing hardware and software to optimize the performance of AI models. This includes developing \textbf{domain-specific architectures (DSAs)} that are tailored for specific AI workloads, such as natural language processing (NLP), computer vision, and reinforcement learning. DSAs will allow hardware to process AI algorithms more efficiently by exploiting the characteristics of the models they are running, such as sparsity in neural networks or the locality of reference in data. Additionally, \textbf{heterogeneous computing architectures}, which combine different types of processing units (e.g., CPUs, GPUs, TPUs, and FPGAs) in a single system, will become more prevalent in edge AI deployments. These architectures enable systems to allocate tasks to the most appropriate processing unit based on the computational and energy requirements of each AI model, optimizing overall system performance.}

\textcolor{black}{As the demand for edge AI grows, there will also be a focus on improving hardware for secure AI processing. This includes developing \textbf{trusted execution environments (TEEs)} that protect sensitive data and AI model integrity in edge devices. TEEs create isolated environments where AI computations can be executed securely, preventing unauthorized access or tampering with data. TEE \cite{Li2022ENIGMA:} will be particularly important in industries such as healthcare, finance, and autonomous driving, where security and privacy are paramount. Hardware support for \textbf{privacy-preserving AI}, such as homomorphic encryption and secure multi-party computation, will also be integrated into edge devices to enable secure, decentralized AI processing without exposing raw data to external threats.}

Edge AI hardware is evolving rapidly to support the increasing complexity and diversity of AI workloads at the edge. \textcolor{black}{This evolution will be driven by advances in specialized hardware accelerators, energy-efficient processing architectures, and secure computation technologies, ensuring that edge AI systems can meet the performance, energy, and security requirements of modern applications.}

\subsection{Heterogeneity}

\textcolor{black}{In edge AI environments, heterogeneity refers to the diverse nature of data, devices, and communication networks across edge nodes. This diversity introduces challenges in federated learning, where edge devices typically possess non-identically distributed (non-IID) data, varied computational capacities, and inconsistent communication bandwidths. As traditional FL approaches often rely on a single global model, they may struggle to capture the diverse patterns and distributions present across different edge devices. To address these challenges, multi-prototype-based federated learning \cite{Qiao2023MP-FedCL:}  has emerged as a promising approach for enhancing model inference by leveraging multiple weighted prototypes rather than relying on a single prototype, which can be incomplete and ambiguous \cite{10048999}.}

\textcolor{black}{A key aspect of this approach involves calculating local prototypes at each edge device, ensuring that the diverse distributions of client data are effectively represented. These prototypes capture the unique characteristics of each client’s data distribution, enabling a more nuanced aggregation process during global model updates. Clustering algorithms like \textbf{k-means} \cite{Qiao2023MP-FedCL:} can be employed locally at each edge device to generate multiple prototypes that correspond to different data clusters within the client’s dataset. By calculating prototypes that reflect the underlying structure of local data, this approach enhances the model’s ability to generalize across heterogeneous client distributions. Furthermore, these multiple weighted prototypes provide a richer representation of client data compared to traditional single-prototype methods, which often oversimplify local data distributions.}

\textcolor{black}{Once local prototypes are calculated, these prototypes are aggregated across devices during the global model update process. Rather than averaging model updates from each device, as in traditional FL approaches, multi-prototype-based FL aggregates the prototypes in a weighted manner, where each prototype’s contribution is proportional to the importance of the corresponding data cluster. This approach improves robustness against non-IID \cite{Zhong2022Optimizing} data distributions by ensuring that the global model is not overly influenced by outliers or overrepresented data points. Instead, the weighted aggregation process captures the full diversity of data across the edge devices, leading to higher test accuracy and better generalization across all devices.}

\textcolor{black}{An important challenge in this approach is ensuring communication efficiency, especially in bandwidth-constrained edge environments. By reducing the need to transmit the entire local model or large datasets, multi-prototype-based FL \cite{Mu2021FedProc:} minimizes communication overhead. Instead of sharing raw data or full model updates, devices only transmit the prototypes and their associated weights. This reduces the amount of information exchanged between the server and the devices, while still enabling effective global model updates. \textbf{Quantization} techniques can also be applied to further reduce the size of transmitted prototypes, allowing for more efficient communication without compromising model performance.}

\textcolor{black}{In addition to improving communication efficiency, multi-prototype-based FL demonstrates significant improvements in both \textbf{accuracy} and \textbf{convergence rates}. The ability to capture more representative patterns from local data allows the global model \cite{Mu2021FedProc:} to converge faster, particularly in heterogeneous environments where traditional FL methods tend to suffer from slow convergence due to non-IID data. The richer and more representative model built through multi-prototype aggregation helps the system achieve higher accuracy in a shorter period, reducing the number of communication rounds required to achieve an optimal model.}

This approach demonstrates significant improvements in accuracy and convergence rates, making it a promising direction for handling heterogeneity in edge AI. \textcolor{black}{The multi-prototype strategy opens new avenues for edge AI applications where data heterogeneity is a major bottleneck, such as in \textbf{personalized healthcare}, where patient data varies significantly across different locations, or in \textbf{smart city infrastructures}, where sensor data may differ drastically based on environmental conditions. As future research progresses, integrating other advanced clustering techniques like \textbf{Gaussian Mixture Models (GMMs)} \cite{Jiao2020EGMM:} or \textbf{spectral clustering} could further enhance the capability of multi-prototype FL systems, enabling even better handling of non-IID data and improving performance in highly heterogeneous environments.}

\subsection{Security}

\textcolor{black}{As edge AI becomes more prevalent, the security of edge devices, data, and communications becomes a critical concern, particularly with the growing threat of quantum computing, which can break classical encryption methods. One of the primary directions for securing edge AI systems is the integration of \textbf{AI-based quantum-safe cybersecurity automation}. Quantum-safe cryptographic techniques are essential for protecting sensitive data from the computational power of quantum computers, which could easily break traditional public-key encryption systems. Algorithms such as \textbf{lattice-based cryptography}, \textbf{hash-based signatures}, and \textbf{code-based cryptography} are being researched as quantum-resistant alternatives that can secure edge AI devices and communications against quantum attacks \cite{Hummelholm2023AIbasedQC}. These quantum-safe solutions ensure that even as quantum computing capabilities advance, edge AI systems remain resilient against cryptographic threats.}

Improving device and sensor security is another critical focus area. \textcolor{black}{Edge devices, by their nature, are distributed and often deployed in insecure environments, making them vulnerable to physical tampering and cyberattacks. Integrating AI-based security mechanisms that can detect abnormal behavior at the device level is essential for enhancing the security of edge networks. For example, \textbf{deep learning-based intrusion detection systems (IDS)} can be implemented at the edge to monitor incoming traffic for potential threats, such as denial-of-service (DoS) attacks or unauthorized access attempts. These IDS systems can utilize \textbf{anomaly detection algorithms}, such as autoencoders or generative adversarial networks (GANs), to identify deviations from normal traffic patterns and flag potential security breaches in real time. Such AI-driven systems can adapt over time, learning from new attack vectors and updating their detection models to address emerging threats. Additionally, lightweight \textbf{blockchain-based solutions} can be integrated to ensure the secure exchange of data between edge devices by creating immutable records of transactions, further reducing the risk of tampering \cite{samriya2023secured}.}

\textcolor{black}{In the context of quantum-safe solutions, edge AI systems must also adopt \textbf{post-quantum cryptographic algorithms} to ensure long-term data security. Post-quantum cryptography (PQC) focuses on developing encryption algorithms that are resistant to both classical and quantum attacks. Integrating PQC into edge AI devices ensures that secure communications are maintained even when quantum computers become widely available. Furthermore, using AI to optimize the implementation of PQC algorithms \cite{C2022Analysis}, such as by reducing their computational overhead, can make these solutions more practical for deployment in resource-constrained edge environments. Research is also focusing on \textbf{quantum key distribution (QKD)} \cite{Wang2020Experimental}, a technique that leverages quantum mechanics to generate provably secure cryptographic keys. QKD can be used to secure communication between edge devices and central servers, ensuring that keys cannot be intercepted or tampered with, even by quantum adversaries.}

The research will focus on developing scalable and efficient cybersecurity systems, including AI-driven automation for threat detection and mitigation and using blockchain for secure communications \cite{samriya2023secured}. \textcolor{black}{A key challenge in securing edge AI is the need for \textbf{scalability}. As the number of connected devices in edge networks increases, cybersecurity solutions must be able to scale to protect millions of devices without introducing significant latency or computational overhead. AI-driven security systems can help address this challenge by automating threat detection and response processes. For example, machine learning models can be trained to detect anomalies in device behavior or network traffic, identifying potential cyberattacks before they can cause damage. In addition to intrusion detection, AI can also automate \textbf{patch management} by identifying vulnerabilities in edge devices and applying security updates in real time, ensuring that devices remain protected against known threats.}

Additionally, creating robust test environments for cybersecurity validation will ensure the effectiveness of these solutions in diverse operational scenarios \cite{yang2022federated}. \textcolor{black}{Test environments, such as \textbf{digital twins} of edge networks, can be used to simulate cyberattacks and validate the effectiveness of AI-driven security solutions. By replicating real-world conditions, these environments enable researchers to fine-tune their algorithms and improve the resilience of edge AI systems. For example, by simulating distributed denial-of-service (DDoS) attacks on a digital twin of an edge network \cite{Yigit2022Digital}, AI-based security systems can be stress-tested and adjusted to ensure their ability to respond to large-scale cyber threats in real time. Furthermore, AI-based systems can be used to predict potential vulnerabilities in edge networks by analyzing historical data and identifying patterns that could lead to future security breaches. This proactive approach to security helps to mitigate risks before they can be exploited by attackers.}

These developments aim to provide a comprehensive security framework for future edge AI systems, ensuring resilience against evolving cyber threats. \textcolor{black}{The future of edge AI security lies in the combination of quantum-safe cryptography \cite{Ahmad2021Enhancing}, AI-driven threat detection, and automation. These technologies will allow edge AI systems to remain secure in the face of both classical and quantum-based cyber threats, ensuring that they can operate reliably in increasingly complex and hostile environments. As the number of connected devices in smart cities, autonomous vehicles, and industrial IoT grows, maintaining robust security at the edge will be critical to safeguarding sensitive data and ensuring the integrity of AI-driven processes.}

\subsection{Privacy}
Privacy enhancement in early health prediction through federated learning would be another interesting area to investigate \cite{fi15110370}. Future directions in this field involve developing more advanced privacy-preserving techniques within federated learning frameworks to keep patient data secure during model training. Improvements in differential privacy and homomorphic encryption are essential for protecting sensitive health information \cite{gill2024ITL}. Additionally, optimizing the communication efficiency between edge devices and central servers will mitigate privacy risks associated with data transmission. Integrating privacy-conscious AI models with real-time health monitoring systems, like wearable devices, can deliver immediate and secure health insights. Collaboration among healthcare providers, AI researchers, and policymakers is vital to creating standardized privacy protocols. Future research should also focus on scalable and adaptive federated learning methods capable of handling diverse and large-scale health data while maintaining high privacy standards.

\subsection{6G and beyond}
In the context of 6G and beyond, Edge AI is set to make several significant advancements; utilizing the ultra-low latency and high bandwidth of 6G networks will improve the deployment of AI models at the edge, facilitating real-time applications such as autonomous vehicles and smart cities \cite{duan2023combining}. The research will prioritize optimizing AI algorithms to meet the requirements of 6G, including dynamic resource allocation and energy efficiency. Combining quantum ML with 6G will enable more complex computations at the edge, enhancing predictive accuracy and decision-making processes. Additionally, improvements in secure edge computing and blockchain technology will address data privacy and security issues, ensuring robust and reliable edge AI systems. These advancements will collectively enhance the performance, scalability, and security of edge AI applications in a 6G environment \cite{cloudAIBus2024}.

\section{Summary and Conclusions}\label{sec:summary}

The systematic review analysis of Edge AI provides a comprehensive overview of the present status of research in edge intelligence and its applications. The significance of these findings lies in the need for Edge AI systems to consider infrastructure, resource management, and the scale of ML models. According to the findings of the study, it is essential to conduct a thorough examination of both the positive and negative aspects of prior research to identify any potential research gaps and to estimate prospective developments and concerns. 

The study emphasises the importance of using a systematic approach to record and evaluates the existing research in the field of Edge AI. Moreover, it highlights the importance of implementing a standardized procedure to reduce the possible impact of discrepancies in the study. The results of this review have the capacity to ignite a new field of investigation in Edge AI and offer direction for prospective research in this field. 
The exhaustive examination of Edge AI offers profound insight into the most current study on edge intelligence and its realistic implementations. Moreover, this emphasises the importance of gaining additional understanding about the key factors that govern the choice of Edge AI infrastructure, as well as the effect of the scale of the model used for ML on efficiency and resource allocation.

To summarize, the comprehensive study on Edge AI is to fully evaluate the various AI methodologies. This study integrates all the feasible methodologies incorporated in edge intelligence or AI at the edge. This review examines the crucial factors that impact the choice of Edge AI infrastructure, such as Cloud, Fog, and Edge computing, and assesses their impact on application efficacy and resource utilization. Furthermore, it investigates the influence of the size of an ML model on the efficacy and resource usage of an Edge AI application. A total of 78 studies have been chosen for this evaluation due to their specific focus on the application of AI in edge computing. In order to enhance our comprehension in the context of AI applied to edge computing, these studies were categorized into multiple domains, including infrastructure, resource management, and ML model sizing, among others. Resource supply, allocation, scheduling, and job deployment are crucial considerations in the field of Edge AI.

\subsection{Open Challenges}

\textcolor{black}{Following facts can be further concluded to improve this survey:}

\begin{itemize}
    
\item \textcolor{black}{EdgeAI Infrastructure Optimization: Future studies can examine Edge, Fog and Cloud systems that can be integrated into EdgeAI to create a hybrid model. In this way, scalable infrastructure solutions for EdgeAI systems can be discussed in detail. The research can focus on optimization techniques for resource allocation and latency in systems with varying workloads.}

\item \textcolor{black}{Security and Privacy in EdgeAI: Considering the heterogeneous structure of the nodes that make up EdgeAI systems, security and privacy issues arise for sensitive data (biometric data). Future research can examine the measures and technical methods taken to ensure the security and privacy of data.}

\item \textcolor{black}{EdgeAI Applications: Future studies can examine real-world EdgeAI applications such as smart cities and IoT-based healthcare systems. In this way, application challenges and solutions can be provided for researchers.}

\end{itemize}

\section*{Declarations}
\begin{itemize}
\item Funding \\
M. Golec is supported by the Ministry of Education of the Turkish Republic. B. Ali is supported by the Ph.D. Scholarship at the Queen Mary University of London. H. Wu is supported by the National Natural Science Foundation of China (No. 62071327) and Tianjin Science and Technology Planning Project (No. 22ZYYYJC00020). F. Cuadrado has been supported by the HE ACES project (Grant No. 101093126). M. Xu is supported by the National Natural Science Foundation of China (No. 62102408),  Guangdong Basic and Applied Basic Research Foundation (No. 2024A1515010251), Shenzhen Industrial Application Projects of undertaking the National key R \& D Program of China (No. CJGJZD20210408091600002).
\item Conflict of interest/Competing interests \\
On behalf of all authors, the corresponding author states that there is no conflict of interest.
\item Ethics approval 
\\
Not Available 
\item Consent to participate
\\
Not Available 
\item Consent for publication
\\
Not Available 
\item Availability of data and materials
\\
Not Available 
\item Code availability 
\\
Not Available 

\item Authors' contributions\\
\textbf{Sukhpal Singh Gill} (Conceptualization: Lead; Data curation: Lead; Formal analysis: Lead;  Investigation: Lead; Methodology: Lead; Validation: Lead; Writing – original draft: Lead)
\textbf{Muhammed Golec} (Conceptualization: Lead; Data curation: Lead; Formal analysis: Lead;  Investigation: Lead; Methodology: Lead; Validation: Lead; Writing – original draft: Lead)
\textbf{Jianmin Hu} (Conceptualization: Lead; Data curation: Lead; Formal analysis: Lead;  Investigation: Lead; Methodology: Lead; Validation: Lead; Writing – original draft: Lead)
\textbf{Minxian Xu} (Conceptualization: Lead; Data curation: Lead; Formal analysis: Lead;  Investigation: Lead; Methodology: Lead; Validation: Lead; Writing – original draft: Lead)
\textbf{Junhui Du} (Conceptualization: Lead; Data curation: Lead; Formal analysis: Lead;  Investigation: Lead; Methodology: Lead; Validation: Lead; Writing – original draft: Lead)
\textbf{Huaming Wu} (Conceptualization: Lead; Data curation: Lead; Formal analysis: Lead;  Investigation: Lead; Methodology: Lead; Validation: Lead; Writing – original draft: Lead)
\textbf{Guneet Kaur Walia} (Conceptualization: Lead; Data curation: Lead; Formal analysis: Lead;  Investigation: Lead; Methodology: Lead; Software: Lead; Validation: Lead; Writing – original draft: Lead)
\textbf{Subramaniam Subramanian Murugesan} (Conceptualization: Lead; Data curation: Lead; Formal analysis: Lead;  Investigation: Lead; Methodology: Lead; Validation: Lead; Writing – original draft: Lead)
\textbf{Babar Ali} (Conceptualization: Lead; Data curation: Lead; Formal analysis: Lead;  Writing – original draft: Lead)
\textbf{Mohit Kumar} (Conceptualization: Lead; Data curation: Lead; Formal analysis: Lead;  Investigation: Lead; Methodology: Lead; Validation: Lead; Writing – original draft: Lead)
\textbf{Kejiang Ye} (Conceptualization: Lead; Data curation: Lead; Formal analysis: Lead;  Investigation: Lead; Methodology: Lead; Validation: Lead; Writing – original draft: Lead)
\textbf{Prabal Verma} (Conceptualization: Lead; Data curation: Lead; Formal analysis: Lead;  Investigation: Lead; Methodology: Lead; Validation: Lead; Writing – original draft: Lead)
\textbf{Surendra Kumar} (Conceptualization: Lead; Data curation: Lead; Formal analysis: Lead;  Investigation: Lead; Methodology: Lead; Validation: Lead; Writing – original draft: Lead)
\textbf{Felix Cuadrado} (Conceptualization: Lead; Data curation: Lead; Formal analysis: Lead;  Investigation: Lead; Methodology: Lead; Validation: Lead; Writing – original draft: Lead)
\textbf{Steve Uhlig} (Conceptualization: Lead; Data curation: Lead; Formal analysis: Lead;  Investigation: Lead; Methodology: Lead; Validation: Lead; Writing – original draft: Lead)

\end{itemize}

\bibliographystyle{IEEEtran}
\bibliography{references}

\begin{IEEEbiography}[{\includegraphics[width=1in,height=1.15in,clip]{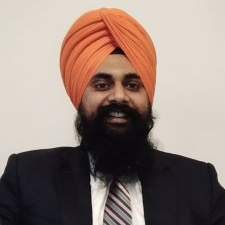}}]{Sukhpal Singh Gill}  (FHEA) is an Assistant Professor of Cloud Computing at the School of Electronic Engineering and Computer Science, Queen Mary University of London, UK. Dr. Gill is serving as an Editor-in-Chief for IGI Global IJAEC and Area Editor for Springer Cluster Computing Journal, also serving as an Associate Editor in IEEE IoT, Elsevier IoT, Wiley SPE, Wiley ETT and IET Networks Journals. He has co-authored 200+ peer-reviewed papers (with Citations 10100+ and H-index 51) and has published in prominent international journals and conferences such as IEEE TCC, IEEE TSC, IEEE TSUSC, IEEE COMST, IEEE TCE, ACM TOIT, IEEE TII, IEEE TNSM, IEEE IoT Journal, Elsevier JSS/FGCS, IEEE/ACM UCC and IEEE CCGRID. He has received several awards, including the Queen Mary University Education Excellence Award 2023, Outstanding Reviewer Award from IEEE IT Professional Magazine 2024, Elsevier Internet of Things Editor’s Choice Award 2024, Elsevier Best Paper Award 2023, Distinguished Reviewer Award from SPE (Wiley), Best Paper Award AusPDC at ACSW 2021. He has edited and authored research various books for Elsevier, Springer and CRC Press. His research interests include Cloud Computing, Edge Computing, IoT and Energy Efficiency. For further information, please visit: \url{http://www.ssgill.me}.
\end{IEEEbiography}

\vspace{-0.55in}
 \begin{IEEEbiography}[{\includegraphics[width=1.0in,trim=0in 0in 0in 0in,clip,keepaspectratio]{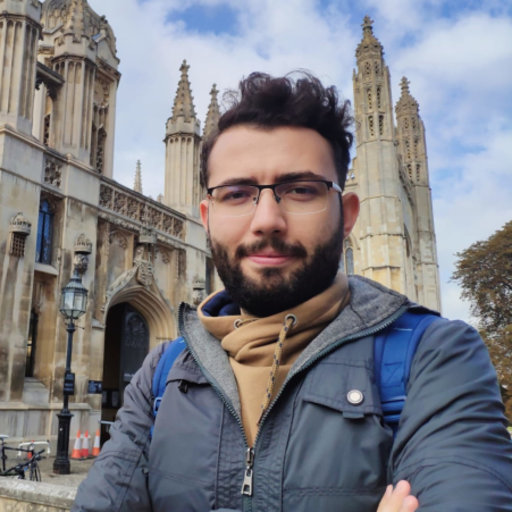}}]{Muhammed Golec} is a PhD student in Computer Science at Queen Mary University of London (QMUL). Earlier, he graduated from QMUL in MSC Computer Science (Distinction) through the Ministry of Education Scholarship. He has published articles in prominent journals and conferences such as IEEE TII and Elsevier IoT, IEEE Consumer Electronics Magazine, and IEEE CCGRID. His research interests include Cloud Computing, Serverless Computing, AI, and Security and Privacy. 
\end{IEEEbiography}
\vspace{-0.6in}
\begin{IEEEbiography}[{\includegraphics[width=1in,height=1.15in,clip]{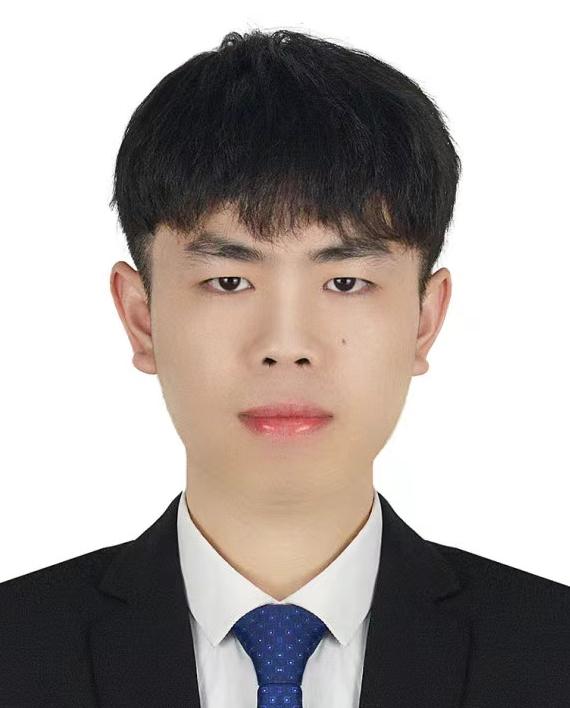}}]{Jianmin Hu}  
is currently a master's student at Shenzhen Advanced Technology Research Institute. He obtained a Bachelor's degree in Software Engineering from Harbin Institute of Technology in 2023. His current research direction is cloud native and system for large language model.
\end{IEEEbiography}
\vspace{-0.6in}
\begin{IEEEbiography}[{\includegraphics[width=1in,height=1.25in,clip]{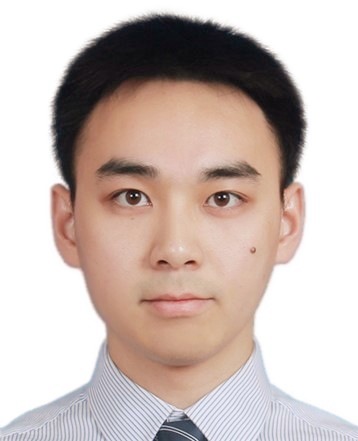}}]{Minxian Xu} (Senior Member, IEEE) is currently an Associate Professor at Shenzhen Institute of Advanced Technology, Chinese Academy of Sciences. He received the BSc degree in 2012 and the MSc degree in 2015, both in software engineering from University of Electronic Science and Technology of China. He obtained his Ph.D. degree from the University of Melbourne in 2019. His research interests include resource scheduling and optimization in cloud computing. He has co-authored 70+ peer-reviewed papers published in prominent international journals and conferences, such as ACM CSUR, IEEE TSC, IEEE TMC, ACM TOIT, ACM TAAS and ICSOC. He was awarded the 2023 IEEE TCSC Award for Excellence (Early Career Award). More information can be found at: minxianxu.info.
\end{IEEEbiography}
\vspace{-0.5in}
\begin{IEEEbiography}[{\includegraphics[width=1in,clip]{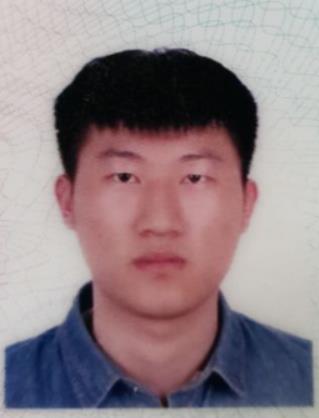}}]{Junhui Du} received the BSc degree in mathematics
from the Nanjing University of Information Science
Technology, China, in 2021. He is currently working
toward his doctor's degree in mathematics at the
Center for Applied Mathematics, Tianjin University,
Tianjin, China. His research interests include Internet
of Things, deep learning, and mobile edge computing.
\end{IEEEbiography}

\vspace{-0.5in}
\begin{IEEEbiography}[{\includegraphics[width=1.0in,trim=0in 0in 0in 0in,clip]{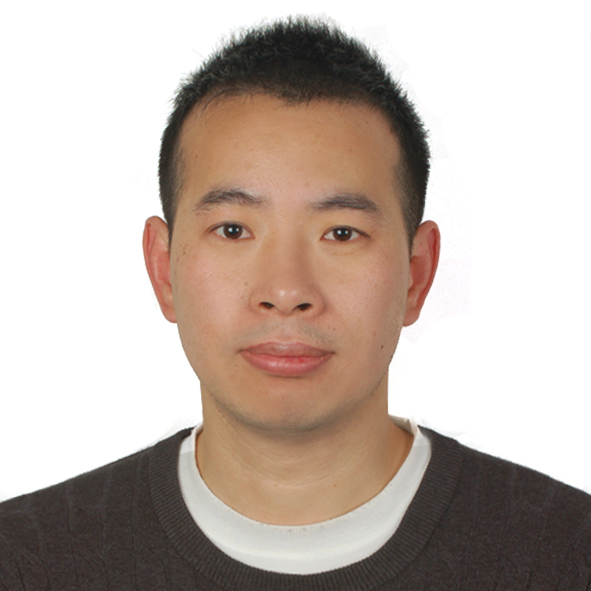}}]{Huaming Wu} (Senior Member, IEEE) received the B.E. and M.S. degrees from Harbin Institute of Technology, China in 2009 and 2011, respectively, both in electrical engineering. He received the Ph.D. degree with the highest honor in computer science at Freie Universit\"at Berlin, Germany in 2015. He is currently a Professor at the Center for Applied Mathematics, Tianjin University, China. His research interests include mobile cloud computing, edge computing, Internet of Things, deep learning, complex networks, and DNA storage.
\end{IEEEbiography}  

\vspace{-0.6in}
\begin{IEEEbiography}[{\includegraphics[width=1in,height=1.15in,clip]{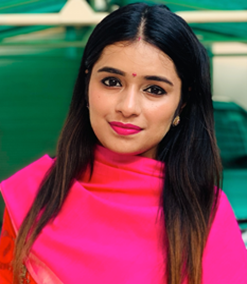}}]{Guneet Kaur Walia} is a Ph.D scholar at the Department of Information Technology, Dr. B. R. Ambedkar National Institute of Technology, Jalandhar, Punjab, India. She successfully completed her Masters in Computer Science Engineering at Punjab Agricultural University, Ludhiana, Punjab, in 2016. Her dedication and enthusiasm for exploring various cutting-edge technologies make her a passionate researcher. She has published in prominent international journals including IEEE TCE and IEEE COMST. Her research interests include Cloud Computing, Edge Computing, Internet of Things (IoT), Resource Management in Edge Computing, and Artificial Intelligence (AI). 
\end{IEEEbiography}

\vspace{-0.6in}
\begin{IEEEbiography}[{\includegraphics[width=1in,height=1.15in,clip]{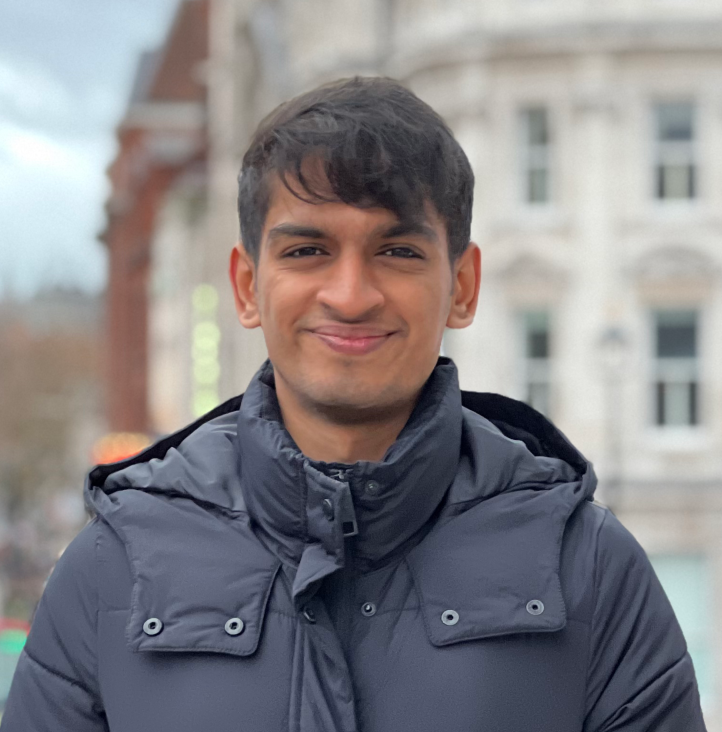}}]{Subramaniam Subramanian Murugesan} is a PhD student in Electronic Engineering at Queen Mary University of London (QMUL), funded by a UKRI EPSRC Doctoral Training Partnership (DTP) studentship. He holds a master's degree in Big Data Science from QMUL. His research focuses on RF emission detection, identification, and localization using Software-Defined Radio (SDR), advanced antennas, AI/ML/DL applications, Cloud \& IoT, software engineering, and edge AI technologies. He has published work in journals such as the IEEE Journal of Biomedical and Health Informatics (JBHI), Cluster Computing (Springer).

\end{IEEEbiography}
\vspace{-0.6in}
\begin{IEEEbiography}[{\includegraphics[width=1in,height=1.15in,clip]{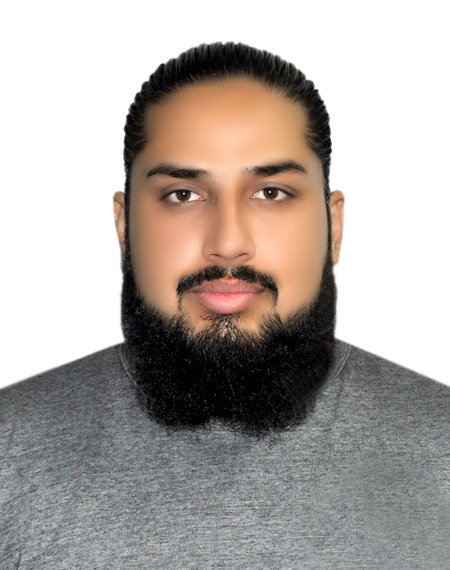}}]{Babar Ali} is a PhD student in the School of Electronic Engineering and Computer Science at Queen Mary University of London (QMUL). He has published PhD findings in journals such as Wiley International Journal of Network Management and Elsevier Internet of Things. His research interests include  Cloud Computing, Fog Computing, IoT, Edge Computing, and Wireless Sensor Networks.
\end{IEEEbiography}
\vspace{-0.6in}
\begin{IEEEbiography}[{\includegraphics[width=1in,height=1.15in,clip]{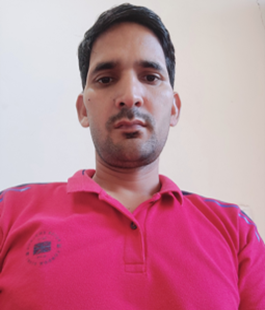}}]{Mohit Kumar} is Assistant Professor in the Department of Information Technology at Dr. B R Ambedkar National Institute of Technology, Jalandhar, India. He received his Ph.D. degree from Indian Institute of Technology Roorkee in the field of Cloud Computing, 2018, and M. Tech degree in Computer Science and Engineering from ABV-Indian Institute of Information Technology Gwalior, India in 2013.  His research topics cover the areas of Cloud computing, Fog/ Edge Computing, Internet of Things, federated learning, Blockchain, and Artificial Intelligence. He has published more than 40 research articles in reputed journals, IEEE Transactions and international conferences. He has been Session chair and keynotes Speaker of many International conferences, webinars, FDP, STC in India. He has guided six M. Tech Thesis and supervising Ph.D. Scholar. He is an active reviewer of several reputed journals and international conferences. He is a member of the IEEE.
\end{IEEEbiography}

\vspace{-0.5in}
 \begin{IEEEbiography}[{\includegraphics[width=1.0in,trim=0in 0in 0in 0in,clip,keepaspectratio]{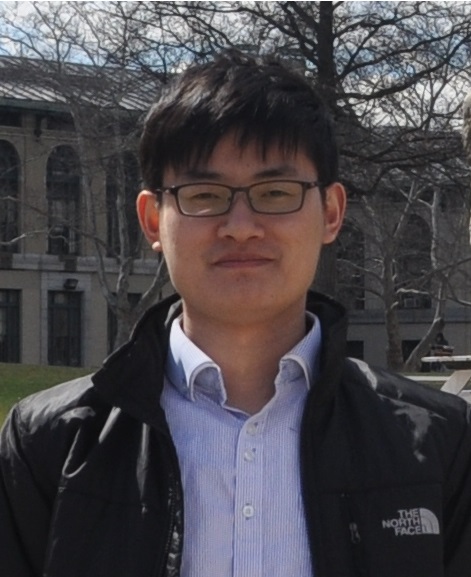}}]{Kejiang Ye} is currently a Professor and the Director of the Research Center for Cloud Computing, Shenzhen Institute of Advanced Technology (SIAT), Chinese Academy of Sciences (CAS). He received his B.S. and Ph.D degrees both from Zhejiang University and was a Post Doctoral Research Associate at Carnegie Mellon University (CMU). His research interests include Digital Technology and Systems (e.g., Cloud Computing, Big Data and Industrial Internet). He is a Senior Member of IEEE and a Distinguished Member of China Computer Federation (CCF).
\end{IEEEbiography}

\vspace{-0.6in}
\begin{IEEEbiography}[{\includegraphics[width=1in,height=1.15in,clip]{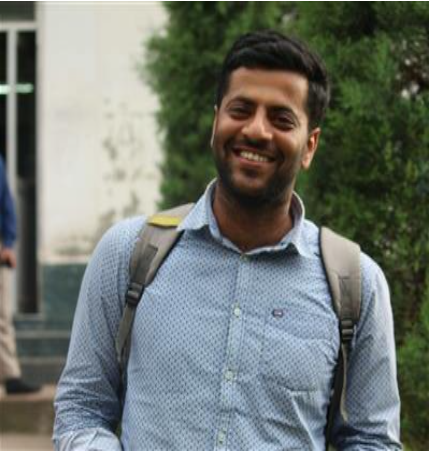}}]{Prabal Verma} is currently working as Assistant Professor in Information Technology Department, National Institute of Technology (NIT), Srinagar, Jammu and Kashmir India. He also worked as an Assistant Professor in the department of Computer Science and Engineering at Thapar Institute of Technology, Patiala. He did his Doctoral degree in Computer Science and Engineering from Guru Nanak Dev University, Amritsar. His work is published in highly reputed publishers like IEEE, Elsevier, Wiley, Taylor and Francis, and Springer. His current working research areas include Internet of Things (IoT) in Healthcare, Big Data and Fog-Cloud computing. He is also an IEEE Member.
\end{IEEEbiography}
\vspace{-0.6in}
\begin{IEEEbiography}[{\includegraphics[width=1in,height=1.15in,clip]{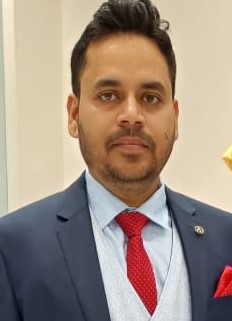}}]{ Surendra Kumar }  is an Assistant Professor at GLA University in Mathura, Uttar Pradesh, India. He has a strong academic background, having received his Master of Computer Application (M. C. A.) from Babasaheb Bhimrao Ambedkar University (A Central University), Lucknow, Uttar Pradesh, India in 2013. He continued his academic pursuits at the same university, earning his Doctor of Philosophy in the Department of Computer Science. He was recently recognized with an International Distinguished Young Researcher Award for 2021-22 from the International Institute of Organized Research. Dr. Kumar has also published several research articles in reputed journals, conferences, and book chapters. Dr. Kumar's academic achievements and research contributions demonstrate his dedication to his field of study and his commitment to advancing knowledge and innovation in the areas of resource management, cloud security, cryptography, and distributed computing, blockchain technology.
\end{IEEEbiography}

\vspace{-0.5in}
 \begin{IEEEbiography}[{\includegraphics[width=1.0in,trim=0in 0in 0in 0in,clip,keepaspectratio]{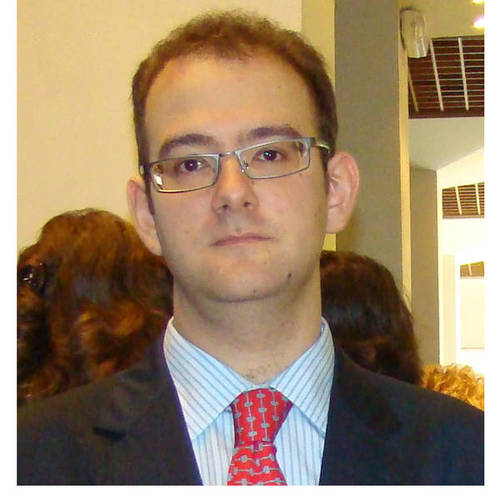}}]{Felix Cuadrado} received a Ph.D. degree in telecommunications engineering from the Universidad Politécnica de Madrid (UPM), Spain, in 2009. He is currently a Senior Distinguished Fellow (Beatriz Galindo scheme) with the Universidad Politécnica de Madrid, a Visiting Reader at the Queen Mary University of London, and a fellow of the Alan Turing Institute. He has numerous publications in top-tier journals and conferences, including IEEE TSC, IEEE TCC, Elsevier JSS, Elsevier FCGS, IEEE ICDCS, and WWW. His research explores the challenges arising from large-scale data-intensive applications through a combination of software engineering, distributed systems, and mathematical approaches.
\end{IEEEbiography}

\vspace{-0.5in}
 \begin{IEEEbiography}[{\includegraphics[width=1.0in,trim=0in 0.2in 0in 0.15in,clip,keepaspectratio]{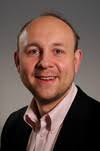}}]{Steve Uhlig} obtained a Ph.D. degree in Applied Sciences from the University of Louvain, Belgium, in 2004. Prior to joining Queen Mary, he was a Senior Research Scientist with Technische Universität Berlin/Deutsche Telekom Laboratories, Berlin, Germany. Starting in January 2012, he is the Professor of Networks and Head of the Networks Research group at Queen Mary, University of London. Between 2012 and 2016, he was a guest professor at the Institute of Computing Technology, Chinese Academy of Sciences, Beijing, China. Current Research interests: Internet measurements, software-defined networking, content delivery.
\end{IEEEbiography}

\end{document}